\documentclass[aps,nofootinbib,longbibliography,superscriptaddress,
tightenlines,notitlepage]{revtex4-1}

\usepackage{amsmath,amsthm,latexsym,amssymb,amsfonts}
\usepackage{graphicx,color,natbib}
\usepackage[export]{adjustbox}
\usepackage[colorlinks=true]{hyperref}
\usepackage{tikz}
\usepackage[normalem]{ulem}
\usetikzlibrary{patterns}

\def\be{\begin{equation}}
\def\ee{\end{equation}}
\def\ba{\begin{eqnarray}}
\def\ea{\end{eqnarray}}

\def\SU{\text{SU}}

\newcommand\nn{\nonumber}
\newcommand{\q}{\quad}
\definecolor{darkgreen}{rgb}{0.0, 0.4, 0.1}

\begin{document}

\title{Tensor network renormalization with fusion charges:\\ applications to 3d lattice gauge theory}

\author{William J. Cunningham}
\email{wcunningham@perimeterinstitute.ca}
\affiliation{Perimeter Institute for Theoretical Physics,\\ 31 Caroline Street North, Waterloo, Ontario, Canada N2L 2Y5}
\author{Bianca Dittrich}
\email{bdittrich@perimeterinstitute.ca}
\affiliation{Perimeter Institute for Theoretical Physics,\\ 31 Caroline Street North, Waterloo, Ontario, Canada N2L 2Y5}
\author{Sebastian Steinhaus}
\email{sebastian.steinhaus@uni-jena.de}
\affiliation{Theoretisch-Physikalisches Institut, Friedrich-Schiller-Universit\"at Jena,\\ Max-Wien-Platz 1, 07743 Jena, Germany}

\begin{abstract}
Tensor network methods are powerful and efficient tools to study the properties and dynamics of statistical and quantum systems, in particular in one and two dimensions. In recent years, these methods were applied to lattice gauge theories, yet these theories remain a challenge in $(2+1)$ dimensions. In this article, we present a new (decorated) tensor network algorithm,  in which the tensors encode the lattice gauge amplitude expressed in the fusion basis.
 This has several advantages: Firstly, the fusion basis does diagonalize operators measuring the magnetic fluxes and electric charges associated to a hierarchical set of regions. The algorithm allows therefore a direct access to these observables. Secondly the fusion basis is, as opposed to the previously employed spin network basis, stable under coarse graining. Thirdly, due to the hierarchical structure of the fusion basis, the algorithm does implement predefined disentangles, that remove short-scale entanglement.

 We apply this new algorithm to lattice gauge theories defined for the quantum group $\text{SU}(2)_{\rm k}$ and identify a weak and a strong coupling phase for various levels $\rm k$. As we increase the level $\rm k$, the critical coupling $g_c$ decreases linearly, suggesting the absence of a deconfining phase  for the continuous group $\text{SU}(2)$. Moreover, we illustrate the scaling behaviour of the Wilson loops in the two phases.
\end{abstract}

\maketitle

\newpage

\tableofcontents

\newpage

\section{Introduction}

Tensor network algorithms~\cite{Levin,GuWen,VidalEntFiltering,HigherOrder,DecTNW,OrusReview} have been successfully employed to study the phase diagram of quantum and classical statistical models, in particular for two--dimensional systems.  These algorithms are especially valuable  for systems to which Monte Carlo methods are not efficient, e.g. due to sign problems.  This includes QCD models, but also quantum gravity models such as spin foams~\cite{ AlexReview,finiteSF,Eckert,Asaduzzaman:2019mtx}, which are based on complex (not Wick rotated) amplitudes. Similarly anyon systems~\cite{Burnell} can feature complex amplitudes.

In this work we present a tensor network renormalization algorithm applicable to three--dimensional lattice gauge systems, three-dimensional quantum gravity models~\cite{DittrichLambda}, as well as the study of anyon condensation.

As opposed to previous work~\cite{DecTNW, Milsted, DelcampSF, Kuramashi:2018mmi, Banuls:2019rao}\footnote{See e.g.~\cite{Tagliacozzo:2014bta,Milsted} and the review~\cite{Banuls:2019rao} for applications of tensor networks in the Hamiltonian descriptions of $(2+1)$--dimensional lattice gauge systems, and e.g. \cite{Pichler:2015yqa,Shimizu:2017onf,Funcke:2019zna,Asaduzzaman:2019mtx} and the review \cite{Banuls:2019rao} and references therein for studies of $(1+1)$--dimensional systems with gauge symmetry.} on coarse-graining algorithms for 3D lattice gauge systems, here we introduce several new features.

First, instead of working with finite groups or cut-offs of Lie groups,  we work with a quantum deformation of $\SU(2)$ known as $\SU(2)_{\rm k}$.  This allows us to work with  finite systems with exact gauge symmetry, while at the same time allowing for a systematic approximation to the undeformed case $\SU(2)$, which is reached for ${\rm k}\rightarrow \infty$.  See~\cite{TopIntertwiner,QgroupSpinnets,QGIntertwiners} for a similar strategy employed for two--dimensional systems. In this work we test the new algorithm for three--dimensional lattice gauge systems with $\SU(2)_{\rm k}$ symmetry and extract the critical couplings for different levels ${\rm k}$. Although in this work we perform simulations only for systems with relatively small ${\rm k}$, these initial investigations indicate that the critical couplings approach zero for growing ${\rm k}$ in a surprisingly fast way.

Second, the tensor network algorithm presented here employs a new (gauge invariant) basis, the fusion basis, which is ideally suited for coarse-graining~\cite{ExTQFT,DelcampFusion}.  The reason for this is that a fusion basis diagonalizes observables (the ribbon operators) which are only quasi-local, that is, associated to a set of regions.  This set is partially ordered by the inclusion relation, which on the lattice translates into a coarse-graining scheme for the plaquettes of the lattice. Different choices for the fusion basis lead to different coarse-graining schemes.

This brings as to the third new feature: transformations between different such choices  for the fusion basis will be an important part of the algorithm presented herein. These transformations reorganize the regions into which the finer degrees of freedom are blocked, and thus can be seen to function as disentanglers~\cite{MERA,MERAPartitionF, DelcampMizera}. In other words, they remove short scale entanglement, at least for the truncation we consider.

Additionally, the use of the decorated tensor network algorithm combined with the fusion basis allows us to keep track of the coarse-graining behaviour of ribbon observables. These combine Wilson loops, which measure the magnetic flux\footnote{We will refer to this magnetic flux also as magnetic charge. The reason is that we will use a description which unifies magnetic flux and electric charge into a so-called dyonic charge, and we will refer to both magnetic and electric components as charge.} through the region surrounded by the loop, and `t Hooft operators, which integrate over the electric flux through the loop, and thus are measuring the electric charge in the region surrounded by the loop. For lattice gauge theories the Wilson loops serve as order parameters, whereas, as we will explain in more detail in the course of this paper, the electric loops allow us to monitor the appearance of electrical charges under coarse-graining.

%%%%%%%%%%%%%%%%%%
%%%%%%%%%%

\vspace{2mm}

The development of tensor network coarse-graining algorithms for three--dimensional gauge systems faces several challenges.  These algorithms are much more computationally demanding for three-dimensional systems as opposed to two-dimensional systems. Therefore, for systems with gauge symmetries it is important to design algorithms which only include the gauge invariant, and therefore physical, degrees of freedom. However, these gauge invariant degrees of freedom cannot be localized to (lattice) sites, as is the case for systems described by standard tensor networks.  The work~\cite{DecTNW}, co-written by two of the current authors, introduced a generalization of tensor networks called decorated tensor networks. This generalization permits the definition of (decorated) tensor network coarse-graining algorithms for Abelian~\cite{DecTNW} and non-Abelian~\cite{DelcampSF} lattice gauge theories and finite group analogues~\cite{finiteSF} of 3D spin foam partition functions~\cite{DelcampSF}, which were successfully tested for systems with finite gauge groups.  These algorithms were based on the spin network basis, which provides a basis for the gauge invariant degrees of freedom for a lattice gauge system. Violations of gauge invariance, according to the Gau\ss~law, lead to electrical charges. Yet, for non-Abelian gauge theories, finite regions can feature electrical charge without featuring a violation of gauge invariance.
For non-Abelian lattice gauge systems we can have gauge invariance violations appearing under coarse-graining, although the initial systems are gauge invariant. Thus, the (gauge invariant) spin network basis is not closed under coarse-graining, and one is forced to truncate these electrical charges which appear without being able to check their relevance for the dynamics of the system.

In contrast, the fusion basis allows for the inclusion, but also systematic suppression, of electric charges, both on lattice scale as well as on coarser scales. This allows us to test different truncation schemes: (a) one where electric charges are allowed on lattice as well as on coarser scales, (b) one where electric charges are truncated at the lattice scale, and (c) one where electric charges are suppressed both at the lattice scale and at coarser scales. We show that the truncation (b) is justified (for the dynamics of Yang-Mills type lattice gauge systems we consider in this work), whereas (c) allows for the extraction of critical couplings but interferes considerably with the functioning of the fusion basis transformations as disentanglers, i.e., short-scale entanglement filters. Indeed, suppressing electric charges at coarser scales does lead to truncations for the fusion basis transformations.

The possibility to either include or suppress the electrical charges in a scale-dependent manner is a main advantage of the fusion basis over the spin network basis. Additionally, as mentioned above, the fusion basis based algorithm does allows us to monitor the magnetic and electric charge observables, which provide order parameters for studying lattice gauge theories, anyon condensation, as well as transitions between an effective cosmological constant in 3D quantum gravity~\cite{DittrichLambda}.

~\\
The paper is organized as follows.
We start with a very short review of tensor network renormalization algorithms and discuss the challenges of dealing with gauge systems in section~\ref{sec:TNW}. We continue by providing the necessary background on the fusion basis in section~\ref{sec:FB}. Section~\ref{Sec:algo} gives both an overview as well as details of the new tensor network coarse-graining algorithm.
We then apply this algorithm to lattice gauge theory with a quantum deformed structure group $\SU(2)_{\rm k}$ in section~\ref{sec:lgt}. This includes the construction of the initial amplitudes, a description of the range of models, as well as a description of various versions of the algorithm, which either truncate or keep electric charge excitations (i.e., torsion degrees of freedom). We also discuss numerical costs and the measures we take to decrease these costs.  We then describe the results of applying the various versions of the algorithms to the lattice gauge theory models. This allows us to draw first conclusions of the behaviour of the critical Yang-Mills coupling with growing level ${\rm k}$. Furthermore, we can compare the different versions of the algorithms and thus learn whether we can truncate the electric charge excitations without affecting results. Lastly, we discuss the expectation values of (Wilson loop) observables, which can be tracked with the coarse-graining algorithm. We conclude with a discussion in section~\ref{discussion}.

Appendices~\ref{app-basics} and~\ref{app-FB} include the necessary background on $\SU(2)_{\rm k}$, as well as the definitions and proofs used for the fusion basis.

The version of the algorithm used to produce the results in this article is available at \url{https://github.com/ssteinhaus/Fusion-basis-coarse-graining}.

\section{Tensor network algorithms for lattice gauge theories}\label{sec:TNW}

We are interested in approximating partition functions for physical systems in an iterative way, that is, via coarse-graining procedures. One way to proceed is to rewrite a given partition function as a contraction of a tensor network, and then to use tensor network renormalization algorithms~\cite{Levin,GuWen,VidalEntFiltering,HigherOrder}. However, we will advertise a generalization of tensor networks which can handle gauge systems, for instance, in a more effective way~\cite{DecTNW,DelcampSF}.  We consider physical systems, whose partition function can be understood as a gluing of amplitudes associated with building blocks. More precisely, the building blocks come with a boundary Hilbert space, and the amplitude is a functional on this boundary Hilbert space. Assuming the boundary Hilbert space admits a basis $b_{i_1,i_2, \ldots}$, where $(i_1,i_2,\ldots)\,,i_k\in I$ label localized or quasi-localized degrees of freedom, the amplitude is given by a function ${\cal A}(i_1,i_2,\ldots)$ of the labels.

Partition functions given via the contraction of a tensor network can be understood easily in this language. Consider a cubical 3D tensor network whose edges $e$ are labelled by indices $i_e \in I$ and whose vertices $v$ carry rank-six tensors $T_{\{i_e|e\supset v\}}$. We can define basic cubic building blocks, which each include one vertex. Each side of the elementary cubes is punctured in the middle by one edge of the network. We associate with a basic building block a boundary Hilbert space $\otimes_s {\cal H}_s$, where $s$ labels the  six  sites (situated where the tensor network edges puncture the boundary of the building block), and the dimension of the site Hilbert space ${\cal H}_s$ is given by the cardinality $|I|$ of the index set $I$. The site Hilbert space ${\cal H}_s$ can be interpreted as a space carring the degrees of freedom associated to the site $s$, which is considered completely localized. By introducing an abstract basis $\{b_i, i\in I\}$ in ${\cal H}_s$, and by numbering the sites by $1,\ldots 6$, the amplitude encoded in the tensor network is given by ${\cal A}(i_1, \ldots i_6)=T_{i_1,\ldots i_6}$.
\begin{figure}[h]
  \includegraphics[width=0.8\textwidth]{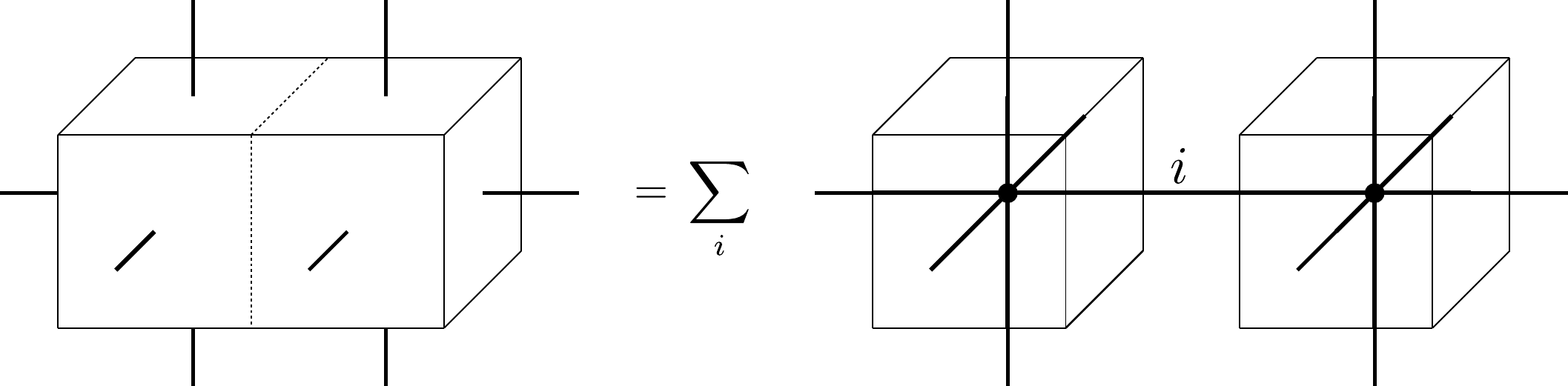}
  \caption{ \label{fig:TNW-cubes}
 Two tensors  associated with cubical building blocks are glued together by summing over the index associated with the shared edge, labeled `$i$' above. The result is a higher rank tensor that describes a building block equipped with more boundary data.
  }
\end{figure}

Gluing two neighbouring basic building blocks (Fig.~\ref{fig:TNW-cubes}) amounts to a summation over the index associated with the corresponding tensor network edge. In terms of the amplitudes, we identify the basis elements (i.e., the indices) of the site Hilbert spaces that are matched to each other, and then sum them.
\ba
{\cal A}'(i_1,i_2,i_3,i_4,i_5,j_2,j_3,j_4,j_5,j_6)=\sum_i {\cal A}_1(i_1,i_2,i_3,i_4,i_5,i) {\cal A}_2(i,j_2,j_3,j_4,j_5,j_6) \q .
\ea
The resulting building block comes with a larger boundary Hilbert space given by a tensor product over ten sites.

Further iterations of the gluing procedure lead to an exponential growth of the boundary Hilbert spaces. Hence, one needs to find a way to truncate the least relevant degrees of freedom in the partition function such that the remainder fits in a boundary Hilbert space of the original (or alternatively a pre-defined) size with the  same localization structure as before. This allows one to iterate the gluing and truncation procedure alternatively while keeping the size of the boundary Hilbert space finite.  In this way, we can compute (approximately) the partition function depending on coarse-grained boundary data. The precise form and physical  interpretation of these data depends on the  truncation procedure.

We have seen that partition functions arising from tensor networks lead to a description in terms of degrees of freedom localized at sites. Yet, gauge systems might not allow for such a localization of their physical, i.e., gauge invariant degrees of freedom~\cite{Donnelly:2011hn,Casini:2013rba,Donnelly,DelcampEntanglement,RielloSubsystems}. A complete localization (or tensor network description) in these cases can be obtained only if one introduces unphysical gauge degrees of freedom, which in a tensor network description corresponds to auxiliary tensors and degrees of freedom, see~\cite{DecTNW}. However, the introduction of auxiliary structures makes the tensor network methods much more expensive to calculate numerically, both in terms of memory and runtime.

The two basic steps in tensor network algorithms, namely, gluing and truncation, can be generalized if we consider boundary Hilbert spaces of a more general structure than the one arising from tensor networks. An example of such generalized structures which deal with lattice gauge theories are the decorated tensor networks~\cite{DecTNW}. These encode systems with  boundary Hilbert spaces of more general structures than the completely localized Hilbert space $\otimes_s {\cal H}_s$ described above.

To be more explicit, let us consider a lattice gauge theory with (finite or compact) structure group $G$. The partition function for a lattice gauge theory on a cubical lattice can  be rewritten as a gluing between amplitudes  associated with the basic cubical building blocks, see~\cite{Eckert,DecTNW,DelcampSF}. There are different ways to cut the spacetime lattice into building blocks, and there are also two different choices for the kind of boundary Hilbert space. The first kind is supporting gauge variant amplitudes and, therefore, unphysical gauge degrees of freedom, and it is given by $\otimes_{l\in {\cal N}} L^2(G)$, where $l$ denotes a link of a network ${\cal N}$ on the boundary and $L^2(G)$ is the Hilbert space of square integrable functions on the group\footnote{We will suppose links coincide with the edges of the building blocks, but other choices are possible.}. Thus, we can localize the degrees of freedom in this Hilbert space to the links of the network.

However, this Hilbert space is unnecessarily large, since the amplitudes are gauge invariant. Therefore, including the gauge degrees of freedom is unnecessary, and it simply makes the coarse-graining algorithm less efficient. The gauge invariant Hilbert space is a subspace of the gauge variant Hilbert space, but it cannot\footnote{This applies to non-Abelian groups. We will see that it is possible modulo a global constraint for  $(2+1)$-dimensional Abelian gauge systems.} be written in the form of a tensor product over localized degrees of freedom.

For the coarse-graining algorithm, we need a basis in this gauge invariant Hilbert space. Such a basis can be characterized by the set of commuting gauge invariant operators which is diagonalized by the basis. These operators also determine how localized the degrees of freedom described by the basis are. A candidate for the set of commuting operators are closed Wilson loops. However, for non-Abelian groups, the set of Wilson loops around the basic plaquettes are insufficient to determine a state. On the other hand, if one considers the set of all possible Wilson loops, there are complicated relations known as Mandelstam identities which have to be imposed~\cite{LollGauge}. This makes an explicit construction of a basis difficult. In fact, we will see that in $(2+1)$ dimensions the fusion basis does provide the diagonalization of a certain subset of Wilson loop operators, but it also includes `t Hooft operators measuring the electric charges.

The spin network basis~\cite{SNW} is gauge invariant, and it provides a diagonalization of gauge invariant combinations of electric flux operators. A coarse-graining algorithm based on the spin network basis was developed in~\cite{DecTNW} for Abelian groups and in~\cite{DelcampSF} for non-Abelian groups. The spin network basis is, however, not stable under coarse-graining: as we will discuss below, coarse-graining can lead to non-vanishing electric charges appearing as gauge invariance violations~\cite{Etera, DelcampSF}. This can be dealt with by projecting to zero electric charge after each coarse-graining step~\cite{DelcampSF}, but this assumes that the corresponding degrees of freedom are not relevant at larger scales\footnote{See also~\cite{EteraExtend} for suggestions on how to extend the spin network basis to capture these additional degrees of freedom.}.

A further issue with the spin network basis is the following: after gluing two building blocks, the resulting spin network needs to be transformed into a different spin network better suited for the truncation step; see~\cite{DelcampSF} for the detailed algorithm. The reason is that the choice of graph on which the spin network is defined determines the localization of the degrees of freedom, and for the coarse-graining one wishes to localize the fine-grained degrees of freedom in a certain way. The relation between the transformation and the corresponding rearrangement of the localization of degrees of freedom is not very transparent in the case of the spin network basis: it subsequently requires a transformation to a holonomy basis and back~\cite{DelcampSF}. Both issues are resolved with the fusion basis. Later we will show its structure is ideally suited for coarse-graining purposes, and that it is also applicable to systems with quantum deformed structure groups, which we employ in this work.

\section{Fusion basis in a nutshell}\label{sec:FB}

The fusion basis arises in $(2+1)$--dimensional anyon systems~\cite{Koenig,Hu}, but it can be also constructed for $(2+1)$--dimensional lattice gauge theories~\cite{DelcampFusion}, as well as for $(2+1)$--dimensional gravity \cite{ExTQFT,DittrichLambda}. The associated algebraic structures, the so-called {\it Drinfeld Doubles},  have been discussed for various physics applications~\cite{BaisDiscreteGauge,Kitaev,Bombin, Wen1, ExTQFT,DelcampFusion}.

In lattice gauge theory, the fusion basis provides gauge invariant states that diagonalize the set of Wilson loop operators around the lattice plaquettes. However, for non-Abelian gauge theories this set of Wilson loop operators does not provide a maximal set of commuting observables. Adding all possible Wilson loops, that is, also loops around arbitrary clusters of plaquettes, one does encounter complicated dependencies between the Wilson loop operators called Mandelstam identities. This leads to the quite involved task of constructing an independent set of (Wilson loop) observables~\cite{LollGauge}.

In $(2+1)$--dimensions this problem is solved by the fusion basis\footnote{See~\cite{DelcampDittrich4D,Delcamp:2018efi} for a discussion of the $(3+1)$--dimensional case with classical structure groups and~\cite{Dittrich4DQ} for the case with a quantum deformed structure group $\SU(2)_{\rm k}$.}.
The  basis is most easily characterized by describing the maximal set of commuting observables, which diagonalize the basis.  In the case of the fusion basis these observables are known as closed ribbon operators~\cite{Kitaev,Wen1,Hu,DelcampFusion}.  The closed ribbon operators are based on a closed path, or loop, and can be understood to measure the magnetic and electric charge contained in the region enclosed by the loop\footnote{Here we only consider non-intersecting loops.}.

In the case of lattice gauge theory with a (not quantum deformed) structure group this ribbon operator includes a Wilson loop operator, which measures the magnetic charge, and a `t Hooft operator which integrates the electric flux along the loop, which measures the electric charge; see~\cite{DelcampFusion} for a detailed description. In the case of a quantum deformed structure group such as $\SU(2)_{\rm k}$, the magnetic Wilson loop and the electric `t Hooft observables are encoded into two connection variables. These connections are non-commutative, and thus force the introduction of the notion of Wilson lines over- or under-crossing each other. The pair of connections lead to two classes of  loop and open line operators, namely those which under-cross or over-cross the networks which characterize a given state. The closed ribbon operators are then given by a parallel pair of an under-crossing and an over-crossing loop operator.

\begin{figure}
  \includegraphics[width=0.6\textwidth]{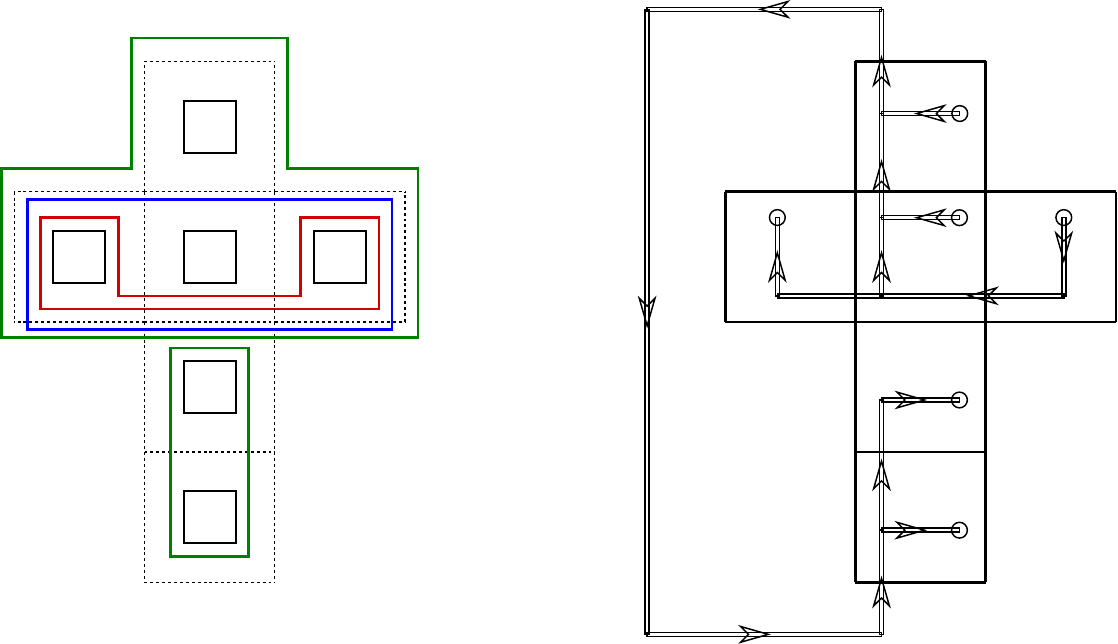}
  \caption{\label{fig:Loops_plaquettes}
  	{\it Left:} Choosing a set of non-intersecting loops for a  state on a cubical building block (with six punctures / plaquettes),  drawn on the unfolded 2D boundary of the cube. We start with loops around the basic plaquettes (in black). We then add loops around pairs of plaquettes, given by the green loop (around two plaquettes) and the red loop. The blue loop  surrounds an effective plaquette made from two basic plaquettes and a  third basic plaquette.  The resulting effective plaquette is fused with another basic plaquette. The (green) loop around this effective plaquette made out of four basic plaquettes is equivalent to the other green loop around two basis plaquettes, since the building block has spherical topology.\\
  {\it Right:} The fusion tree encodes the same choice of commuting ribbon observables as the set of loops does. For instance, if two plaquettes are directly fused together, the fusion basis diagonalizes the ribbon operator surrounding these two plaquettes. Fusing another plaquette to the pair of plaquettes diagonalizes the ribbon operator encircling the three plaquettes, etc. From the fusion tree one can also read off equivalent observables: a ribbon operator surrounding two  fused plaquettes is equivalent to the operator surrounding the remaining four plaquettes.}
\end{figure}

Ribbon operators which cross each other generally do not commute. Hence, the fusion basis is characterized by a set of non-crossing, and therefore commuting, closed ribbon operators. Restricted to a surface with spherical topology\footnote{See~\cite{Koenig,DelcampFusion,Delcamp:2018efi} on how to generalize the basis for non-trivial topologies.},
this set consists of ribbon operators associated with the following sets of loops: (a) those around the basic plaquettes of the lattice, and (b) those around coarse-grained plaquettes defined by the fusion procedure. Each basic step involves only the coarse-graining of two (possibly already coarse-grained) plaquettes. The second set of loops is given by those around all the coarse-grained or fused plaquettes, which arise in this manner. Note that the loops should not cross over each other. Fig.~\ref{fig:Loops_plaquettes}  gives an example for a set of loops constructed in this manner, as well as the encoding of this set into a three-valent (fusion) tree. To construct the tree,  we map the basic plaquettes of the lattice to the leaves of the tree, and to each fusion of two (possibly already fused) plaquettes we associate a branch of the tree. We also choose orientations for the leaves and branches of the tree - these encode an orientation for corresponding ribbon loop operator, and in our case a choice of phase factors for the basis states.

The fusion tree can describe more general structures than (regular) lattices. In the context of (extended) topological quantum field theory, one introduces so--called defect excitations by allowing topology-changing punctures in the underlying spatial manifold~\cite{Balsam, Koenig, ExTQFT}. The fusion basis provides a basis of states on such manifolds with punctures. The leaves of the fusion tree are identified with the punctures. The fusion basis diagonalizes the ribbon operators which go around single punctures as well as the operators encircling certain clusters of punctures, as encoded in the fusion tree. Thus, the basic plaquettes can be identified with punctures; see~\cite{Hu,DelcampFusion} for  constructions of explicit mappings. Hereafter, we use the terms plaquettes and punctures interchangeably.

Next, we specify the data associated with the fusion tree. These data encode the eigenvalues of the ribbon operators. As each branch of the tree is associated with an operator, each branch carries a label. These labels are given by objects in a certain fusion category: for a finite group this is the (fusion) category of representations of the Drinfeld Double of the group. For $\SU(2)_{\rm k}$ this is the Drinfeld centre (also referred to as Drinfeld Double) of the fusion category of representations of $\SU(2)_{\rm k}$. (See Appendix~\ref{app-basics} for some essential basics on $\SU(2)_{\rm k}$.)

In physical terms, the objects in this fusion category describe the electric and magnetic charges measured by the ribbon operators. The fusion of two charges $\rho_1,\rho_2$, that is, the set of resulting charges, is described by the so-called fusion product of the corresponding objects
 \ba
 \rho_1 \times \rho_2 =\sum_{\sigma_3} N^{\sigma_3}_{\rho_1\rho_2} \, \sigma_3 \q ,
 \ea
 where $\sigma_3$ labels the charges in the fusion product.
$N_{\rho_1\rho_2}^{\sigma_3}$ are the fusion coefficients, which we assume to be zero or one; otherwise, we would have further degeneracy labels at the vertices of the fusion tree.  Thus, we have at each vertex of the fusion tree a condition on the labels of the three adjacent branches, namely that the associated fusion coefficient should be non-vanishing, $N^{\sigma_3}_{\rho_1\rho_2}=1$. In this way, the labelled fusion tree encodes the closed ribbon operators which are diagonalized by the fusion basis, along with the eigenvalues of these ribbon operators.

In general, the fusion basis also describes non-gauge invariant states. The violation of gauge invariance is equivalent to the violation of the Gau\ss~ constraint and, therefore, the presence of electric charges at the plaquettes\footnote{To locate these electric charges in the middle of the plaquettes one can imagine that, e.g., on a square lattice one has an open link that starts in one corner of the given plaquette and ends in the middle of the plaquette.}.

For such non-gauge invariant states we need to specify one additional piece of information, which is only attached to the endpoint of the leaves of the tree.  For a given leaf $l$ this is a `tail' label $s_l$, which denotes a basis element in the irreducible representation of the Drinfeld Double associated with the leaf. The tail labels are necessary to specify the action of open ribbon operators, which start and end at the mid-point of the plaquettes, and are in general not gauge invariant (at their endpoints).

In this work, we consider only the initial amplitudes which are gauge invariant. As we will discuss later, coarse-graining can lead to the emergence of electric charges, which would make the tail labels necessary. However, we will define a truncation in which the amplitudes have a trivial dependence on the tail labels. This allows us to neglect the tail indices in our coarse-graining algorithm, in turn leading to a significant reduction in the numerical costs of the algorithm.

To construct a fusion basis, we have to choose a fusion scheme for the plaquettes of the lattice.  Different choices for this fusion scheme lead to different fusion trees and associated fusion bases.  Two fusion trees can be transformed into each other via a set of basic transformation steps. Here we will need the so-called F-move,
\begin{equation}\label{Fmove}
  \includegraphics[scale=0.5,valign=c]{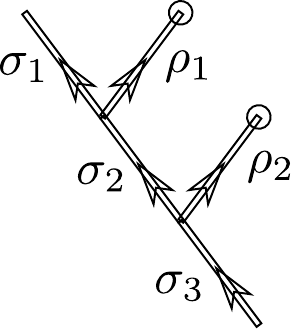} \,\,\,=\,\,\, \sum_{\sigma_4}
 \mathbb{F}^{\rho_1 \, \sigma_1 \, \sigma_2}_{\sigma_3 \, \rho_2 \, \sigma_4} \,% F^{j_2 \, j_1 \, b_1}_{b_2 \, j_3 \, b'_1}
  \; \includegraphics[scale=0.5,valign=c]{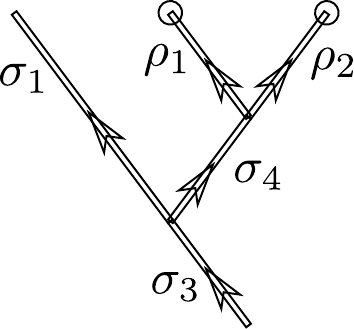} \; .
\end{equation}
and a transformation that we will refer to as R-move:
\begin{equation} \label{eq:pulling_trafo}
  \includegraphics[scale=0.5,valign=c]{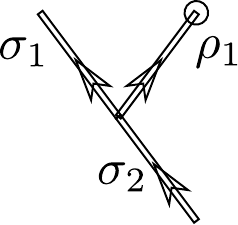}\, \,\,=\,\,\,
 \mathbb{R}^{\rho_1 \sigma_1}_{\sigma_2}
  \includegraphics[scale=0.5,valign=c]{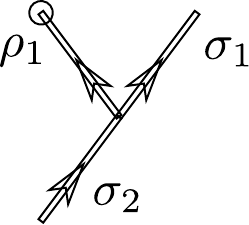} \; .
\end{equation}
Note that for $\SU(2)_{\rm k}$ the recoupling symbols ${\mathbb F}$ and ${\mathbb R}$ do not depend on the orientations of the leaves and branches. The reader can find the definitions for ${\mathbb F}$ and ${\mathbb R}$ in Appendix \ref{app-FB}.

With a choice of fusion basis for a given building block, that is, for the associated boundary Hilbert space, we can express the amplitude associated with this building block. We can understand the amplitude as defining a state $\mathcal{A}$ in the boundary Hilbert space. This allows us to make the choice of fusion basis explicit: for a cubical building block with six plaquettes, we write
\begin{align}
  &\mathcal{A}=
   \sum_{ \rho_l,\sigma_b} \mathcal{A}(\{ \rho_l, \sigma_b\})
  \; \includegraphics[scale=0.5,valign=c]{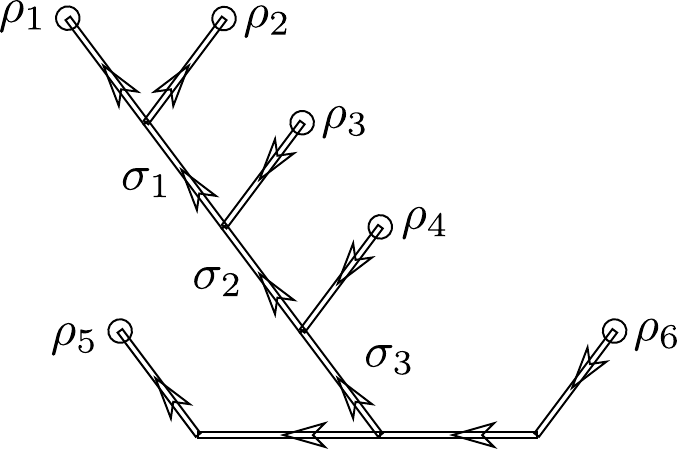} \; \; .
\end{align}
Here $\mathcal{A}(\{ \rho_l,\sigma_b \})$ is the amplitude associated to the building block, $\rho_l$ are the (Drinfeld Double) representations attached to the leaves $l$, and $\sigma_b$ are the (Drinfeld Double) representations attached to the branches $b$. (As mentioned previously, we assume a trivial dependence on the tail indices $s_l$; therefore, we drop these indices all together.) The notation makes it obvious how the amplitudes change under a change of fusion basis. Under an F--move, we have
\begin{align} \label{eq:tree_trafo}
 \mathcal{A}\,\,\,=\, & \sum_{\rho_1,\rho_2,\sigma_1,\sigma_2,\sigma_3,\ldots} \includegraphics[scale=0.5,valign=c]{Drawings/tree_trafo_1} \;\;\; \mathcal{A}(\rho_1,\rho_2;\sigma_1,\sigma_2,\sigma_2,\ldots) \nonumber \\
  = & \sum_{\rho_1,\rho_2,\sigma_1,\sigma_4,\sigma_3,\ldots} \includegraphics[scale=0.5,valign=c]{Drawings/tree_trafo_2}
  \;\; \sum_{\sigma_2} \mathcal{A}(\rho_1,\rho_2;\sigma_1,\sigma_2,\sigma_3,\ldots)
 \,\,  \mathbb{F}^{\rho_1 \, \sigma_1 \, \sigma_2}_{\sigma_3 \, \rho_2 \, \sigma_4} \,
  \nonumber \\
  =: & \sum_{\rho_1,\rho_2,\sigma_1,\sigma_4,\sigma_3,\ldots} \includegraphics[scale=0.5,valign=c]{Drawings/tree_trafo_2}
\;\;  \mathcal{A}(\rho_1,\rho_2;\sigma_1,\sigma_4,\sigma_3,\ldots) \; .
\end{align}
Remembering that the fusion tree encodes a fusion ordering for the plaquettes, we can interpret this transformation as a change of how we coarse-grain, and accordingly we can organize the degrees of freedom of the system. In fact, these F-move transformations can be seen as (dis-)entanglers~\cite{Koenig2009ExactER,TimeEvol}.

In the following, we will briefly review three classes of examples: Abelian structure groups, non-Abelian structure groups, and finally the quantum deformed structure group $\SU(2)_{\rm k}$, which we employ later in this paper.

~\\
\noindent
{\bf Abelian structure group:} We consider a $(2+1)$--dimensional lattice gauge theory with finite Abelian group ${\mathbb Z}_n$. As mentioned above, the ribbon operator includes a Wilson loop whose possible eigenvalues are labelled by the conjugacy classes of the structure group, i.e., by ${\mathbb Z}_n$. This Wilson loop measures the magnetic flux through the region surrounded by the loop, which represents the curvature of the lattice connection. The other part of the ribbon operator is given by a product of group translation operators\footnote{For non-Abelian groups these translation operators need to be parallel transported, via an adjoint action, to a common reference system. This is not necessary for Abelian groups.} that act on the lattice links which are crossed by the ribbon.  These translation operators measure the electric charge (or the violation of the Gau\ss~constraint). The resulting eigenvalues are again\footnote{Technically, we would have the Poincare dual of ${\mathbb Z}_n$, which is ${\mathbb Z}_n$.} labelled by ${\mathbb Z}_n$.  Correspondingly, the irreducible representations of the Drinfeld Double of ${\mathbb Z}_n$ are  labelled by ${\mathbb Z}_n \times {\mathbb Z}_n$ and are one-dimensional. Thus,  tail labels do not appear for Abelian structure groups.

For an Abelian structure group, both the magnetic and electric charges satisfy very simple fusion rules. With $\rho=(z^m,z^e)\in {\mathbb Z}_n\times {\mathbb Z}_n$ the fusion rules are given by
\ba
(z^m_1,z^e_1) \times (z^m_2,z^e_2) = \delta^n_{z^m_1+z^m_2,z^m_3}\,  \delta^n_{z^e_1+z^e_2,z^e_3}\,\, (z^m_3,z^e_3)
\ea
where $\delta^n_{p,q}=1$ if $p=q \, \text{mod} \, n$ and is vanishing in all other cases.

Thus, the measurement results of coarser ribbon operators are completely determined by the measurement results of the ribbon operators around the basic plaquettes\footnote{For a spherical topology we actually need all but one of these plaquettes.}. It follows that the labels of the fusion tree are entirely determined by the labels associated with the leaves. We can restrict to gauge invariant states, i.e., states for which the basic electrical charges associated with the plaquettes are trivial. In the Abelian case, the electrical charges associated with coarser regions are then also trivial.

~\\
\noindent
{\bf Non-Abelian structure group:} The situation is more involved for finite\footnote{See~\cite{Koornwinder:1998xg} for a construction of the Drinfeld double representations for $\SU(2)$.} non-Abelian groups; see~\cite{DelcampFusion} for a detailed construction of the fusion basis and ribbon operators. The irreducible representations of the Drinfeld Double are labelled by $\rho=(C,R)$ where $C$ denotes a conjugacy class of the group and $R$ an irreducible representation of the stabilizer of (one of the representatives of) the conjugacy class $C$. The dimension of a representation $\rho=(C,R)$ is given by $\text{dim}\rho=|C|\times \text{dim}R$, where $|C|$ denotes the number of group elements in $C$.

The fusion rules are now more intricate: the fusion rule for the magnetic part can be deduced from the interpretation of the Wilson loop. A loop around two plaquettes with conjugacy classes $C_1$ and $C_2$ respectively, can yield all conjugacy classes of representatives of the form $c_1 \cdot c_2$, where $c_1\in C_1$ and $c_2 \in C_2$. Thus, specifying the measurements of coarser Wilson loops provides extra information in addition to the measurements of Wilson loops around the basic plaquettes. But, we also see that if the magnetic charges around the basic plaquettes are all trivial (i.e., the conjugacy classes are given by the identity element), this will also hold for the magnetic charges associated with coarser regions.

The situation is different for the electric charge: having vanishing electric charges for the basic plaquettes (but non-vanishing magnetic charges) does not exclude non-vanishing electric charges for the coarser plaquettes. These ``Cheshire charges''~\cite{BaisDiscreteGauge} signify a non-trivial interaction between the magnetic and electric parts of the Drinfeld Double. It can be explained by the fact that the part of the operator which measures the electric charge involves parallel transport~\cite{DittrichGeillerFlux}. In the presence of curvature (i.e., magnetic charge) the parallel transport is non-trivial and can induce an electric charge.  This is an issue for coarse-graining, as one a priori needs to keep track of the electric charges, even if one starts with a gauge invariant state, i.e., a state without electric charges associated with the basic plaquettes.   The fusion basis provides us with such a tracking device, whereas the spin network basis is not naturally suited for this task.

~\\
\noindent
{\bf Structure group $\SU(2)_{\rm k}$:} Later in this paper we work with a more abstract generalization of groups and Drinfeld Doubles of groups. Specifically, we use a quantum deformation of $\SU(2)$, with the deformation parameter a root of unity $q= \exp(2\pi \i /({\rm k}+2))$~\cite{Yellowbook,Biedenharn}.  See Appendix~\ref{app-basics} for some basic details, and~\cite{ExTQFT} for an extensive exposition, which discusses in particular the various ribbon operators and their lattice gauge theoretic interpretations.

The origin of the deformation can be understood in the following way: the Hilbert space $L^2(SU(2))$, which underlies lattice gauge theory, can be obtained as a quantization of the phase space $T^*\SU(2)$, which has a compact configuration space $\SU(2)$, but comes with a flat non-compact momentum space $T^*_g\SU(2) \simeq {\mathbb R}^3$.  A quantum deformation at a root of unity leads to a replacement of  $T^*\SU(2)$ with $\SU(2) \times \SU(2)$; see~\cite{Riello:2017iti}. Thus, both the configuration space and momentum space are now compact and curved.  Therefore, the ribbon operators, which provide both configuration space and momentum space information, have discrete and bounded spectra.

The ribbon operator eigenvalues, that is, the charges measured by the ribbon operators, are described by the Drinfeld Double  of $\SU(2)_{\rm k}$, which is given by $\SU(2)_{\rm k}\otimes \overline{\SU(2)_{\rm k}}$. The bar above the second factor indicates that this copy of $\SU(2)_{\rm k}$ comes with a complex conjugated braiding structure, as opposed to the first factor.

The fusion rules for the Drinfeld Double of $\SU(2)_{\rm k}$ are indeed given by ``doubling'' those of the fusion category $\SU(2)_{\rm k}$. The fusion category $\SU(2)_{\rm k}$ has irreducible objects (i.e., irreducible representations or charges) labelled by half integers $j=0, \tfrac{1}{2},\ldots, \tfrac{\rm k}{2}$. The fusion rules can be seen as a deformation of the $\SU(2)$ rules, so that only  representations $j\leq {\rm k}/2$ appear:
\ba\label{frulek1}
j_1 \times j_2 \,=\,  \sum_{j_3=|j_1-j_2|}^{\text{min}(j_1+j_2,{\rm k}-(j_1+j_2) ) }  \,\,\, j_3 \q .
\ea
The sum uses integer steps, so the coupling condition $j_1+j_2+j_3\in {\mathbb N}$ holds. In addition to the usual $\SU(2)$--coupling rule $j_3\leq j_1+j_2$, the quantum deformation leads to the coupling condition $j_1+j_2+j_3 \leq {\rm k}$.

We denote the irreducible representations of the Drinfeld Double by $\rho=(j,\overline{j})$. The fusion rule for the double are given by
\ba\label{frulek2}
(j_1,\overline{j_1}) \times (j_2,\overline{j_2}) \,=\,  \sum_{j_3=|j_1-j_2|}^{\text{min}(j_1+j_2,{\rm k}-(j_1+j_2) ) } \,\,\,
 \sum_{\overline{j_3}=|\overline{j_1}-\overline{j_2}|}^{\text{min}(\overline{j_1}+\overline{j_2},{\rm k}-(\overline{j_1}+\overline{j_2}) ) } \,\,\, (j_3,\overline{j_3}) \q .
\ea

Furthermore, we have non-trivial tail labels $s$, which for a leaf with label $(j,\bar{j})$ can take values in $s=|j-\bar{j}|,|j-\bar{j}|+1, \ldots, \text{max}(j+\overline{j} ,{\rm k}-(j+\overline{j}) )$. In other words, the tail carries a representation label which arises from the fusion product of $j$ and $\bar{j}$.

The  interpretation of the Drinfeld Double representation labels in term of charges is the following:  the sum $(j+\bar{j})$ specifies the magnetic charge, whereas the difference $|j-\bar{j}|$ characterizes the electric charge. States with vanishing electric charges for the basic plaquettes carry  labels $\rho=(j,j)$ at their fusion tree leaves, and they have an associated tail label $s=0$. However, the fusion of two charges $(j_1,j_1)$ and $(j_2,j_2)$ can lead to a charge $(j_3,\bar{j_3})$ with $j_3\neq \bar{j_3}$ and, therefore, to the ``Cheshire charge'' phenomenon as we saw with the non-Abelian group.

In this paper, we use gauge invariant amplitudes, which are only non--vanishing for leaf labels $\rho=(j,j)$ and tail index $s=0$. The coarse-graining procedure leads to non-trivial electric charges at the leaves. As mentioned above, we will use a truncation that assumes a trivial dependence on the tail indices. Thus, we will omit the tail indices hereafter.

\section{The coarse-graining algorithm}\label{Sec:algo}

\subsection{Sketch of the algorithm} \label{sec:algorithm}

The first basic piece in a coarse-graining algorithm is the iterative gluing of smaller building blocks into larger ones. The gluing process implements a summation over the variables and, thus, amounts to evaluating the partition function of the system. The larger building blocks which arise from the gluing carry increasing amounts of data. The second essential piece is, therefore, a truncation of these data, or in other words, a coarse-graining process. For the gluing and truncation we must specify the following:
\begin{itemize}
 \item {\it Gluing:} Given our cubical building blocks each with six sides (hence, six plaquettes), we must specify a basis in the boundary Hilbert space in which to express the amplitudes. We then have to determine how to glue. That is, in identifying two faces of opposite cubes we need to specify which basis labels on these two cubes we need to identify with each other, and over which (identified) labels we need to sum. This defines the amplitude of the new building block, expressed in a specific basis, which is also determined by the gluing process. As this basis is associated with a building block with now ten plaquettes, it carries more data than the initial basis for the building blocks with six plaquettes.

 \item {\it Transformation of the fusion basis:}    The basis for the ten-plaquette state, which arises from the gluing, is {\it not} well-suited for the truncation step. The reason is that this basis does not diagonalize the ribbon operators around the pairs of plaquettes, which we want to coarse-grain into effective plaquettes. Therefore, we apply basis transformations, i.e., unitary maps, which transform the basis arising from the gluing to a basis in which the ribbon operators around to-be-coarse-grained pairs of plaquettes are diagonalized. These transformations amount to a reorganization of the degrees of freedom in the fusion scheme, so that this scheme is adjusted to the intended coarse-graining.

 The transformations can be seen as disentanglers, which also appear in the MERA algorithm for two--dimensional systems~\cite{MERA, MERAPartitionF} (but have not been implemented yet for $(2+1)$--dimensional systems).  Such disentanglers are supposed to minimize the entanglement  between plaquettes belonging to different to-be-coarse-grained pairs. The difference from the disentanglers in the MERA algorithm and the transformations here is that the MERA disentanglers are dynamically determined, e.g., by a minimization procedure, and thus can change from one coarse-graining step to the next. In contrast, the transformations here are defined from the outset.  The works \cite{DecTNW,DelcampSF,Milsted} introduce similar transformation, that reorganize the degrees of freedom in a spin network basis. \cite{Milsted}~also argues that these transformations act as disentanglers. One can argue, however, that the fusion basis transformations have a more transparent interpretation in terms of reorganizing the regions into which the finer degrees of freedom are blocked.

  We conjecture that the transformations do indeed decrease the entanglement between fine degrees of freedom located in different coarser plaquettes. We later see that this conjecture is justified: a version of our algorithm in which these transformations are restricted leads to the appearance of short-range entanglement, which is supposed to be removed by the disentanglers.

 \item {\it Truncation:}  With the ten-plaquette amplitude transformed to a fusion basis adjusted to the intended coarse-graining scheme, we can proceed now to the truncation step. The labels of the fusion basis can be sorted into two sets: those labels associated with coarse degrees of freedom (or more precisely coarse observables), which we want to keep, and those labels associated with the finer degrees of freedom, e.g., the labels associated with finer plaquettes, which we wish to discard.  To implement the truncation,  we need to specify how to compute an amplitude that depends only on the coarse labels given the amplitude that depends on all labels.  To this end, we will construct an embedding map (also known as an isometry) from the boundary Hilbert space associated with the coarser building block to the boundary Hilbert space associated with the finer building block.
 One can interpret this embedding map as assigning a local vacuum state to the finer degrees of freedom~\cite{Dittrich12,TimeEvol}. In the next iteration, one does not need to sum these finer degrees of freedom anymore. A key feature of tensor network algorithms is that this embedding map is defined from the amplitudes themselves, that is, the truncation is informed by the dynamics of the system.

\end{itemize}

The coarse-graining procedure proceeds by gluing two cuboids into a larger cuboid. That is, the lattice constant is only changed in one direction. After completing a coarse-graining step in one direction, we rotate the lattice to coarse-grain one of the other another directions. This rotation can be implemented by fusion tree transformations. The amplitude then appears in a fusion basis which matches the initial fusion basis after a rotation of the cube. By applying the exact same gluing, transformations, and truncations as in the previous coarse-graining step, we obtain a coarse-graining in a different direction. A complete coarse-graining in all three directions is accomplished after a round of three such coarse-graining steps.

\subsection{Fusion basis for the cubes and gluing procedure}\label{Sec:gluing}

We now explain the gluing step in more detail. As mentioned above, we always glue two cubic building blocks together, which we now refer to as `left' and `right' building blocks.  To specify the gluing procedure we first have to choose a fusion basis for each of the building blocks.
\begin{figure}
  \includegraphics[width=0.45\textwidth]{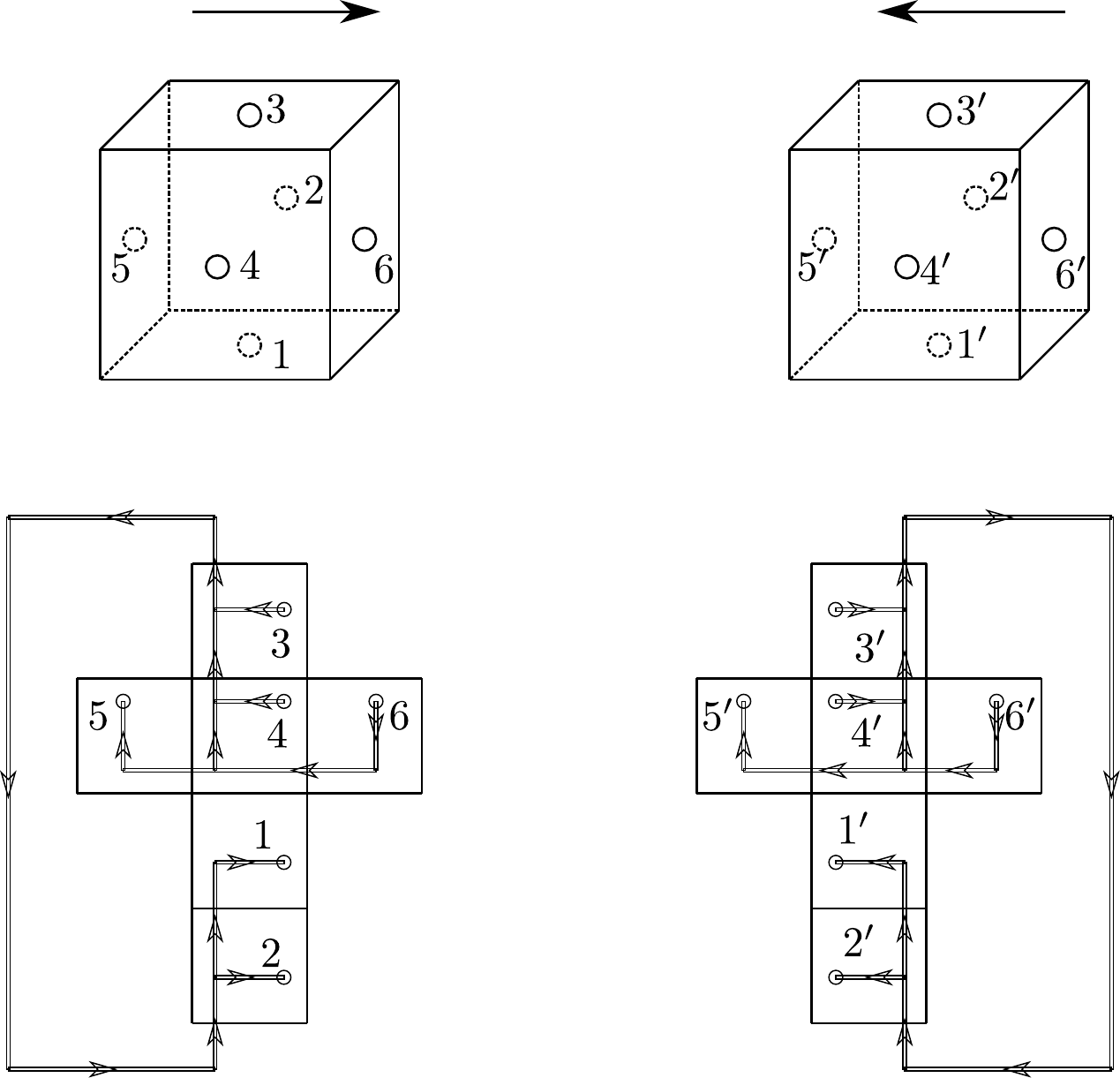}
  \caption{ \label{fig:cube_states}
  Two cubes and their respective fusion trees.
  }
\end{figure}

Fig.~\ref{fig:cube_states} shows the fusion trees for both cubes
and how the cubes are glued together. Since the fusion tree
is defined via a planar graph, it is convenient to draw the tree on the
unfolded cube. The two cubes are glued together by identifying data of
punctures $6$ on the left and $5'$ on the right. Note that punctures $5'$ and
$6$ have opposite orientations of over- and under-crossing branches to ensure that
the orientation matches upon gluing. This in turn ensures the cancellation of phase factors
in the respective amplitudes\footnote{Changing the orientation of a puncture gives rise to a phase factor. Gluing punctures with opposite orientation cancels out the phases.}. After gluing, the remaining tasks are to fuse punctures
$1,\dots,4$ to punctures $1',\dots,4'$, respectively. Note that punctures
corresponding to opposite sides of the cube, e.g. $1$ and $3$, are chosen with
opposite orientation as well, in order to prepare them for gluing in subsequent iterations of the
algorithm.

We express the amplitudes for each cube using the following choices for a
fusion basis:
\begin{align} \label{LR1}
   &\mathcal{A}^{\text{left}} =
   \sum_{ \{\rho_l,\sigma_b\}} \mathcal{A}^{\text{left}}(\{\rho_l,\sigma_b \})
  \; \includegraphics[scale=0.5,valign=c]{Drawings/tree_left} \; \; ,\;
%\end{align}
%\begin{align}
   &\mathcal{A}^{\text{right}} =
  \sum_{ \{\rho_{l'},\sigma_{b'}\}} \mathcal{A}^{\text{right}}(\{\rho_{l'},\sigma_{b'} \})
  \; \includegraphics[scale=0.5,valign=c]{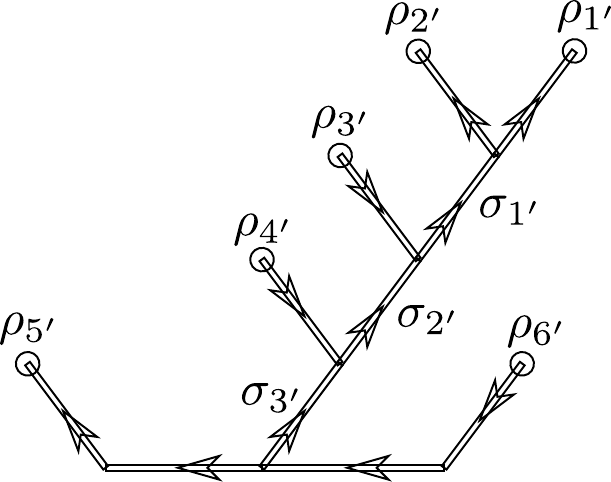} \; \; .
\end{align}
For each of the six punctures, i.e., for each leaf of the fusion tree, we have a label $\rho_l$.  Moreover, we have three branches labeled by representations $\sigma_b$. Note that the state of the right cube is obtained from the state of the left cube by  `pulling'  punctures over the pair of strands, and this is implemented by an R-move, as decribed in~(\ref{eq:pulling_trafo}). Thus, the amplitude components of left and right cube are related by phase factors resulting from this transformation.

\begin{figure}
  \includegraphics[width=0.275\textwidth]{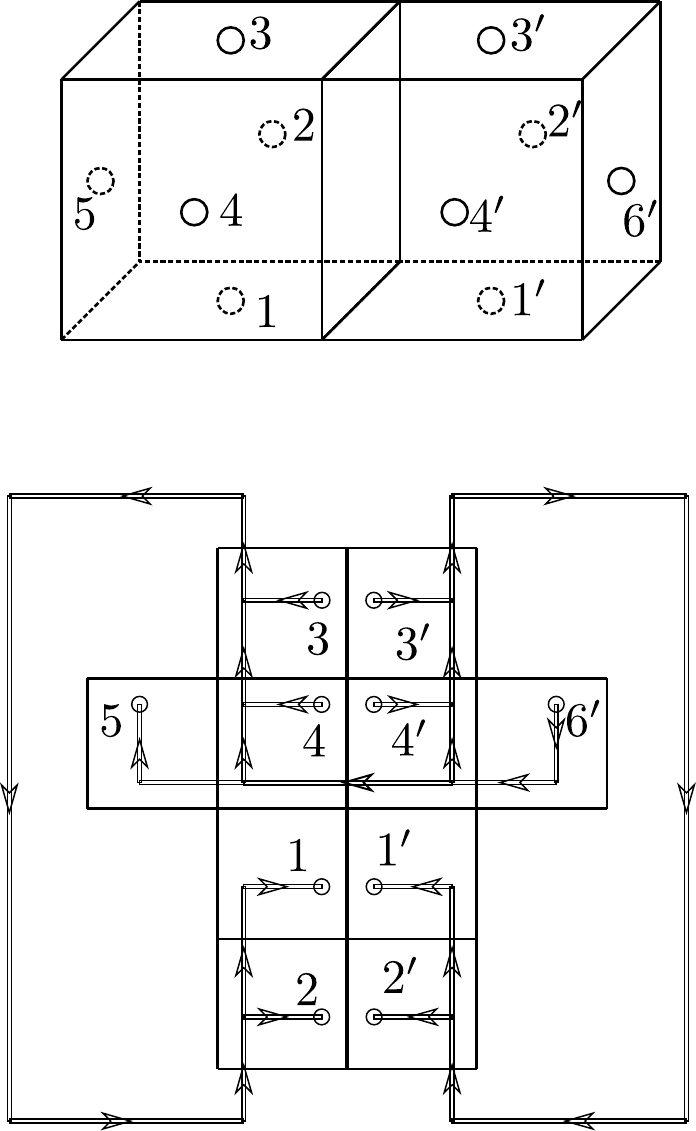}
  \caption{ \label{fig:cube_states-glued}
  The figure shows the gluing of two cubes and the resulting fusion tree. Here we readily see that this fusion tree is not suitable for the intended coarse-graining, since it does not diagonalize ribbon operators around the to-be-coarse-grained pairs of plaquettes.
  }
\end{figure}

\vspace{2mm}
Gluing the left and right cube defines a new state associated with the glued building block.  As shown in Fig.~\ref{fig:cube_states}, puncture $6$ of the left cube is identified with puncture $5'$ of the right cube. Hence, we identify the labels $\rho_6$ with $\rho_{5'}$. The identified representation labels $\rho_6=\rho_{5'}=\sigma_4$ define a branch label $\sigma_4$ for the resulting fusion tree. This is accompanied by a weight factor determined by the (quantum) dimension of $\sigma_4$.  This weight for the glued state can be derived from a similar gluing procedure using a spin network basis~\cite{DelcampSF}. The resulting state (Fig.~\ref{fig:cube_states-glued}) is given by
\begin{align} \label{eq:glued_state}
  & \mathcal{A}^{\text{glued}} := \nonumber \\
  & \sum_{\{ \rho_l,\sigma_b,\rho_{l'},\sigma_{b'},\sigma_4 \}}\!\!\! \!\! \!\!{\cal D}^2
  \; \frac{ \delta_{\sigma_4,\rho_6}  \delta_{\sigma_4,\rho'_{5}}  }{v_{o(\sigma_4)}v_{u(\sigma_4)} }
   \; \mathcal{A}^{\text{left}}(\{\rho_l,\sigma_b \}) \;
  \mathcal{A}^{\text{right}}(\{\rho_{l'},\sigma_{b'} \})
   \; \includegraphics[scale=0.5,valign=c]{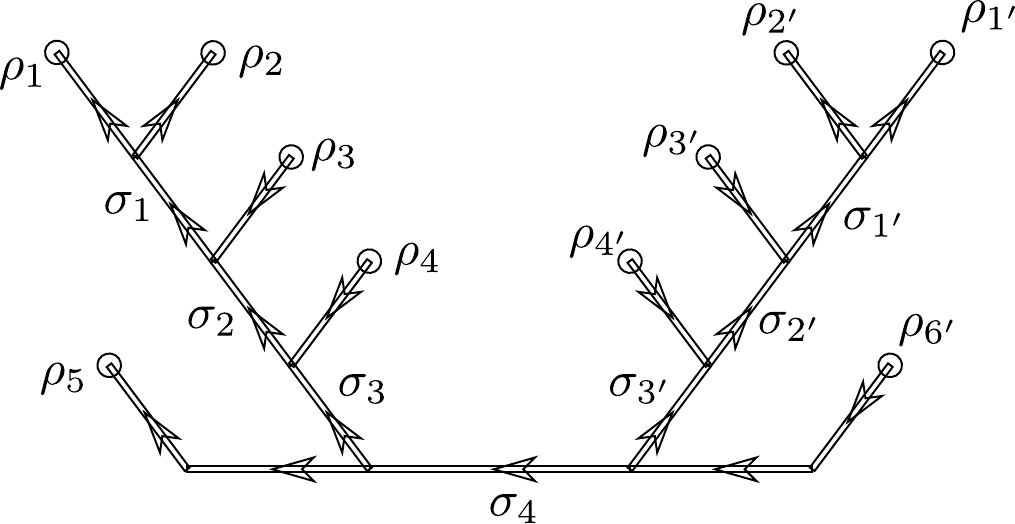} \; .
\end{align}
Here $v_{o(\sigma_4)}=v_{j_4},v_{u(\sigma_4)}=v_{\overline{j_4}}$ are the square roots of the quantum dimensions associated with the $\SU(2)_{\rm k}$ representations appearing in $\sigma_4=(j_4,\overline{j_4})$ and ${\cal D}$ is the total quantum dimension of $\SU(2)_{\rm k}$. (See Appendix~\ref{app-basics}.)
The sum is over the labels  $\sigma_b$ with $b=1,\ldots,3$, $\sigma_{b'}$ with $b'=1',\ldots,3'$, $\rho_l$ with $l=1,\ldots,6$, and $\rho_{l'}$ with $l'=1',\ldots 6'$, as well as over $\sigma_4$.

This implies the amplitude function for the glued building block amounts to
 \ba
 &&{\cal A}^{\text{glued}}(\rho_1,\ldots,\rho_5,\rho_{1'},\ldots, \rho_{4'},\rho_{6'}; \sigma_{1},\ldots,\sigma_3,\sigma_{1'},\ldots, \sigma_{3'},\sigma_4)\,=\
 \nn\\
 &&\q\q\q\q\q
  \; \frac{ {\cal D}^2 }{v_{o(\sigma_4)}v_{u(\sigma_4)} }
   \; \mathcal{A}^{\text{left}}(\rho_1,\ldots,\rho_5,\sigma_4;\sigma_1,\ldots,\sigma_3 ) \;
  \mathcal{A}^{\text{right}}(\rho_{1'},\ldots,\rho_{4'},\sigma_4,\rho_{6'};\sigma_{1'},\ldots,\sigma_{2'} ) \q .
 \ea
Note that in this gluing procedure we do not have a summation appearing\footnote{If the amplitudes include a non-trivial dependence on the tail indices, a summation over $s_6=s_{5'}$ does appear.}. The reason is that the pair of identified indices $\rho_{6}=\rho_{5'}$ are converted into a branch index $\sigma_4$ for the glued state. This corresponds to the fact that one can still define a ribbon operator on the surface of the glued building block that goes around the glued (and thus bulk) plaquettes and whose measurement is encoded in the new branch index $\sigma_4$.

\subsection{Fusion basis transformations}

The glued building block has now ten instead of six plaquettes, and thus carries more data. To avoid an exponential growth in the data, we need to truncate it. We do this by using the built-in coarse-graining feature of the fusion basis: we summarize neighbouring plaquettes on the same face of the glued building block in effective plaquettes. Concretely, we need to summarize $l$ and $l'$, with $l=1,\ldots 4$, with an effective plaquette $\tilde l$.

However, the fusion tree resulting from the gluing (\ref{eq:glued_state}) is not well suited for this task: the ribbon operators measuring
curvature and torsion around the pairs $(l,l')$ of plaquettes are not diagonalized by this basis. As a result, we successively transform the fusion basis, and with it the underlying fusion tree, via F- and R-moves as defined in (\ref{Fmove}) and (\ref{eq:pulling_trafo}).

\begin{figure}[h!]
  \includegraphics[width=0.975\textwidth]{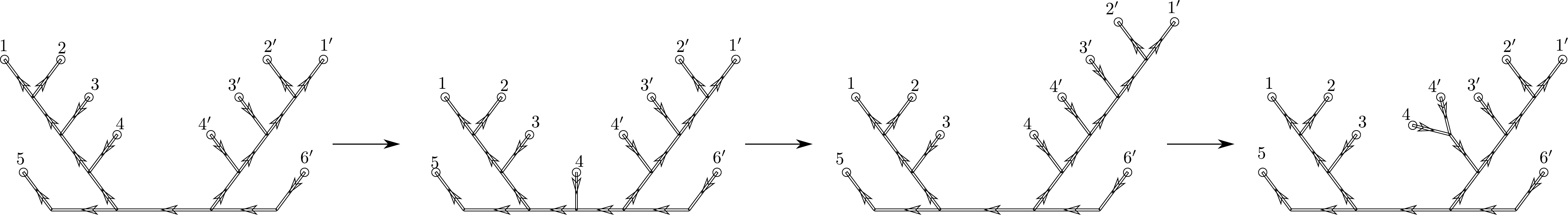}
  \caption{\label{fig:trafo_step_1}
  	First step of the fusion tree transformations. We move puncture $4$
   and attach it to puncture $4'$.  }
\end{figure}

\begin{figure}[h!]
  \includegraphics[width=0.975\textwidth]{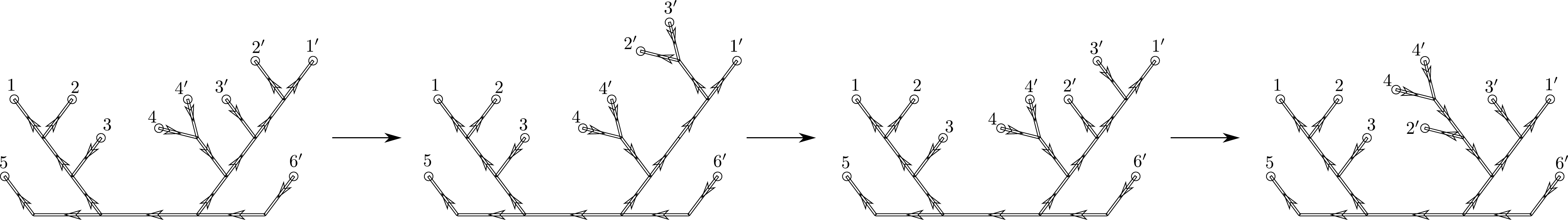}
  \caption{\label{fig:trafo_step_2}
  	Second step of the fusion tree transformations. We move puncture $2'$
  past several punctures, in particular past the fused pair of punctures $(4,4')$.}
\end{figure}

\begin{figure}[h!]
  \includegraphics[width=0.975\textwidth]{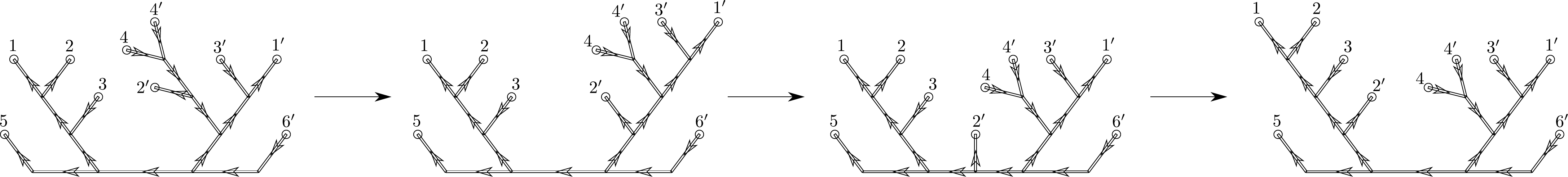}
  \caption{\label{fig:trafo_step_3}
  	Third step of the fusion tree transformations. We move puncture $2'$
  to the other side of the fusion tree.}
\end{figure}

\begin{figure}[h!]
  \includegraphics[width=0.975\textwidth]{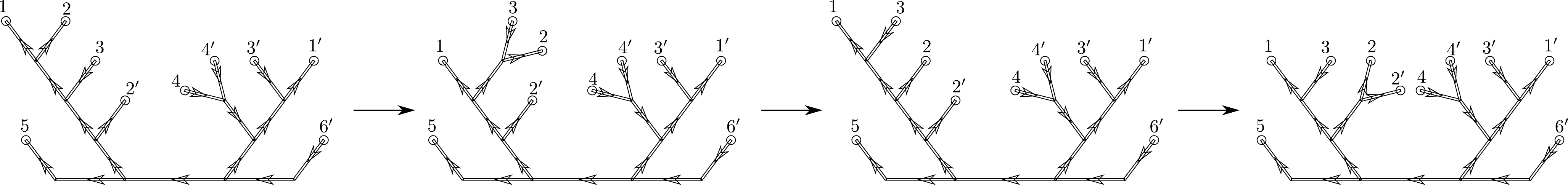}
  \caption{\label{fig:trafo_step_4}
  	Fourth step of the fusion tree transformations. We move puncture $2'$
  over and attach it to puncture $2$.}
\end{figure}

We first transform the tree for the gluing building block (\ref{eq:glued_state}) into a new tree, so that the puncture pairs $(2,2')$ and $(4,4')$, which lay on opposite faces, can be coarse-grained. The necessary transformations are split into several steps as depicted in Figs.~\ref{fig:trafo_step_1}--\ref{fig:trafo_step_4}\footnote{In the actual algorithm, some of these transformations
are performed before gluing the cubes together, since they are not affected by
the gluing, and they are faster to implement for the smaller fusion trees with only six
punctures instead of ten.}.

The first step consists of moving puncture $4$ over to
the other side of the fusion tree and attaching it to puncture $4'$, see
 Fig.~\ref{fig:trafo_step_1}.
Second, we move puncture
$2'$ over the other punctures on its side of the tree, in particular past the
fused pair of punctures $4$ and $4'$; see Fig.~\ref{fig:trafo_step_2}. Besides F-moves, this also requires the R-moves defined in Eq.~(\ref{eq:pulling_trafo}).
In the third step, we continue to move puncture $2'$ to the other side of the
fusion tree; see Fig.~\ref{fig:trafo_step_3}.
The final step is to move puncture $2$ past puncture $3$, and then to fuse it
directly to puncture $2'$; see Fig.~\ref{fig:trafo_step_4}.

\begin{figure}
	\includegraphics[width=0.35\textwidth]{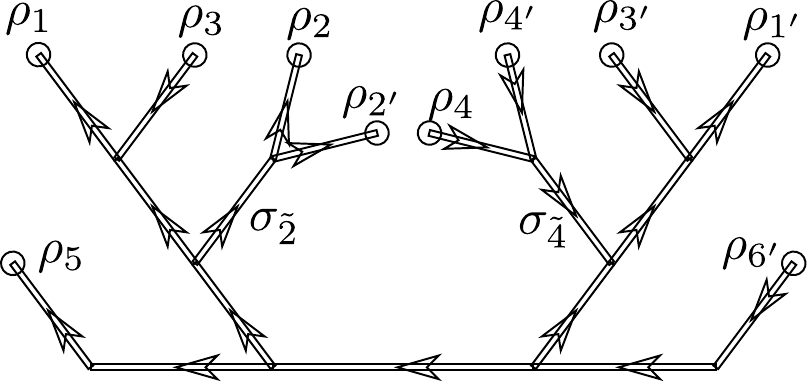}
	\caption{\label{fig:state_b_svd}
	The  fusion tree for the ten-puncture state, adjusted  for coarse-graining punctures $(2,2')$ and $(4,4')$. The ribbon operators surrounding these pairs act diagonally in this fusion basis, and its eigenvalues are determined by the labels $\sigma_{\tilde{2}}$ and $\sigma_{\tilde{4}}$, respectively.
	}
\end{figure}

After this series of transformations, we reach a fusion basis (Fig.~\ref{fig:state_b_svd}) which diagonalizes the ribbon operators around the pairs of punctures $(2,2')$ and $(4,4')$, respectively. Using the same principle as in (\ref{eq:tree_trafo}), the amplitude for the glued building block can be expressed with respect to this new basis. This allows us to truncate the amplitude by coarse-graining the pairs of punctures $(2,2')$ and $(4,4')$  into effective punctures $\tilde 2$ and $\tilde 4$, respectively. We explain this truncation procedure in Sec.~\ref{Sec:Truncation}.

\begin{figure}[h!]
  \includegraphics[width=0.975\textwidth]{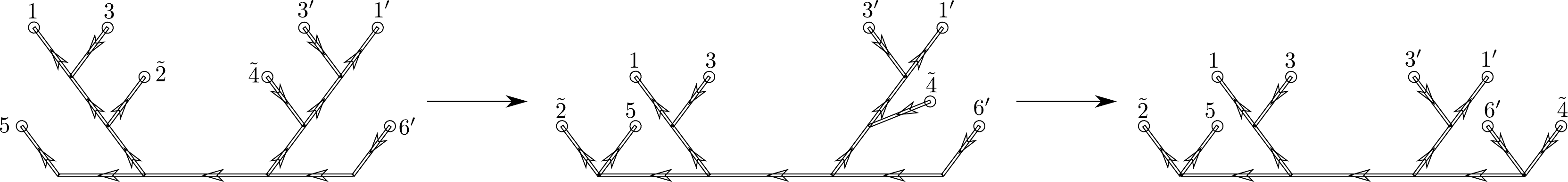}
  \caption{\label{fig:trafo_8p_step1}
  	Step five of the fusion tree transformations. We pull the effective punctures
  $\tilde{2}$ and $\tilde{4}$ across the strands and attach them to the
  punctures $5$ and $6'$, respectively.}
\end{figure}

\begin{figure}[h!]
  \includegraphics[width=0.975\textwidth]{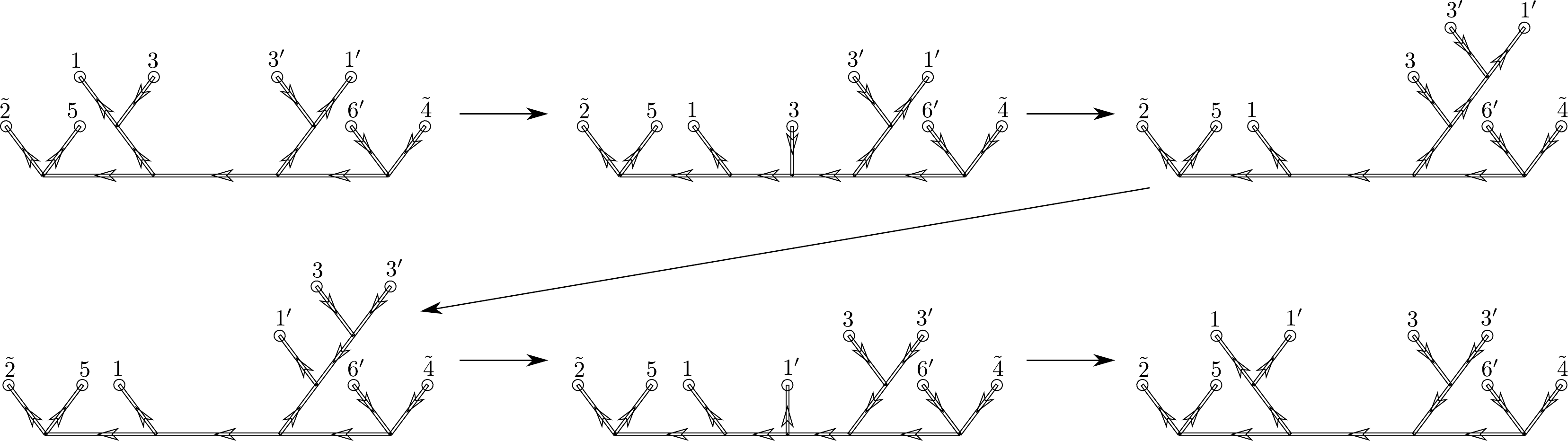}
  \caption{\label{fig:trafo_8p_step2}
  	Step six of fusion tree transformations. We first move puncture $3$
  to the other side of the fusion tree and attach it to $3'$. Then, we
  move $1'$ past the fused punctures $3,3'$ to the other side of the tree and
  attach it to $1'$.}
\end{figure}

%~\\
The truncated amplitude is now based on a building block with eight punctures. We now perform a further truncation corresponding to a coarse-graining of the pairs of punctures $(1,1')$ and $(3,3')$ into effective punctures $\tilde 1$ and $\tilde 3$, respectively. We perform the fusion tree transformations once more\footnote{It saves computational time to do these transformations after the first truncation step, as the truncated amplitudes involve less data.} (Fig. \ref{fig:trafo_8p_step1} and \ref{fig:trafo_8p_step2}).

First, we move the effective punctures $\tilde 2$ and $\tilde 4$ next to punctures $5$ and $6'$, respectively, in order to attach $\tilde 2$ to $5$ and $\tilde 4$ to $6'$.   These transformations
are also in preparation for the next iteration of the algorithm, in which the
punctures $\tilde{2},\tilde{4}$ are glued.

Next, we move punctures $3$ and $1'$ such that they are attached to $3'$ and $1$, respectively (Fig.~\ref{fig:trafo_8p_step2}).

This allows us to further truncate the amplitude, since we fused the pairs of punctures $(1,1')$ and $(3,3')$ into effective punctures $\tilde 1$ and $\tilde 3$, respectively. The resulting amplitude is now based on a building block with six punctures.

~\\

To start a new iteration of the coarse graining algorithm we have to first bring the
fusion tree back to the same form as for the original states (\ref{LR1}).  Having coarse-grained in a given direction (here the `left'-`right' horizontal direction) we next want that the next iteration step coarse grains in a different direction so that three consecutive iterations amounts to a coarse graining in all three directions. Each of these steps can be satisfied with another round of fusion basis transformations.

\begin{figure}[h!]
  \includegraphics[width=0.975\textwidth]{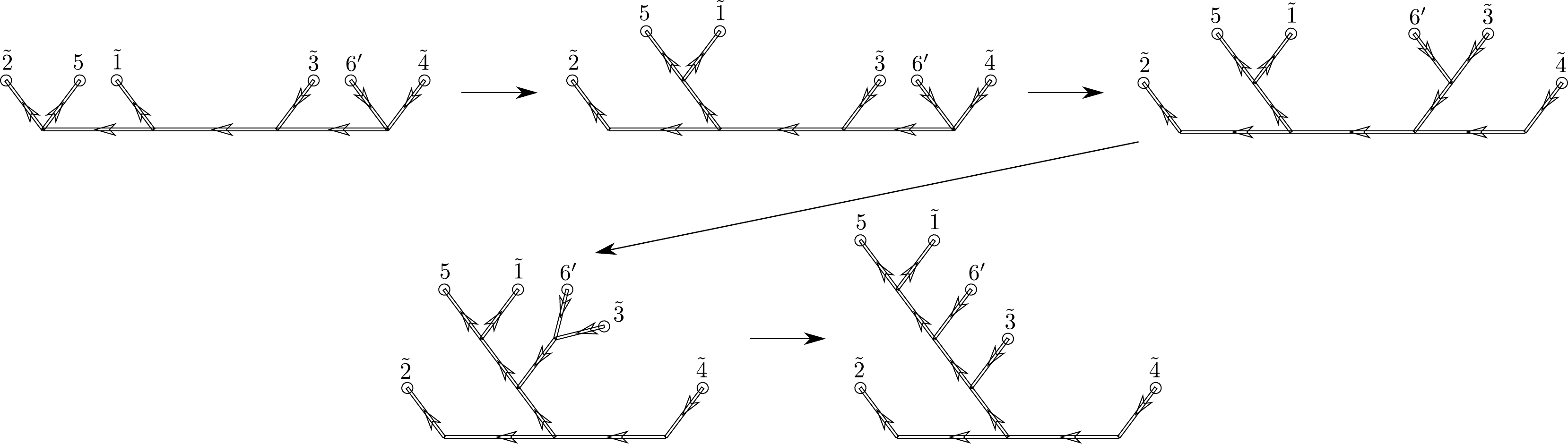}
  \caption{\label{fig:trafo_6p_ending}
  	Step seven. We change the tree back
  to the original form, preparing it for subsequent iterations.}
\end{figure}

We attach puncture $5$ to puncture $\tilde{1}$, then $6'$ to $\tilde{3}$, and then
pull $\tilde{3}$ over the strands. After that, we attach the combined punctures
$6'$ and $\tilde{3}$ to the combined punctures $5$ and $\tilde{1}$. Eventually,
we attach $6'$ directly below puncture $\tilde{1}$ to arrive back at the original
fusion tree (Fig.~\ref{fig:trafo_6p_ending}).

We have not yet completed the iteration, as we have to bring the
fusion tree back to the same form as for the original state. At the same time, we have to make sure that we coarse-grain the
amplitude in all spatial directions and eventually return to an amplitude
associated with a coarser cuboid.

The freedom to choose a fusion tree allows us to kill two birds with one stone.
We transform the fusion tree~\eqref{eq:6p_state_end} to the original
form in such a way that subsequent iterations of the same algorithm automatically coarse
grain punctures in all three space-time directions. This is straightforwardly
achieved by ordering the punctures according to the following scheme.

First, we attach puncture $5$ to puncture $\tilde{1}$, then $6'$ to $\tilde{3}$, and
then pull $\tilde{3}$ over the strands. After that, we attach the combined punctures
$6'$ and $\tilde{3}$ to the combined punctures $5$ and $\tilde{1}$. Eventually,
we attach $6'$ directly below puncture $\tilde{1}$ to arrive back at the original
fusion tree (Fig.~\ref{fig:trafo_6p_ending}).

\begin{figure}[h!]
  \includegraphics[width=0.975\textwidth]{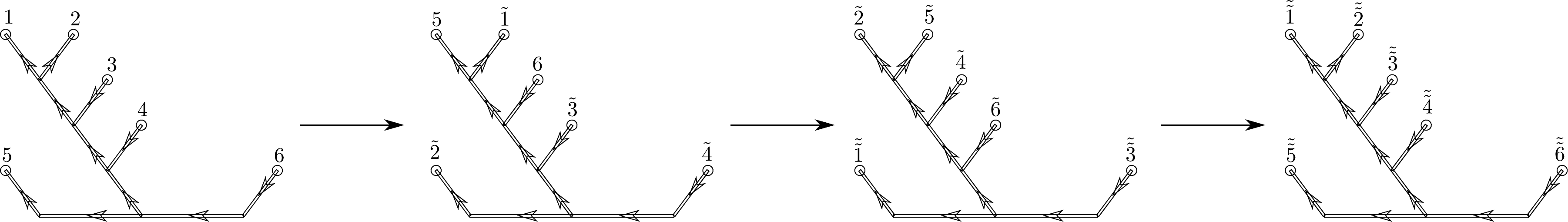}
  \caption{\label{fig:full_iteration}
  	We need three iterations of the algorithm to coarse-grain the
  amplitude associated with one cube into the one for a coarse cube.}
\end{figure}

~\\
Fig.~\ref{fig:full_iteration} shows how three consecutive iterations of the coarse graining algorithm do indeed perform a coarse graining in all three directions.

We started by gluing of a `left' cube with punctures $1,\ldots, 6$ to a `right' cube with punctures $1', \ldots 6'$, that is, a gluing in the $x$-direction. This gluing proceeded by identifying the puncture $6$ with $5'$. The corresponding label becomes a branch label. Furthermore, we coarse-grain  the pairs of punctures $(1,1'), \ldots, (4,4')$ into effective punctures $\tilde 1, \ldots, \tilde 4$, respectively.  We also have the punctures $5$ and $6'$ (renamed into $6$) from the original cubes. These punctures appear in the second fusion tree in Fig.~\ref{fig:full_iteration} (which coincides with the last tree appearing in Fig.~\ref{fig:trafo_6p_ending}).

We imagine that we perform this coarse-graining in a 3D cubical lattice, i.e., for each row with fixed $(y,z)$ coordinates we glue a cube positioned at $x=n$ to the neighbouring cube at $x=n+1$.

In the next iteration, we glue the resulting building block, with punctures $\tilde 1, \ldots, \tilde 4$ and $5,6$, to a neighbouring (now in the $y$-direction) building block . We again use primes to indicate punctures of this neighbouring building block. The gluing identifies the punctures $\tilde 4$ and $\tilde 2'$, and it coarse-grains the pairs of punctures $(5,5'),(\tilde 1,\tilde 1')$ and $(6,6'),(\tilde 3,\tilde 3')$ into effective punctures $\tilde 5, \tilde{\tilde 1}, \tilde 6, \tilde{\tilde 3}$, respectively. Additionally, we have the punctures $\tilde 2$ and $\tilde 4'$ (renamed into $\tilde 4$) from the previous building blocks.

The third iteration implements a gluing in the $z$-direction, as we now identify the puncture $\tilde{\tilde 3}$ with $\tilde{\tilde 1}'$ from the neighbouring building block. The (double-tilded) labels at the resulting fusion tree indicate that all punctures have undergone two coarse-grainings. Indeed, these punctures represent two-dimensional plaquettes and a coarse-graining in  three spatial directions in a cubical lattice leads to a coarse graining of quadruples of neighbouring plaquettes into one effective plaquette.  Dropping the tildes from the labels, we see that after three iterations of the algorithm we regain a fusion basis of the original form and with the original labeling.

Note that these fusion tree transformations are costly, in particular for larger
quantum group levels ${\rm k}$: an  F-move requires us to sum over a double index $\sigma=(j,\overline{j})$, but this sum
has to be computed for all possible labels of the new fusion tree. Thus, it is
imperative to keep these transformations to a minimum to avoid unnecessary
numerical costs, in particular for fusion trees with many punctures.

\subsection{Effective punctures from singular value decompositions}\label{Sec:Truncation}

We now describe the truncation procedure, which is based on replacing pairs of plaquettes, situated on the same face of the building block, with one effective plaquette.  In our algorithm, this is applied first to the pairs $(4,4')$ and $(2,2')$. These lie on opposite faces of the glued building block, and as we will see, will be dealt with by one truncation step. A second truncation coarse grains the pairs $(3,3')$ and $(1,1')$, which also lie on opposite faces.

\begin{figure}
  \includegraphics[width=0.3\textwidth]{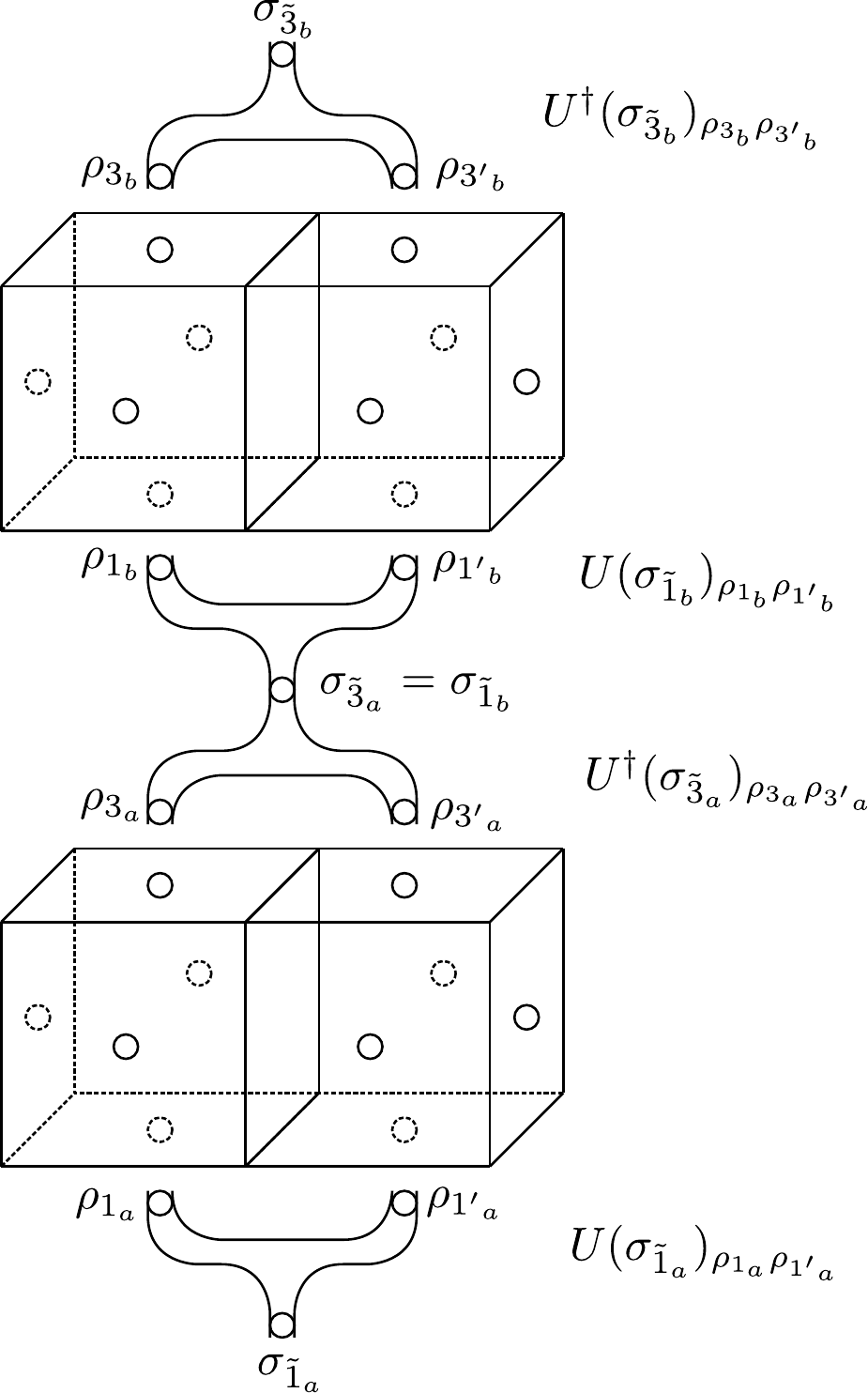}
  \caption{ \label{fig:Embedding-Map_Z}
  Inclusion of embedding maps from the lattice perspective. In order to change the partition function  as little as possible, we implement maps $U^\dagger U$ between neighbouring building blocks that coarse grain two punctures into one effective puncture. Seen from a single ten-puncture state, this implies gluing the map $U^\dagger$ to one coarse face while gluing $U$ to the opposite side.
  }
\end{figure}

Let us consider the first  truncation from the  lattice perspective (as opposed to considering just one building block): After the first gluing we pick a building block $a$ from the lattice. This building block has on its front side (Fig.~\ref{fig:cube_states}) a pair of plaquettes  $(4_a,4'_a)$, and on its back side a pair $(2_a,2'_a)$. In the lattice, the pair $(4_a,4'_a)$ is  identified with the pair $(2_b,2'_b)$ from a neighbouring building block $b$. For illustration, see also Fig.~\ref{fig:Embedding-Map_Z}, where we show the analogous situation for puncture pairs $(1,1')$ and $(3,3')$.

The lattice partition function is defined by summing the product of the cube amplitudes over the variables. Here, variables on faces that are identified with each other also need to be identified\footnote{Note that in our case we can have variables, which are associated with a cluster of neighbouring plaquettes situated on a cluster of  neighbouring faces. This applies to the labels of (non-leaf) branches in the fusion tree for the cubical building blocks. These labels encode the eigenvalues of ribbon operators encircling the given cluster of plaquettes. Let us consider the case of a pair of plaquettes on two neighbouring faces. We imagine that the two building blocks adjacent via this pair of faces are first glued, and that the resulting amplitude is expressed in a fusion basis which diagonalizes the ribbon operator going around the pair of identified plaquettes. In the summation procedure, one identifies and sums over the labels associated to the single plaquettes as well as the labels associated with the pair of plaquettes.}.
For the identified face, this leads to the identifications $\rho_{4_a}=\rho_{2_b}=:\rho$ (together with $s_{4_a}=s_{2_b}$) and $\rho_{4'_a}=\rho_{2'_b}=:\rho'$ (together with $s_{4_a}=s_{2_b}$). Additionally, we have $\sigma_{\tilde4_a}=\sigma_{\tilde 2_b}=:\sigma$.
Truncating means summing over fewer variables -- we replace the sum over the two identified $\rho$--variables and the $\sigma$--variable, with a sum over the $\sigma$--variable only.
This leaves us with the  $\rho$--variables on each of the building blocks. To explain how to deal with these variables, we insert a unity $\delta_{\rho_a\rho_b}\delta_{\rho'_a\rho'_b}$  into the summation over the $\rho$--variables and split this unity into a product of $\sigma$--dependent unitary maps
\ba
\delta_{\rho_a\rho_b}\delta_{\rho'_a\rho'_b}=\sum_A U(\sigma)_{(\rho_a{\rho'}_a) \, A} (U(\sigma)^\dagger)_{A \, (\rho_b{\rho'}_b)} \q
\ea
where we have introduced an auxiliary label $A$ (and an associated auxiliary Hilbert space).
Isolating the two building blocks and the summation over $\sigma,\rho,\rho'$ in the partition function, we summarize  all other indices on building block $a$ by $I_a$ and on $b$ by $I_b$. The summation then becomes
\ba\label{PartiTrunc}
{\cal Z}(I_a,I_b)= \sum_{\sigma} \sum_{(\rho_a{\rho'}_a),(\rho_b{\rho'}_b) } {\cal A}^a(I_a,(\rho_a{\rho'}_a),\sigma)  \sum_A U(\sigma)_{(\rho_a{\rho'}_a) \, A} (U(\sigma)^\dagger)_{A \, (\rho_b{\rho'}_b)}   {\cal A}^b(I_b,(\rho_b{\rho'}_b),\sigma)  \q .
\ea
The truncation\footnote{We are discussing a truncation that keeps the same amount of data as was on the original building blocks. See the end of this section for different methods to achieve higher-order truncations.} consists of replacing the summation over the index $A$ with $A=1$
\ba\label{Repl}
\sum_A U(\sigma)_{(\rho_a{\rho'}_a) \, A} (U(\sigma)^\dagger)_{A \, (\rho_b{\rho'}_b)}  \q \rightarrow  \q U(\sigma)_{(\rho_a{\rho'}_a) \, 1} (U(\sigma)^\dagger)_{1 \, (\rho_b{\rho'}_b)}  \q .
\ea
Thus, a crucial ingredient which determines the quality of the truncation is the choice of the unitaries $U(\sigma)$. Before coming to this choice, let us explain what this replacement achieves for our coarse-graining algorithm. The main point is that we can  do the summations over $(\rho_a{\rho'}_a)$ and $(\rho_b{\rho'}_b)$ locally on each building block. That is, we define the truncated amplitudes
\ba
{\cal A}^{a}_{\rm tr}(I_a,\sigma_a) &=&\sum_{ (\rho_a{\rho'}_a)}  {\cal A}^a(I_a,(\rho_a{\rho'}_a),\sigma_a)  U(\sigma_a)_{(\rho_a{\rho'}_a) \, 1} \q ,\nn\\
{\cal A}^{b}_{\rm tr}(I_b,\sigma_b) &=&\sum_{ (\rho_b{\rho'}_b)} (U(\sigma_b)^\dagger)_{1 \, (\rho_b{\rho'}_b)}   {\cal A}^b(I_b,(\rho_b{\rho'}_b),\sigma_b) \q .
\ea

In the lattice, each building block has a front and a back side, and thus functions both as an $a$--building block and a $b$--building block. That is, going back to the coarse graining of plaquette pairs $(2,2')$ and $(4,4')$ into $\tilde 2$ and $\tilde 4$, we have
\ba\label{truncA}
{\cal A}_{\rm tr}(I,\sigma_{\tilde{2}},\sigma_{\tilde{4}})= \sum_{ (\rho_2\rho_{2'}),(\rho_4\rho_{4'}) }(U(\sigma_{\tilde 2})^\dagger)_{1 \, (\rho_2{\rho}_{2'})} \,\, {\cal A}(I, \rho_2, \rho_{2'}, \rho_4,\rho_{4'},\sigma_{\tilde 2},\sigma_{\tilde 4} )  \,\,U(\sigma_{\tilde 4})_{(\rho_4{\rho}_{4'}) \, 1}
\ea
where $I$ denotes all labels of the glued building block, excluding the set $( \rho_2, \rho_{2'}, \rho_4,\rho_{4'},\sigma_{\tilde 2},\sigma_{\tilde 4})$. We have reached a truncated amplitude, which depends on fewer variables. The  maps $U(\sigma)_{(\rho\rho'1)}$, therefore, do define a coarse-graining\footnote{We also can read these maps the other way around by defining embeddings from a coarser (boundary) Hilbert space to a finer (boundary) Hilbert space. It assigns to the additional degrees of freedom described by $( \rho_2, \rho_{2'}, \rho_4,\rho_{4'})$ in the finer Hilbert space a localized notion of a $(\sigma_{\tilde 2},\sigma_{\tilde 4})$-dependent vacuum state~\cite{Dittrich12,TimeEvol}.}.

We come now to the choice of the $\sigma$--dependent unitary $U(\sigma)_{(\rho\rho')A}$. This choice is informed by our goal of minimizing the error made by the replacement (\ref{Repl}) in the summation (\ref{PartiTrunc}) for the partition function ${\cal Z}$.  This minimization is achieved when we define $U$ via a singular value decomposition (SVD) of the amplitude itself\footnote{We use the $a$--building block to extract $U$. We could have also used the $b$--building block. For a method that compares both possibilities and then chooses the one which minimizes the summation error, see~\cite{HigherOrder}.}:
\ba\label{SVD1}
 {\cal A}((I  \sigma_{\tilde{2}} \rho_2 \rho_{2'}    ),\sigma_{\tilde 4},(\rho_4 {\rho}_{4'})) &=& \sum_{B,A} V(\sigma_{\tilde 4})_{(I  \sigma_{\tilde{2}} \rho_2 \rho_{2'} )  B}\,\,\text{diag}( \lambda_I(\sigma_{\tilde 4}))_{BA} \,\, (U(\sigma_{\tilde 4})^\dagger)_{A (\rho_4{\rho}_{4'})} \q ,
\ea
where $V(\sigma)$ and $U(\sigma)$ are unitary matrices and $\text{diag}(\lambda_I(\sigma))$ is a rectangular diagonal matrix with entries $\lambda_{1}(\sigma)\geq \lambda_{2}(\sigma) \geq \cdots \geq 0$.
To define the SVD, we understand the amplitude ${\cal A}(I,\sigma,(\rho{\rho'}))$ as a $\sigma_{\tilde 4}$--dependent matrix $(M(\sigma_{\tilde 4}))_{LR}$ with left-index given by $L=(I  \sigma_{\tilde{2}} \rho_2 \rho_{2'})$ and a right-index given by $R=(\rho_4 {\rho}_{4'})$. We apply an SVD to this matrix in order to obtain the coarse-graining map $U(\sigma_{\tilde 4})_{(\rho_4{\rho}_{4'}) \, 1}$ used in~(\ref{truncA}).  The singular values $\lambda(\sigma_{\tilde 4})_{AA}$ can be used to control the quality of the approximation.

After coarse-graining the pair of punctures $(2,2')$ and $(4,4')$ into effective punctures $\tilde 2$ and $\tilde 4$, respectively, the truncated amplitude is expressed with respect to the following fusion tree
\begin{align}
  & \mathcal{A}_{\rm tr}
  = \sum_{\{\rho_l, \sigma_b\}} \mathcal{A}_{\rm tr} ( \{\rho_l; \sigma_b\}) \;
  \includegraphics[scale=0.5,valign=c]{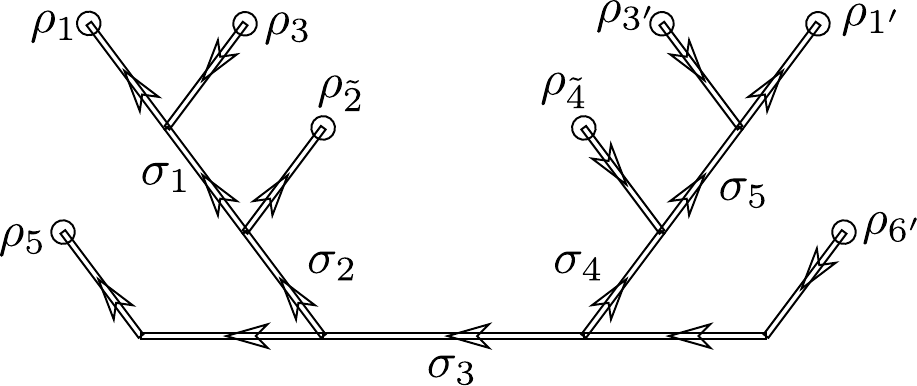} \; \; ,
\end{align}
where we renamed $\sigma_{\tilde 4}$ as $\rho_{\tilde 4}$, since it is now an index attached to a leaf. This is the amplitude that in the coarse graining algorithm must undergo steps five and six of the fusion basis transformations (Figs.~\ref{fig:trafo_8p_step1} and~\ref{fig:trafo_8p_step2}).

After these fusion basis transformations, we apply a second truncation in which we coarse grain the pair of punctures $(1,1')$ and $(3,3')$ into effective punctures $\tilde 1$ and $\tilde 3$. The procedure is the same as for the coarse-graining of $(2,2')$ and $(4,4')$. After this second truncation we reach a building block with the original number of punctures:
\begin{align} \label{eq:6p_state_end}
  & \mathcal{A}_{\rm tr tr}
  = \sum_{\{\rho_l, \sigma_b\}} \mathcal{A}_{\rm tr tr}(\{\rho_l, \sigma_b\}) \;
  \includegraphics[scale=0.5,valign=c]{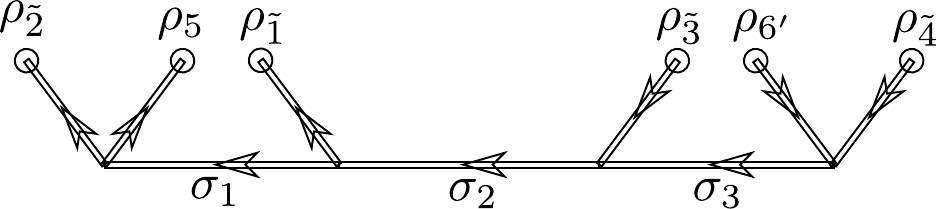} \; .
\end{align}
This amplitude needs to be transformed to a tree of the initial form, which is described in Fig.~\ref{fig:trafo_6p_ending}.

~\\
The truncation procedure via SVD is similar to other tensor network algorithms~\cite{Levin,GuWen,HigherOrder}. However, in our case we rewrite the amplitude as a matrix which depends on some of the variables, e.g., $\sigma_{\tilde 4}$ in~(\ref{SVD1}). The SVD needs to be performed for all allowed values of $\sigma_{\tilde 4}$. That is, we perform the truncation step for a fixed value of the associated observable, which is the ribbon operator going around the coarse grained pair of plaquettes. This allows us to continue interpretating these variables as the measurements of a coarse observable. Such ``passive'' variables appearing in the SVD were introduced for the decorated tensor network algorithm~\cite{DecTNW, DelcampSF}, where they also served as a way to maintain control over coarse observables. This allows us to compute expectation values of such coarse observables easily, and these can serve as order parameters.

We employ a truncation that brings us back to building blocks carrying the same amount of data as the initial ones.  The truncation can be improved by truncating the sum on the left side of~(\ref{Repl}) at a larger value of $A$. (Note, however, that the memory requirements, outlined in Sec.~\ref{numcost}, grow steeply with the level ${\rm k}$ for the truncation discussed here. The growth is polynomial in ${\rm k}$, but comes with a large power.) This introduces additional degeneracy labels on top of the fusion basis labels. To keep the observables and fusion basis structure intact it is better to permit building blocks with more plaquettes, e.g., four plaquettes on each of the faces. For non-Abelian groups (or quantum groups), this replaces labels $(\rho_l,s_l)$ with $(\rho_{la},s_{la})$, where $a=1,\ldots,4$, and three further representation labels $\sigma_{l1},\sigma_{l2},\sigma_{l3}$, which arise from fusing the four plaquettes.

\section{Application to  lattice gauge theories} \label{sec:lgt}

Having described the coarse graining algorithm, we now focus on its application to lattice gauge theory with a (quantum deformed) structure group $\SU(2)_{\rm k}$.  The fusion basis algorithm offers several advantages for lattice gauge theories. In particular, the coarse graining scheme allows easy access to the magnetic (or curvature) and electric (or torsion) charges throughout the coarse-graining procedure. The magnetic charges provide order parameters which distinguish between the weak- and strong-coupling phase. Although there might be no electric charges excited on the lattice level, such electric charges can appear in the coarse-grained region. Employing the fusion basis allows us to test different truncations that suppress such electric charges to various degrees.

\subsection{Amplitudes for quantum deformed lattice gauge theory}\label{SecS:amplitudes}

In this section, we define the amplitudes for Yang-Mills lattice gauge theory with a quantum-deformed gauge group $\SU(2)_{\rm k}$. A popular choice in the case of undeformed Lie groups is the Wilson action, which is expressed in terms of the plaquette holonomies $h_l=g_{e_1}g_{e_2}g_{e_3}g_{e_4}$, where $g_{e_i}$ are the group elements associated with the cyclically ordered edges enclosing the (square) plaquette $l$.  But in the quantum-deformed case, we do not have access to the holonomies. We will consider instead amplitudes based on the heat kernel action~\cite{Menotti:1981ry}.
%reference for heat kernel action?
For example, for $\SU(2)$ the amplitude is defined as
\ba\label{LGAmp1}
{\cal A}_\beta(\{g_e\})&=&\prod_l {\cal A}_l(h_l)  \q, \q\q\text{with}\q {\cal A}_l(h_l)= \sum_{k_l} d_{k_l} \exp\left(-\frac{1}{\beta} k_l(k_l+1)\right) \, \chi_{k_l}(h_l) \q .
\ea
The amplitude is a product of local amplitude factors associated with each of the plaquettes $l$. The plaquette factor is expanded into group characters $\chi_{k_l}(h_l)$, which are evaluated on the plaquette holonomy $h_l$. In light of generalizing these amplitudes to the quantum-group case, we note that this character $\chi_{k_l}(h_l)$ coincides with the value of the plaquette Wilson loop in the representation $k_l$.  The expansion coefficients are given by the product of the dimension $d_{k_l}=2k_l+1$ and the exponential $\exp(-\frac{1}{\beta} k_l(k_l+1))$, where $C_{k_l}=k_l(k_l+1)$ is the value of the Casimir for the representation $j_l$.

The state described by the amplitude~\eqref{LGAmp1} can be obtained by starting from the state $\psi_{\rm SC}(\{g_e\})\equiv 1$ and multiplying it by a factor ${\cal A}_l$ for each plaquette $l$. These plaquette operators are sums over insertions of Wilson loop operators  with certain weights.

Here $\psi_{\rm SC}$ is the state describing the strong coupling limit. This state is obtained in the limit  $\beta\rightarrow 0$, where only the terms with $k_l=0$ survive.  For $\beta\rightarrow \infty$, the exponential factor goes to one and the plaquette amplitude defines the $\SU(2)$--delta--function ${\cal A}_l(h_l) =\sum_{k_l} d_{k_l}  \, \chi_{k_l}(h_l)\,=\, \delta(h_l)$.
The relation between $\beta$ and the usual Yang-Mills lattice coupling constant $g$ is given by $\beta\sim\frac{1}{ag^2}$, where $a$ is the lattice constant.

To adopt this amplitude to the quantum-group case, we also start from the state $\psi_{\rm SC}$ describing the strong coupling limit. In a spin network basis (which also exist for the quantum-group case), the amplitude is only non--vanishing if all representation labels are trivial. To obtain a quantum-deformed version of~(\ref{LGAmp1}), we also apply local plaquette operators to this state
\ba
{\cal A}_\beta &=& {\cal D} \left(\prod_l  {\bf B}_l(\beta) \triangleright \right) \psi_{\rm SC} \, ,\q \text{with} \q\q  {\bf B}_l(\beta)=\frac{1}{\cal D}\sum_{k_l} v^2_{k_l}   \exp\left(-\frac{1}{\beta}[ k_l][k_l+1]\right) \, {\bf W}^{k_l}_l
\ea
where $v_k^2=(-1)^{2k}d_k$ is the signed quantum dimension, $[j]$ is the quantum number, ${\cal D}$ is the total quantum dimension of $\SU(2)_{\rm k}$ (see Appendix~\ref{app-basics}), and ${\bf W}^{k}_l$ is an operator that inserts a Wilson loop\footnote{The Wilson loop operator coincides with the ribbon operator ${\bf R}^{k0}_l$ or ${\bf R}^{0k}_l$. A (closed) ribbon operator ${\bf R}^{k_ok_u}$ inserts two parallel Wilson loops labelled with $k_o$ and $k_u$ into a state $\psi$, which itself may be defined via a network of labeled (spin network) strands. The loop labelled with $k_o$ over-crosses all strands in $\psi$ and the loop labelled with $k_u$ under-crosses all strands in $\psi$. We assume $\psi$ has vanishing electric charge (i.e., it is torsionless) at the punctures: there is a region around each puncture without any pre-existing strands. Thus, an over-crossing and an under-crossing Wilson loop around a puncture lead to the same action.} in the representation $k$ around the puncture $l$.

Going from the Lie group to the quantum group, we replace the quantum dimension with the signed quantum dimension, the exponential $\exp(-\frac{1}{\beta} k(k+1))$ with $\exp(-\frac{1}{\beta}[ k][k+1])$ and the Wilson loop operator with its quantum-deformed equivalent. With these choices (in particular by using the signed quantum dimension) we regain the strong coupling state (modulo a normalization) for $\beta \rightarrow 0$ and, as we will see below, the weak-coupling state for $\beta\rightarrow \infty$.

\begin{figure}
	\includegraphics[width=0.6\textwidth]{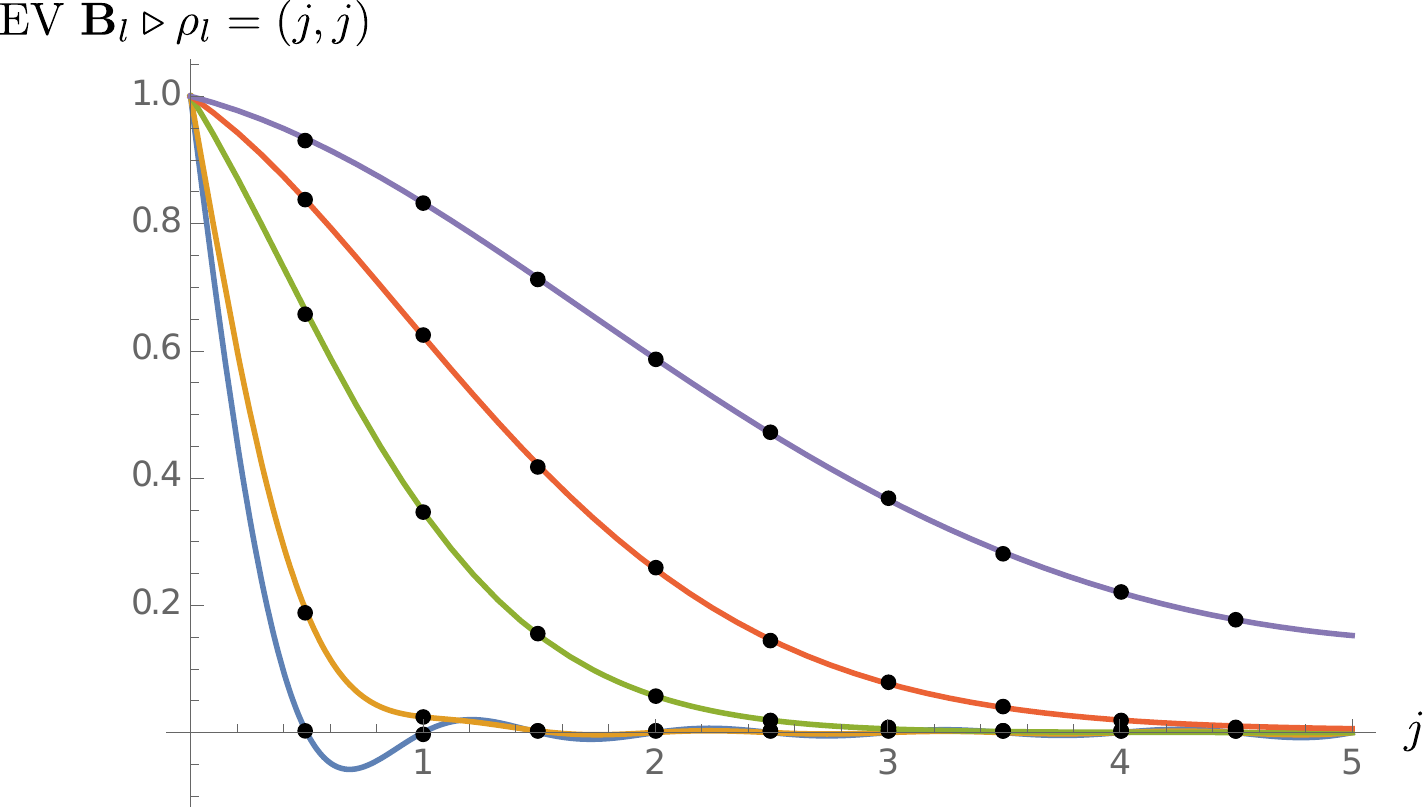}
	\caption{
		The (normalized) eigenvalues of the operator ${\bf B}_l$ acting on a leaf state $\rho_l = (j,j)$ for different values of $\beta$ and for ${\rm k}=10$. The eigenvalues define the plaquette weights and are here normalized to be equal to 1 for $j=0$.  From top to bottom we show $\beta = \frac{1}{2}$, $\beta = 1$, $\beta = 2$, $\beta = 10$ and $\beta = \infty$. Note that the weights are only evaluated at half-integer $j$  marked in black. Thus, the bottom curve  represents the weak coupling limit, as the weight is non-vanishing only for $j=0$.
		\label{fig:heat_kernel_weight}
	}
\end{figure}

Now the operators ${\bf W}^k_l$ and, therefore, the operators ${\bf B}_l$, are diagonalized in the fusion basis. For states without electrical charge at the leaf $l$, we have a leaf label $\rho=(j,j)$ and the Wilson loop operator ${\bf W}^k_l$ acts as
\ba\label{tHk}
{\bf W}^k_l \triangleright \includegraphics[scale=0.7,valign=c]{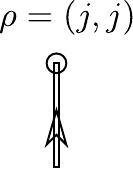} &\,=\,&   \frac{S_{kj}}{S_{0j}}  \includegraphics[scale=0.7,valign=c]{Drawings/punc_no_tor} \,\, ,\nn\\
{\bf B}_l (\beta)\triangleright \includegraphics[scale=0.7,valign=c]{Drawings/punc_no_tor} &\,=\,&\left(\frac{1}{\cal D}\sum_{k} v^2_{k}   \exp\left(-\frac{1}{\beta}[ k][k+1]\right) \,  \frac{S_{kj}}{S_{0j}}\,\, \right) \includegraphics[scale=0.7,valign=c]{Drawings/punc_no_tor} \; ,
\ea
where $S_{kj}$ is the S-matrix associated with the fusion category $\SU(2)_{\rm k}$ (see Appendix~\ref{app-basics} for its explicit definition). In particular, we have $S_{0j}=v_j^2/{\cal D}$. The S-matrix is unitary and for $\SU(2)_{\rm k}$ has real entries, i.e., $\sum_k S_{k0}S_{jk}=\delta_{j0}$. Thus, we see that we indeed regain the vacuum state in the limit $\beta \rightarrow \infty$. Fig.~\ref{fig:heat_kernel_weight} shows plots for the eigenvalues of ${\bf B}_l (\beta)$ for a range of $\beta$'s.

Therefore, it is easiest to define the heat kernel states ${\cal A}(\beta)$ in the fusion basis. To this end, we need to express the strong coupling state $\psi_{SC}$ in the fusion basis. This is given by~\cite{DittrichLambda}
\ba
\psi_{SC}( \{(j_l,j_l) ;(i_b,\overline{i_b})\})\,=\, \frac{1}{{\cal D}^{\sharp l-1}}\text{CoupCond}( \{(j_l,j_l) ;(i_b,\overline{i_b})\})  \prod_l v_{j_l}     {\cal P}^{\pm}_l(j_l) \prod_b \delta_{i_b \overline{i_b}}
\ea
where $\sharp l$ denotes the number of punctures and ${\cal P}^{\pm}_l(j_l)$ are local phase factors that depend on the orientation $\pm$ of the state around the puncture $l$. See the end of this section for the explicit phase factors with respect to the six-puncture fusion basis that we use for the cubical building blocks. The function $\text{CoupCond}$ is equal to one if the labels $( \{(j_l,j_l) ;(i_b,\overline{i_b})\})$ of the fusion tree satisfies all coupling conditions and is vanishing otherwise.
Thus the heat kernel amplitude is given by
\ba\label{HKAqd}
{\cal A}_\beta( \{(j_l,j_l) ;(i_b,\overline{i_b})\}) &=&{\cal D}^2 \text{CoupCond}( \{(j_l,j_l) ;(i_b,\overline{i_b})\}) \prod_b \delta_{i_b \overline{i}_b} \, \prod_l  \left(
\frac{v_{j_l} {\cal P}^{\pm}_l(j_l) }{{\cal D}^2}\sum_{k_l} v^2_{k_l}   \exp\left(-\frac{1}{\beta}[ k_l][k_l+1]\right) \,  \frac{S_{k_l j_l}}{S_{0j_l}}\,\,
\right).\nn\\
\ea
Note that the phase factors ${\cal P}^{\pm}_l(j_l)$ cancel out if one glues two building blocks by identifying a pair of plaquettes (identified with leaves), as these have to come with opposite orientation.

~\\
{\bf Strong coupling state for the cuboid with six plaquettes:}
In the fusion basis appearing for the `left' cubical building block in~(\ref{LR1}), the strong coupling state appears as
\ba
\psi_{SC}(j_1,\ldots,j_6; (i_1,\overline{i_1}), \ldots, (i_3,\overline{i_3}) )&=&  \frac{1}{{\cal D}^5} \delta_{i_1\overline{i_1}}\delta_{i_2\overline{i_2}}\delta_{i_3\overline{i_3}} (R^{j_2j_1}_{i_1})^*(R^{i_3j_5}_{j_6})^* R^{j_4 i_3}_{i_2} R^{j_3 i_2}_{i_1}  v_{j_1}v_{j_2} \cdots v_{j_6} \nn\\
&=& \frac{1}{{\cal D}^5} \delta_{i_1\overline{i_1}}\delta_{i_2\overline{i_2}}\delta_{i_3\overline{i_3}} \delta_{j_1j_2i_1}\delta_{i_3j_5j_6}\delta_{j_4i_3i_2} \delta_{j_3i_2i_1}  \nn\\
&&\q\q\q\q\q
\,\,\prod_{l=1,2,5} (-1)^{j_1} q^{\frac{1}{2} j_l(j_l+1) } \,\,\prod_{{l'}=3,4,6} (-1)^{-j_{l'}} q^{-\frac{1}{2} j_{l'}(j_{l'}+1) }  \q .
\ea
The  leaf indices are given by $\rho_l=(j_l,j_l)$ with $l=1,\ldots,6$ and the branch indices by $\sigma_b=(i_b,\overline{i_b})$ with $b=1,2,3$.
The $R$-symbols are defined as $R^{ab}_c=(-1)^{c-a-b} (q^{c(c+1)-a(a+1)-b(b+1)})^{1/2}$ if $(a,b,c)$ satisfy the $\SU(2)_{\rm k}$ coupling conditions (that is, if $\delta_{abc}=1$), and they vanish if the coupling conditions are not satisfied ($\delta_{abc}=0$). Here $q=\exp(2\pi \i/({\rm k}+2))$ is a pure phase. As $c-a-b$ is an entire number (due to coupling conditions) we also have $(-1)^{c-a-b}=(-1)^{-c+a+b}$.  Hence, the phase factors only depend on the representation labels associated with the leaf indices $j_l$, whereas the dependence on $i_b$ cancels out. Note that the coupling conditions for the $j_l$ and $i_b$ indices are included in the definition of the $R$--symbols.

\subsection{On the appearance of electrical charges  under coarse graining}

We defined the heat kernel amplitudes in Sec.~\ref{SecS:amplitudes} by applying only plaquette operators to the strong coupling state. A more elaborate possibility is to add Wilson loop operators\footnote{For loops around several plaquettes or punctures one needs to specify whether one has an under-crossing or over-crossing Wilson loop or a combination of both, i.e., a ribbon operator.} that surround pairs or even larger clusters of plaquettes.  Such more non-local constructions would appear if one tries to approximate better the vacuum state associated with a given coupling.

As we see in a moment, the heat kernel amplitudes as defined in Sec.~\ref{SecS:amplitudes} do not feature excitations of electrical charge labels or torsion, even if we consider ribbon operators around several plaquettes. We also show the alternative definition, which employs more non-local operators, leads to such excitations.
Furthermore, we show that by gluing states for larger building blocks, we might also encounter such electrical charge or torsion excitations. This is the reason why torsion excitations might appear in the coarse-graining algorithm.

The definition of the heat kernel amplitudes in~\eqref{HKAqd}
includes the condition that for all fusion basis labels $\rho_l$ and $\sigma_b$ we have $\rho_l=(j_l,j_l)$ and $\sigma_b=(i_b,i_b)$.  As electric charge excitations are characterized by labels $(j,\overline{j})$ with $j\neq \overline{j}$, we see that we do not have such excitations, neither for the leaves nor for the branches. Yet, the amplitudes are specified with respect to a particular fusion tree in a way that it could happen that such excitations appear after a tree transformation. For an F-move, given by
\begin{equation}\label{Fmove2}
  \includegraphics[scale=0.5,valign=c]{Drawings/tree_trafo_1} \,\,\,=\,\,\, \sum_{\sigma_4}
 \mathbb{F}^{\rho_1 \, \sigma_1 \, \sigma_2}_{\sigma_3 \, \rho_2 \, \sigma_4} \,% F^{j_2 \, j_1 \, b_1}_{b_2 \, j_3 \, b'_1}
  \; \includegraphics[scale=0.5,valign=c]{Drawings/tree_trafo_2} \; ,
\end{equation}
the transformation involves the sum
\ba\label{NoTorsion1}
 \mathcal{A}'_\beta(\rho_1,\rho_2;\sigma_1,\sigma_4,\sigma_3,\ldots) &=& \sum_{\sigma_2}  \,\,  \mathbb{F}^{\rho_1 \, \sigma_1 \, \sigma_2}_{\sigma_3 \, \rho_2 \, \sigma_4} \,\,\, \mathcal{A}_\beta(\rho_1,\rho_2;\sigma_1,\sigma_2,\sigma_3,\ldots)\nn\\
 &=&
  \sum_{i_2, \overline{i_2}}
  F^{j_1 \,i_1 \, i_2}_{i_3 \, j_2 \, i_4}
 F^{\overline{j_1} \,\overline{ i_1} \,\overline{ i_2}}_{\overline{i_3} \,\overline{ j_2} \,\overline{ i_4}} \,\,
 \mathcal{A}_\beta((j_1,\overline{j_1}),(j_2,\overline{j_2});(i_1,\overline{i_1}),(i_2,\overline{i_2}),(i_3,\overline{i_3}),\ldots)\nn\\
&=& \left(
 \sum_{i_2, \overline{i_2}}   F^{j_1 \,i_1 \, i_2}_{i_3 \, j_2 \, i_4}
 F^{\overline{j_1} \,\overline{ i_1} \,\overline{ i_2}}_{\overline{i_3} \,\overline{ j_2} \,\overline{ i_4}}  \,\, \delta_{i_2\overline{i_2}}  \,\, \delta_{j_1\overline{j_1}} \delta_{j_2\overline{j_2}} \delta_{i_1\overline{i_1}}\delta_{i_3\overline{i_3}}
                 \right)
  {\cal A}^{\rm red}_\beta(  j_1,j_2,\ldots)
  \nn\\
  &=&
\delta_{i_4 \overline{i_4}}   \,\, \delta_{j_1\overline{j_1}} \delta_{j_2\overline{j_2}} \delta_{i_1\overline{i_1}}\delta_{i_3\overline{i_3}}   \,\, \delta_{i_3i_1i_4}\delta_{j_1j_2i_4}
   \,  {\cal A}^{\rm red}_\beta(  j_1,j_2,\ldots)
\ea
To go from the first to the second line we use the expression for the Drinfeld Double ${\mathbb F}$--symbol in terms of the $\SU(2)_{\rm k}$ $F$-symbols (see Appendix~\ref{app-FB}). From the second to the third line we employ the condition that the initial lattice gauge theory amplitude function $\mathcal{A}_\beta$ takes the form
\be
\mathcal{A}_\beta((j_1,\overline{j_1}),(j_2,\overline{j_2});(i_1,\overline{i_1}),(i_2,\overline{i_2}),(i_3,\overline{i_3}),\ldots)= \delta_{i_2\overline{i_2}}  \,\, \delta_{j_1\overline{j_1}} \delta_{j_2\overline{j_2}} \delta_{i_1\overline{i_1}}\delta_{i_3\overline{i_3}} {\cal A}^{\rm red}_\beta(  j_1,j_2,\ldots)\q .
\ee
 To arrive at the fourth line we use the tetrahedral symmetry properties as well as the orthogonality relation for the $F$-symbols (see Appendix~\ref{app-basics}).

The transformed amplitude $\mathcal{A}'_\beta$ again has vanishing electric excitations for the leaves and branches, and also has a trivial dependence (apart from $\delta_{i\overline{i}}$ factors) on the branch indices. For the  R-transformations (see Eq.~\ref{eq:pulling_trafo}) we note that these are trivial for states without electric charges, and thus cannot lead to the appearance of such charges. We conclude that the initial lattice gauge theory amplitudes have vanishing electric charge for all plaquettes and clusters of plaquettes.

Equation~(\ref{NoTorsion1}) would change if the amplitudes have a more non-trivial dependence on the labels $(i_b,\overline{i_b})$, even if we still have the $\delta_{i_b \overline{i_b}}$ factor. We obtain such a dependence if we glue two heat kernel amplitudes, as described in Sec.~\ref{Sec:gluing}. There the punctures $6$ of the left cube and the puncture $5'$ of the right cube are identified with each other. That is, the corresponding labels $\rho_6=\rho_{5'}=:\sigma_4$ are set equal and now serve as a branch label for the glued tree. But, as the punctures $6$ and $5'$ carry non-trivial weights $\lambda_{j_6}$ and $\lambda_{j_5'}$, these are now associated with the branch label $\sigma_4=(i_4,\overline{i_4})$. Summing over $(i_4,\overline{i_4})$ we do not only have the Kronecker-Delta $\delta_{i_4\overline{i_4}}$, but also a weight $\lambda^2_{i_4}$. This weight renders the orthogonality relation for the $F$-symbol non-applicable, and thus it may cause the appearance of electrical excitations $i_b \neq \overline{i_b}$ for the transformed tree.

Because the coarse-graining algorithm consists of iterations of gluings, tree transformations, and truncations (in which branch labels are transformed to puncture labels), we can expect the appearance of electrical charges or torsion for the effective punctures.

Electric charges do not appear in the weak coupling limit, since this describes a state where both magnetic and electric charges are suppressed. For the strong coupling limit we have all possible magnetic charges excited, but the electrical charges are suppressed. This feature is preserved under tree transformations (and gluings) as the weights $\lambda_j$ are trivial in the strong coupling limit.

Note that we could have also defined initial amplitudes which do not feature electric charges at the punctures (because the initial amplitudes are gauge invariant), but do so for branches, i.e., for certain clusters of plaquettes. The same argument that showed electric charges appear under coarse-graining also shows that such charges can appear after tree transformations if we include more non-local Wilson-loop operators in the construction of the amplitudes.  The work~\cite{DittrichLambda} also constructs gauge invariant states that can be interpreted as being peaked on (geometrically) homogeneous curvature, but also show electrical charge excitations for clusters of plaquettes.
The appearance of electrical charge or torsion excitations for coarser regions, which include magnetic or curvature excitations, can be explained with the help of a geometric interpretation of the ribbon operators, see~\cite{DittrichGeillerFlux}.

\subsection{Range of models}\label{Sec:parameters}

Before describing the results of the coarse-graining algorithm, we summarize the parameters of the model and introduce further modifications.
Choosing the heat kernel amplitudes~(\ref{HKAqd}) as initial amplitudes for the coarse-graining procedure, we have two parameters:
\begin{itemize}
  \item {\bf Level ${\rm k}$ of the quantum group $\text{SU}(2)_{\rm k}$}:
  The level ${\rm k}$ of the quantum group determines the maximal admissable representation label
  $j_{\text{max}} = \frac{\rm k}{2}$ of the system. Thus, larger ${\rm k}$ implies higher-dimensional boundary Hilbert spaces. The level ${\rm k}$ can be understood as an effective (inverse) discretization length for the group manifold $\SU(2)$: the eigenvalues  of the Wilson loop operator correspond to a discretization of the class angle $\theta \in [0,\pi]$ of $\SU(2)$ as $\theta\sim \pi (2j+1)/({\rm k}+2)$.

  In the limit of ${\rm k}\rightarrow \infty$ one regains the classical Lie group $\SU(2)$. For the minimal choice ${\rm k}=1$ one has an Abelian fusion algebra, where only the representations $j=0$ and $j=\tfrac{1}{2}$ (both with quantum dimension equal to one) are allowed.

  \item {\bf Coupling constant $g$:} We will use
  $g=1/\sqrt{\beta}$ as the lattice coupling constant. The limit $g \rightarrow 0$
  describes the lattice weak coupling limit, in which all magnetic (or curvature) and electric (or torsion) excitations are suppressed. The weak coupling amplitudes coincides with the amplitudes of the Tuarev-Viro model, which describes a topological field theory~\cite{turaev-viro}.
  The dual state is reached in the strong coupling limit $g \rightarrow \infty$,
  where all curvature charges are excited (with a constant probability distribution), but with torsion suppressed. Both the weak and strong coupling limits constitute fixed points of the coarse-graining algorithm. (In fact, both limits describe topological, that is, triangulation-invariant state sum models.  However, the strong coupling limit leads to a trivial model.)
\end{itemize}
Since we work with finite systems, we
expect each of these limiting cases to come with an extended phase (for fixed ${\rm k}$ defined by the values of $g$ for which the models flow under coarse-graining to the corresponding fixed point) and a phase
transition separating them. Our goal is to find this phase transition (depending
on ${\rm k}$ and $g$) and study its properties using our coarse-graining algorithm.
Before we discuss the results in detail, we briefly mention three different
implementations of our coarse-graining algorithm.

As explained in Sec.~\ref{sec:algorithm}, we define new effective punctures
through a singular value decomposition. We perform such a
singular value decomposition for each label $\rho_{\tilde{l}}=(j_{\tilde l},\overline{j_{\tilde l}})$ that the new effective puncture can have.

We have implemented two versions of these algorithms:
\begin{itemize}
  \item {\bf Torsion:}
  We allow the effective puncture labels $\rho_{\tilde{l}}$ to take $j_{\tilde l}\neq \overline{j_{\tilde l}}$, that is, we allow for torsion to appear. (Note that torsion also appears for the branches, and it can be measured by ribbon operators going around several plaquettes.)

  \item {\bf  No torsion at punctures (NTP):}
  As the name suggests, we suppress torsion for the effective punctures. Note that we still allow torsion on the branch labels, so we permit branch labels $i_b\neq \overline{i_b}$. Also, we still perform an SVD for effective puncture labels with $j_{\tilde l}\neq \overline{j_{\tilde l}}$. This allows us to compare the highest singular values for these labels with those coming from labels with $j_{\tilde l}= \overline{j_{\tilde l}}$, and in this way lets us estimate the relevance of the torsion degrees of freedom.
  However, for the subsequent definition of the coarse-grained amplitude, we implement the condition $j_{\tilde l}= \overline{j_{\tilde l}}$ for the labels  associated with the effective punctures. This truncation is equivalent to projecting back to the  lattice gauge theory Hilbert space of gauge invariant wave functions after each coarse graining step.

  \item  {\bf Fully truncated torsion (FTT):}
 One can further truncate the NTP version, by forbidding also torsion excitations on all branches, i.e., by requiring $i_b = \overline{i_b}$ for all branch labels $\sigma_b=(i_b,\overline{i_b})$.  This truncation constitutes a restriction of the gauge invariant lattice gauge theory Hilbert space. The truncation leads to considerabe simplifications of the fusion tree transformations, e.g., the R-moves become trivial. Further, the number of configurations which need to be computed and saved in each step is reduced considerably, making the algorithm more economical in terms of computational time and memory usage.
\end{itemize}

These three versions of the algorithm allow us to test whether neglecting torsion degrees of freedom changes the results of the coarse graining algorithm significantly. Although the fusion basis algorithm allows for a systematic treatment of torsion, it does naturally increase the required amount of computational resources in a very significant way. We discuss this in  more detail in the next section. The algorithm presented here allows us to test various truncations, which then can lead to huge computational savings.

Furthermore, we consider a variant of the model where we allow only a restricted set of representation labels to appear. We use the condition that the set of integer representations $j \in {\mathbb N}$ (as opposed to half-integers) is closed under fusion. That is, an initial amplitude, which is only non-vanishing for integer labels, will keep this property under coarse-graining. Note that the representation labels here are labels of the fusion basis.
Allowing only integer labels can be understood as allowing fewer values of the curvature class angle, for instance, according to $\theta\sim \pi (2j+1)/({\rm k }+2)$.
\begin{itemize}
  \item {\bf Integer representation labels only model:}
We only allow for integer representation labels throughout ($j_l,\overline{j_l}\in {\mathbb N}$ and $i_b,\overline{i_b}\in \mathbb{N}$). This lets us write a simpler code which neglects non-integer representations, which in turn
  reduces the total number of configurations.
  For the amplitude, we use the same plaquette weights as in~(\ref{HKAqd}), but we allow only integer arguments $j_l$ and $i_b$. Note that the sum over the representation label $k_l$ appearing in~(\ref{HKAqd}) still is taken over half-integers\footnote{Summing over only integer $k_l$ we obtain a plaquette amplitude ${\cal A}_l(j_l)$, which is symmetric around $j_{\rm max}/2$.}.
  The strong coupling limit of this model differs from the one for the general model because we have only all integer labels appearing\footnote{In the spin network basis, this strong coupling limit  of the integer model describes an amplitude which is only non-vanishing for representations $j=0$ and $j=j_{\rm max}$. The appearance of $j_{\rm max}$ in addition to $j=0$ is due to the restriction of integer representations in the fusion basis.}. We test the integer models with the Torsion and NTP version of the algorithm.
\end{itemize}

In the next section we briefly outline optimizations which make the coarse-graining algorithm more feasible to implement.

\subsection{Remarks on the numerical implementation and costs}\label{numcost}

For tensor network coarse-graining algorithms, the building blocks carrying the most data, i.e., the most
configurations to store, set the limiting factor in terms of memory consumption
and operational costs. For our algorithm, these are the building blocks with ten punctures. Ignoring coupling conditions and allowing for the existence of tail indices, we have 20 $j$--labels and 10 $s$--labels for the punctures, as well as $14$ $i$--labels for the branches. At value ${\rm k}$, each $j,i$ and $s$ label can have $({\rm k}+1)$ values, giving $({\rm k}+1)^{44}$ combinations. Already for ${\rm k}=2$ this would lead to the use of $10^{13}$ GB memory, which is not a feasible amount even in modern clusters.

Fortunately, there are two steps which drastically reduce these memory requirements: first, our initial amplitudes do not depend on the tail labels $s_l$, and we use a truncation that does not introduce a dependence on these labels. Therefore, we can ignore these labels altogether, leading to a reduction  in the number of labels from 44 to 34. A second significant reduction results from implementing the coupling rules for the fusion basis labels. Doing so, the ${\rm k}=2$ example only requires 43 GB of memory.

Let us explain this in more detail. At each three-valent vertex of the fusion tree, the three associated pairs of labels must satisfy coupling rules. This puts restrictions on the admissible labelings of the fusion tree, and it leads to a significant reduction in the number of allowed configurations. The labels for the allowed configurations can be represented by a ``super-index.'' Such super-indices have been already employed in simulations of models with global and local symmetries~\cite{Spinnet1,QgroupSpinnets,QGIntertwiners,DelcampSF}. There are various choices in constructing and assigning super-indices, which allows us to adjust the structure of these indices in a way best suited for the next task in the algorithm.

Using super-indices allows us to store only the admissible configurations, which we do in a vectorized format, since the contiguous data storage offers a more optimal memory access pattern compared to that for multidimensional array storage. Moreover, this allows us to write the fusion tree
transformations in a single loop which is straightforwardly parallelizable. For ${\rm k}=2$, these steps leads to a reduction in the required memory (for the ten--puncture building block) from around $10^8$ GB to 43 GB.

Further steps can be taken to reduce the required computational resources, in turn allowing coarse-graining of models at larger values of ${\rm k}$. One possibility is to truncate torsion degrees of freedom, that is, to employ the NTP or the FTT version of the algorithm. Another is to consider models where only integer labels are allowed, as we discussed in the previous section.
 To give an impression of how the algorithm scales for various levels
${\rm k}$, we show the memory cost for storing the amplitude for a ten-puncture state in Table~\ref{tab:memory_cost}.

The required runtimes scale in a similar way: For the Abelian models (${\rm k}=1$ half-integer and ${\rm k=2}$ integer) one full iteration takes a few seconds on a modern laptop. For higher levels we use HPC resources on the MP2B cluster run by Calcul Qu\'ebec (Compute Canada), which provide compute nodes with up to 512 GB of memory. We run most of our simulations using a single core to keep memory cost low. 15 full iterations of the NTP algorithm for ${\rm k} = 3$ take roughly two days, and 15 full iterations of the FTP algorithm for ${\rm k} = 4$ take roughly five days. Increasing the level $\rm k$ further leads to a significant increase of computational times, e.g., a third of a full iteration in the NTP algorithm for ${\rm k} = 4$ takes more than a week. This increase  is due to the increase in the number of fusion basis transformations, as well as the increase in time for singular value decompositions of larger tensors. While it is possible to accelerate the former by CPU-level multithreading, this also increases the memory costs in a prohibitive way. The largest system we studied was the NTP integer algorithm for ${\rm k} = 5$ (using $0.8 \text{ GB})$. Thus, we estimate that with the current algorithms, models that use up to 1 GB of memory can be simulated realistically.

To go towards larger quantum groups in future work, we are considering a redesign of the coarse-graining algorithm which employs smaller intermediate building blocks. These require less memory, and they use a smaller number of fusion basis transformations and singular value decompositions. Moreover, we expect it to be possible to implement such an algorithm on modern multi-GPU systems, where already one can find half a terabyte of pooled GPU memory on a single compute node. Because the algorithm is built on linear algebra operations, GPUs are the optimal solution, provided that memory access can still be made efficient. Another more speculative possibility is to use deep autoencoders to extrapolate results for larger $\rm k$ using lower-dimensional (i.e., compressed) representations of the data. This would allow one to trade memory for accuracy, but the exact tradeoff is unknown.

The version of the algorithm used to produce the results in this article is available at \url{https://github.com/ssteinhaus/Fusion-basis-coarse-graining}.

\begin{table*}
  \begin{tabular}{| c | c | c | c | c | c | c | c |}
    \hline
    & ${\rm k}=1$ & ${\rm k}=2$ & ${\rm k}=3$ & ${\rm k}=4$ & ${\rm k}=5$ & ${\rm k}=6$ & ${\rm k}=7$ \\
    \hline
    Torsion & $4 \, \text{MB}$ & $43 \, \text{GB}$ & $67 \, \text{TB}$ &  & & & \\
    \hline
    NTP & $8 \, \text{KB}$ & $2 \, \text{MB}$ & $0.3 \, \text{GB}$ & $13 \, \text{GB}$ &
    $425 \, \text{GB} $ & $9.3 \, \text{TB}$ &   \\
    \hline
    FTP & $8 \, \text{KB}$ & $0.8 \, \text{MB}$ & $0.03 \, \text{GB}$ & $0.66 \, \text{GB}$ &
    $8.8 \, \text{GB} $ & $83.5 \, \text{GB}$ &   \\
    \hline
    Torsion, Integer & & $4 \, \text{MB}$ & $0.26 \, \text{GB}$ & $460 \, \text{GB}$ &
    $20 \, \text{TB} $ &  &  \\
    \hline
    NTP, Integer & & $ 8 \, \text{KB}$ & $0.5 \, \text{MB}$ & $0.03 \, \text{GB}$ & $0.8 \, \text{GB}$ & $18 \, \text{GB}$ & $280 \, \text{GB}$ \\
    \hline
  \end{tabular}
  \caption{ An overview of approximate memory requirements for a ten-puncture state in various setups. Note that the ${\rm k}=1$ model (and the ${\rm k}=2$ integer model) is Abelian. Abelian models do not lead to the generation of torsion under coarse-graining, and one can savely employ the no-torsion algorithm.
  \label{tab:memory_cost}}
\end{table*}

\subsection{Phase diagram and phase transitions}

\subsubsection{Order parameters}

Although we characterize the initial amplitudes in a given model just by the coupling constant $g$, the coarse-graining algorithm leads to much more general forms of the amplitude. This is a typical feature of tensor network methods, and it allows the amplitude to develop all possible kinds of correlations between the degrees of freedom associated with the basic building blocks. This makes tracking of the coarse-graining flow generally difficult.

However, as explained in Sec.~\ref{Sec:Truncation}, we use a set-up for the SVD procedure which is similar to the one used for decorated tensor network coarse-graining~\cite{DecTNW}. In particular, the singular value decomposition is performed for fixed values of the label that characterizes the coarse-grained degree of freedom. In our case, this is the magnetic and electric charge label $\sigma_{\tilde l}$ associated with the effective puncture $\tilde l$. In the truncation, we keep the largest of the singular values $\lambda_{11}(\sigma_{\tilde l})$. The set of the squared singular values $\{(\lambda_{11}(\sigma_{\tilde l}))^2\}_{\sigma_{\tilde l}}$ gives (after normalization) the probability distribution of the charges $\sigma_{\tilde l}$ according to the truncated amplitude. That is, the set of singular values encodes the expectation values for the ribbon operators around $\sigma_{\tilde l} $ with respect to the (cube boundary) state as defined  by the coarse-grained amplitude. %Thus tracking the set of singular values $\{\lambda_1(\sigma_{\tilde p})\}_{\sigma_{\tilde p}}$ gives us essential information on the behaviour of the amplitude under coarse graining.

Furthermore, the effective puncture $\tilde l$ represents a plaquette with double the side lengths of the previous (full)\footnote{That is, one coarse-graining step is performed in each of the three spatial directions, and thus the lattice constant is doubled in all three directions.} iteration. The  singular values taken from subsequent coarse-graining iterations encode the expectation value of ribbon operators around larger and larger plaquettes, thus allowing us to track the behaviour of the effective amplitude at larger and larger scales.

Generically, the amplitude will flow to one of the two lattice gauge theory fixed points, either the weak coupling limit or the strong coupling limit. These are characterized by the following singular value distributions:
\begin{itemize}
  \item {\bf Weak coupling limit:} All magnetic (curvature) and electric (torsion) charges are suppressed. The amplitude is only non-vanishing if all labels $(j,\overline{j})=(0,0)$ take trivial values. The singular values $\lambda_1(\sigma)$ are also vanishing for all $\sigma$, except for $\sigma=(j,\overline{j})=(0,0)$.
  \item {\bf Strong coupling limit:}  Here electric (torsion) charges are still suppressed, i.e., the amplitude is only non-vanishing if all labels $(j,\overline{j})$ satisfy $j=\overline{j}$, but all purely magnetic charges are excited. The singular values are given by $\lambda_1((j,j))=1$. For the integer-only models we have a strong coupling limit where only magnetic charges with integer labels $(j,j)$ are excited.
\end{itemize}

 In summary, keeping track of the singular values $\lambda_1(j,j)$ allows us to conclude whether magnetic (curvature) charges are suppressed or not. Singular values $\lambda_1(j,\overline{j})$ with $j\neq \overline{j}$ indicate the absence or presence of electric (torsion) excitations.

\subsubsection{Phase transitions}
The initial amplitudes are characterized by the coupling constant $g$. Under coarse-graining, the models flow (eventually) to either the weak or strong coupling limit. For each fixed type of model, in particular for fixed level ${\rm k}$,  there is a critical coupling $g_c$, such that  for $g\in\left[0,g_c\right)$ the amplitudes flow to the weak coupling limit and for $g\in (g_c, \infty)$ they flow to the strong coupling limit. The first interval of couplings defines the weak coupling (deconfined) phase and the second the strong coupling (confined) phase.

\begin{figure}
	\includegraphics[width=0.6\textwidth]{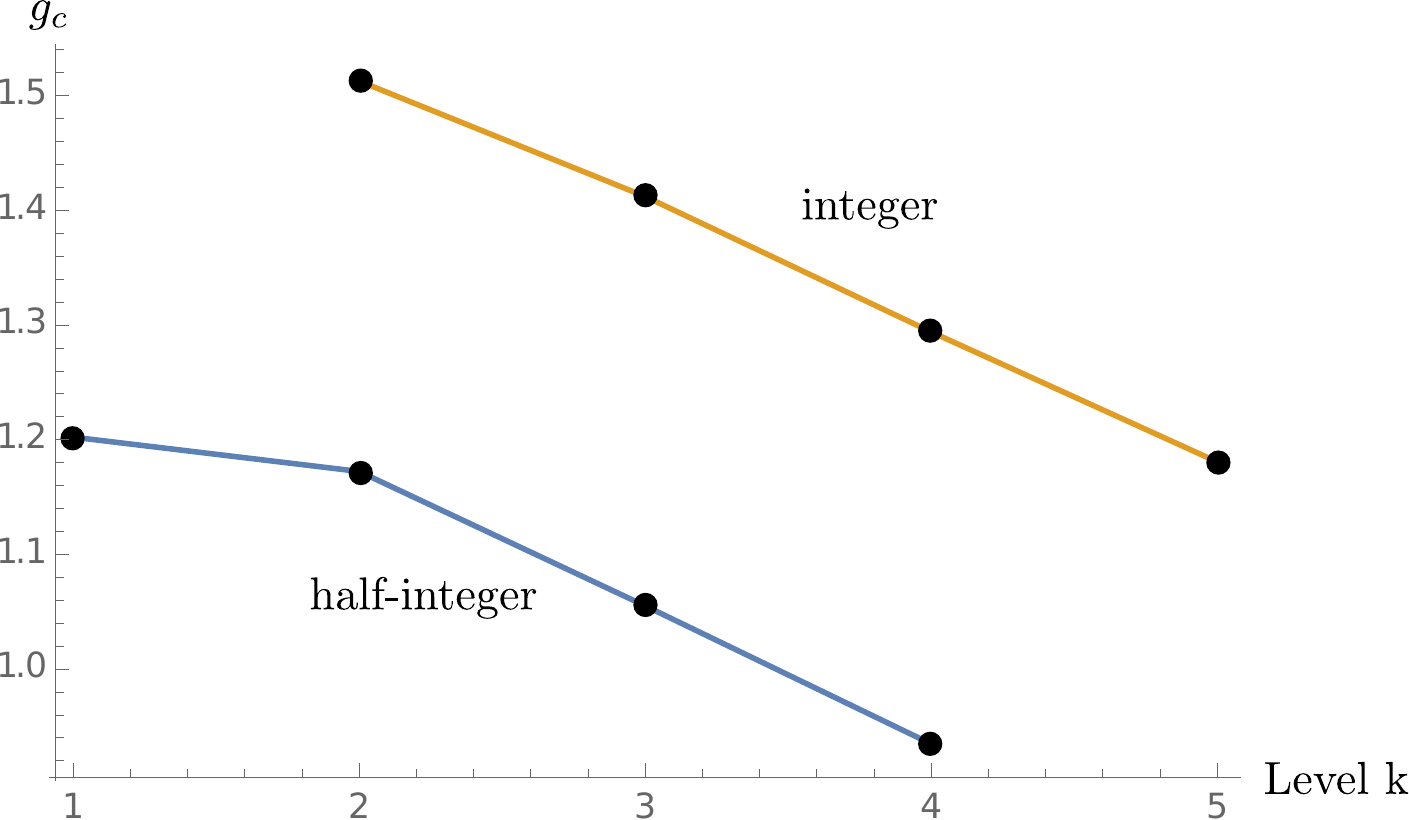}
	\caption{ \label{fig:critical_g}
		We show the critical coupling $g_c$ for the half-integer and integer models for different levels $\rm k$. The plotted values were obtained with the fully truncated torsion (FTT) version of the algorithm for the half-integer models, and with the no torsion at punctures (NTP) version of the algorithm for the integer model. We observe an almost linear decrease of $g_c$ with growing level $\rm k$. Note the plotted lines do not represent a fit.
	}
\end{figure}

Table~\ref{tab:critical_g} lists the critical couplings for different models and levels ${\rm k}$. We plot these critical values as functions of level ${\rm k}$, both for the half-integer and integer models in Fig.~\ref{fig:critical_g}.
\begin{table*}
  \begin{tabular}{| c | c | c | c | c | c | c |}
    \hline
    & ${\rm k}=1$ & ${\rm k}=2$ & ${\rm k}=3$ & ${\rm k}=4$ & ${\rm k}=5$ & ${\rm k}=6$ \\
    \hline
    Torsion $g_c$ & $ 1.2022$ &  &  &  & & \\
    \hline
    No Torsion at Punctures (NTP) $g_c$ & $ 1.2022$ & $1.1734$ & $1.0599$ & &
     &  \\
    \hline
    Fully Truncated Torsion (FTT) $g_c$ & $ 1.2022$ & $1.1723$ & $1.0551$ & $0.9347 $ & &
       \\
    \hline
    Torsion, Integer Only $g_c$ & & $1.5128$ & $1.4124$ &  &
     &  \\
    \hline
    NTP, Integer Only $g_c$ & & $1.5128$ & $1.4121$ & $1.29401$ & $1.1807$ &  \\
    \hline
  \end{tabular}
  \caption{ Summary of critical couplings for various models and levels ${\rm k}$. Missing entries indicate either the model cannot be defined, like integer models for ${\rm k}=1$, or the simulations were too costly to perform. Note that the ${\rm k}=1$ and ${\rm k}=2$ integer-only models are Abelian and do not generate torsion. Thus, finding the same critical coupling when truncating torsion provides a consistency check for the algorithms.
  \label{tab:critical_g}}
\end{table*}

We observe the values for the critical couplings decrease both for the models including half--integer labels, as well as for the models with integers only.  Furthermore, at least for the models we could coarse-grain with the version of the algorithm including torsion, the critical couplings found in the Torsion and NTP versions are essentially the same. There is only a difference for the ${\rm k}=3$ integer model in the fifth digit. Changing from the NTP to the FTT version we find also only small differences in the critical couplings in the fourth digit.

The decrease in the critical coupling with increasing level ${\rm k}$ is expected: larger levels give a better approximation to the classical group $\SU(2)$, which arise in the limit ${\rm k}\rightarrow \infty$.  The conjecture for $\SU(2)$ is that the value of the critical coupling is at $g_c=0$. Our results show a nearly linear decrease in the critical couplings, starting at ${\rm k}=2$.

If this linear dependence on coupling persists for larger ${\rm k}$, it would lead to a vanishing critical coupling for surprisingly small ${\rm k}$. We estimate this occurs for the half-integer models between ${\rm k}=11$ and ${\rm k}=12$ and for the integer models between ${\rm k}=15$ and ${\rm k}=16$. But, we would clearly need simulations for  larger levels ${\rm k}$ to see whether this linear decrease does indeed persist. In fact, one would expect that $g_c=0$ is approached rather in an asymptotic limit instead by a linear decrease.

\subsubsection{ ${\rm k}=3$ model}

Let us consider in more detail the ${\rm k}=3$ model. The representation labels $j$ takes values $j=0,\frac{1}{2},1$ and $j_{\text{max}}=\frac{3}{2}$. Due to the computational requirements, we employed the NTP version of the algorithm. This variant still allows for torsion degrees of freedom at the branches. For this reason, we can still find (and thus plot) the singular values for labels with torsion excitations on the effective punctures.  These can give an indication of the relevance of the torsion excitations. See, however, the discussion in Sec.~\ref{trunctorsion}, which indicates that truncating torsion does not produce any significant changes.
\begin{figure*}
  \includegraphics[width=0.475\textwidth]{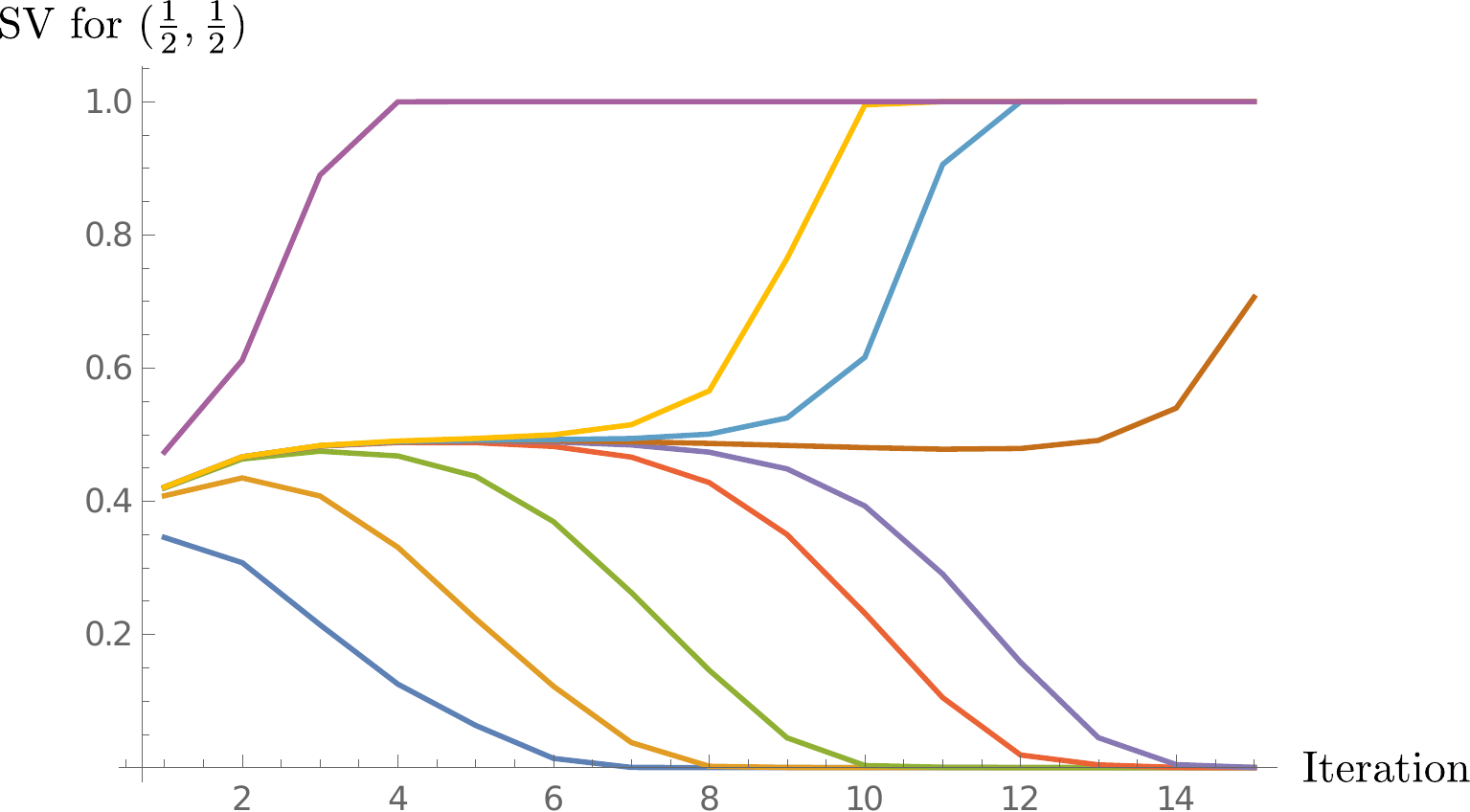}
  \includegraphics[width=0.475\textwidth]{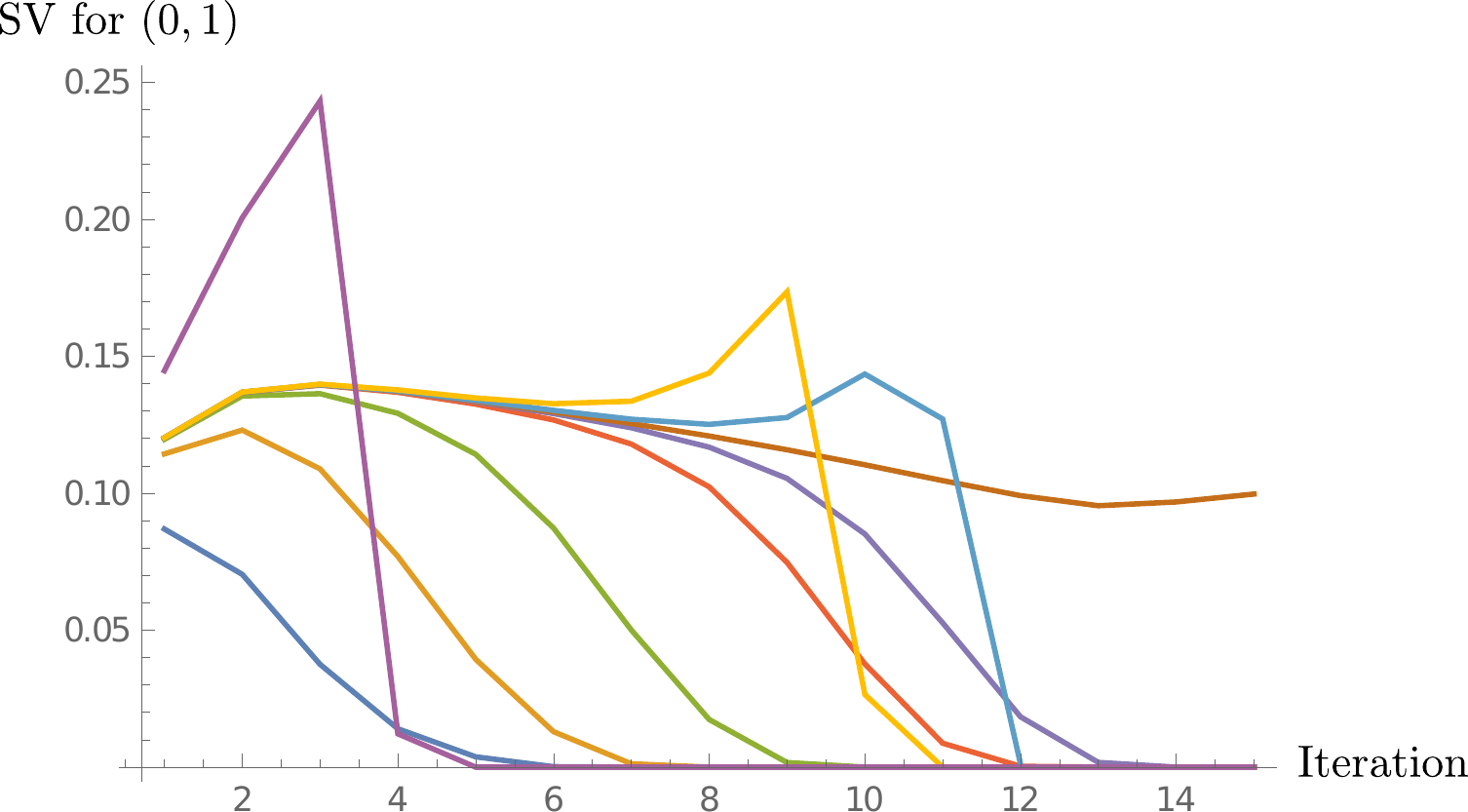}
  \caption{Singular values associated with effective punctures in the
  ${\rm k}=3$ model with the NTP version of the algorithm.  Matching colours describe the
  same initial coupling constant $g$ in both plots.
  {\it Left:} Singular values for the effective puncture $(\frac{1}{2},\frac{1}{2})$ over several
  iterations. From top to bottom, we have the following initial coupling constants $g$:
  $1.1$, $1.06$, $1.05996$, $1.05995$, $1.05994$, $1.0599$, $1.059$, $1.05$, $1.0$. {\it Right:} Plot of the
  singular value for effective puncture $(0,1)$ for several iteraitons.}
  \label{fig:k=3_no_torsion}
\end{figure*}

 In Fig.~\ref{fig:k=3_no_torsion}, we plot the singular values associated
with the effective puncture labels $(\frac{1}{2},\frac{1}{2})$ and $(0,1)$. The first value indicates a flow to either the weak or strong coupling limit. The second value gives an indication of the relevance of the truncated torsion degree of freedom. (Note that due to the coupling rules, charges with label $(0,\tfrac{1}{2})$, for instance, cannot be excited if torsion is suppressed in the initial amplitude.)

We see that the $(\frac{1}{2},\frac{1}{2})$ singular values flow to zero for $g<g_c$ and to one for $g>g_c$. The other $(j,j)$ singular values with $j>0$ show the same  behaviour. Tuning the coupling towards the phase transition, the singular values first flow to some plateau; that is, after some initial convergence remain constant over a number of iterations before eventually flowing to one of the two fixed point values. This plateau indicates the appearance of an almost scale invariant amplitude, and it hints at a higher order phase transition.  In this regime, one expects that the truncated singular values (i.e., the second largest singular value $\lambda_{2}(\sigma_{\tilde l})$ for each $\sigma_{\tilde l}$) are significant, which is indeed the case: for  the diagonal channels $\sigma_{\tilde l} = (j,j)$ the second largest singular value is up to $\sim 60\%$ the largest  one in the same channel.

The $(0,\tfrac{1}{2})$ singular values show a sharp decline for non-critical couplings. However, near the critical couplings these singular values remain significant for a number of iterations. Note that in every iteration we truncate the torsion excitations at the effective punctures. Thus, the torsion is generated anew in each coarse-graining step. A similar behaviour can be observed for the ${\rm k}=2$ model.

\subsubsection{Integer-only models}

\begin{figure*}
  \includegraphics[width=0.475\textwidth]{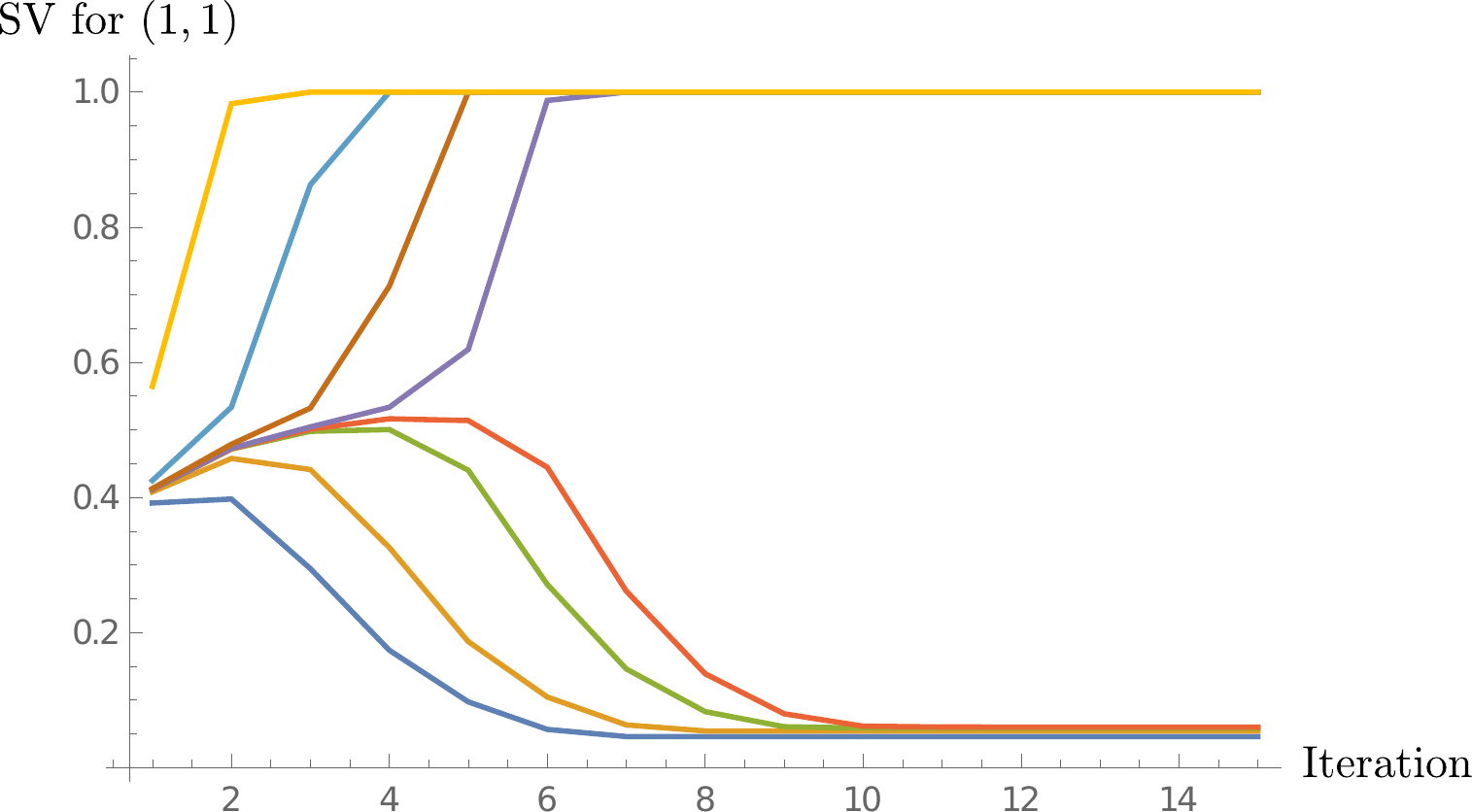}
  \includegraphics[width=0.475\textwidth]{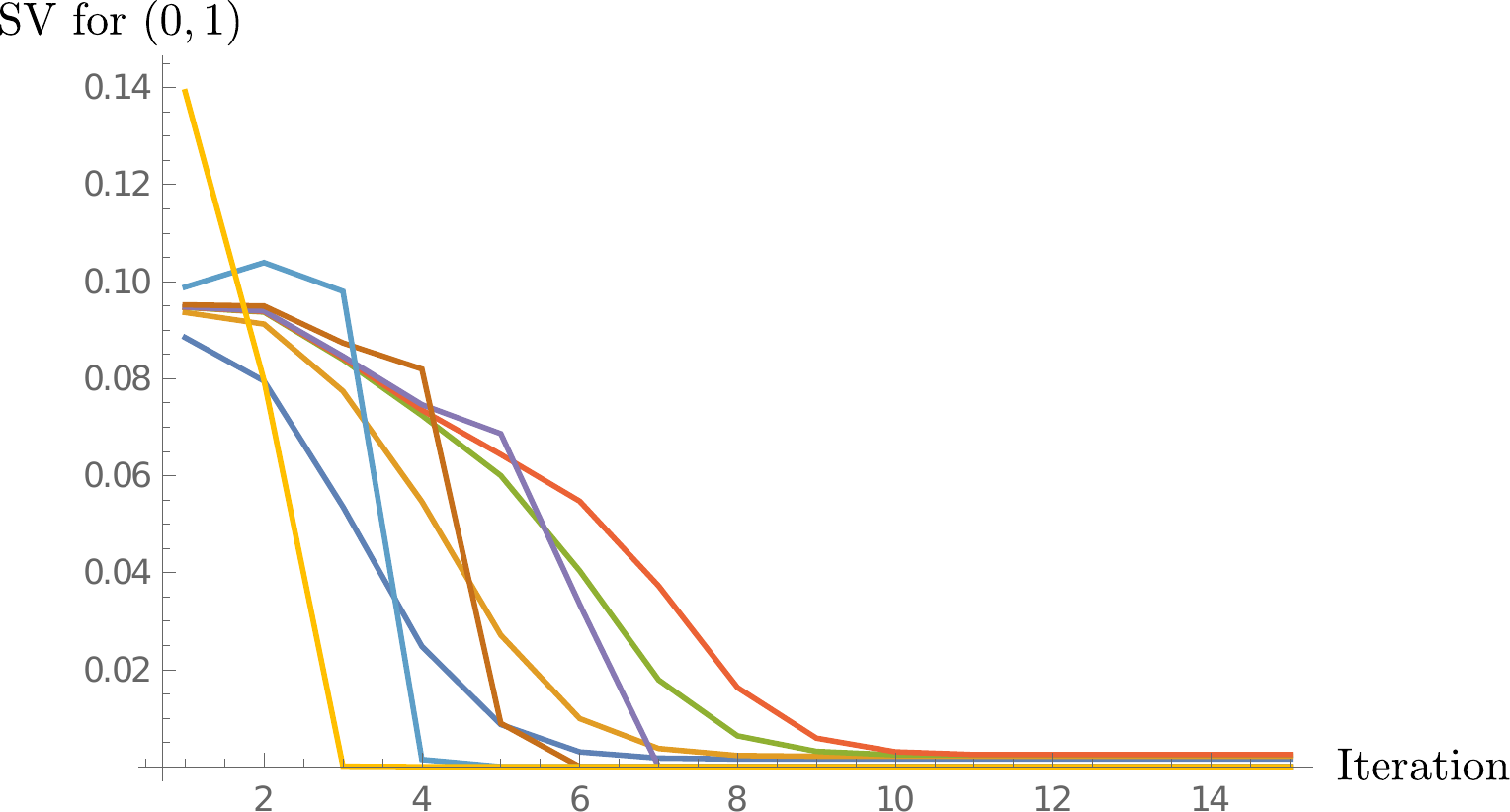}
  \caption{Singular values associated with effective punctures in the
  ${\rm k}=3$ integer-only model with the NTP algorithm.  Matching colours describe the
  same initial coupling constant $g$ in both plots.
  {\it Left:} Singular value for effective puncture $(1,1)$ over several
  iterations. From top to bottom, we have the following initial coupling constants $g$:
  $1.5$, $1.42$, $1.413$, $1.4122$, $1.4121$, $1.412$, $1.41$, $1.4$. {\it Right:} Plot of the
  singular value for effective puncture $(0,1)$.}
  \label{fig:int_k=3_no_torsion}
\end{figure*}
\begin{figure*}
  \includegraphics[width=0.475\textwidth]{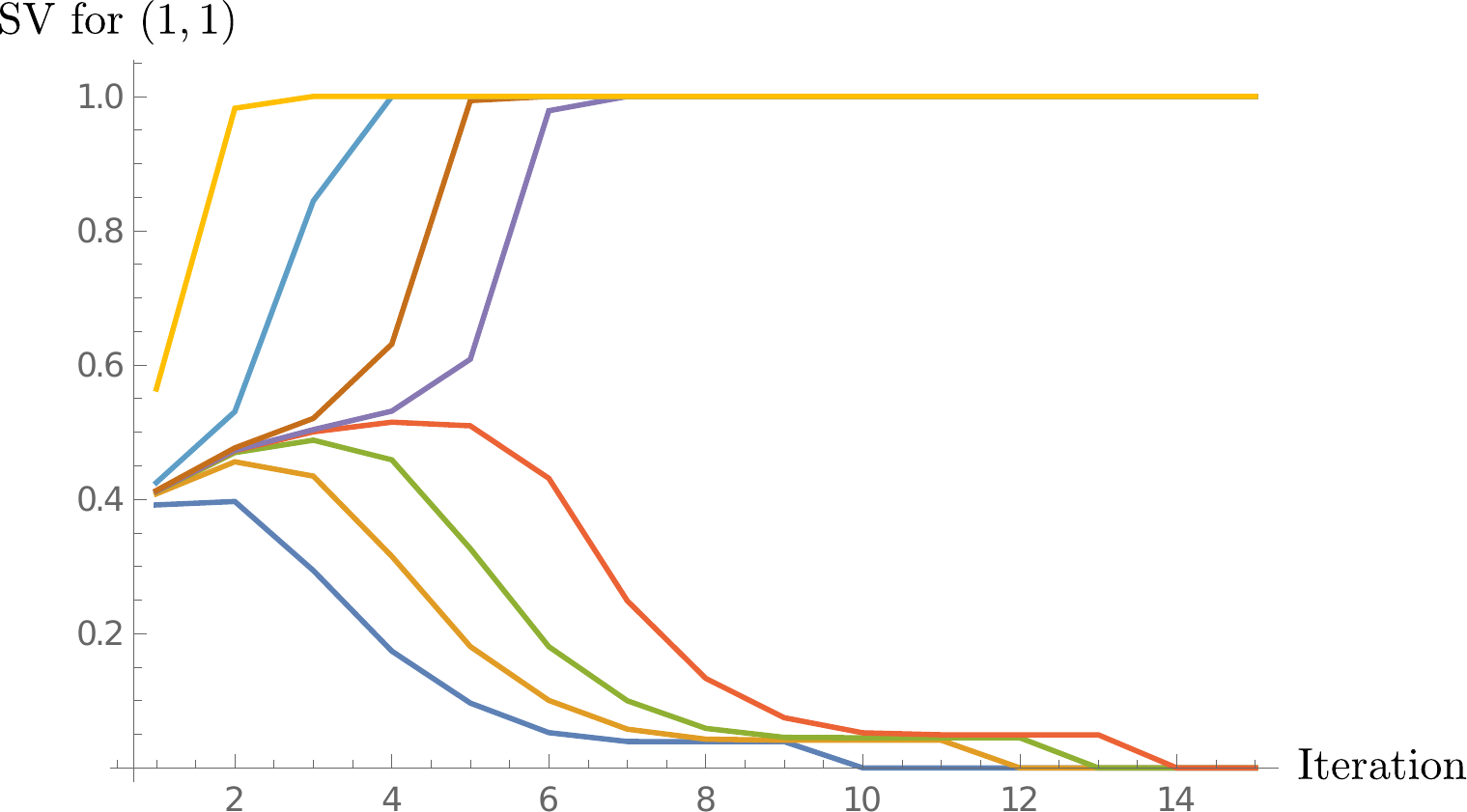}
  \includegraphics[width=0.475\textwidth]{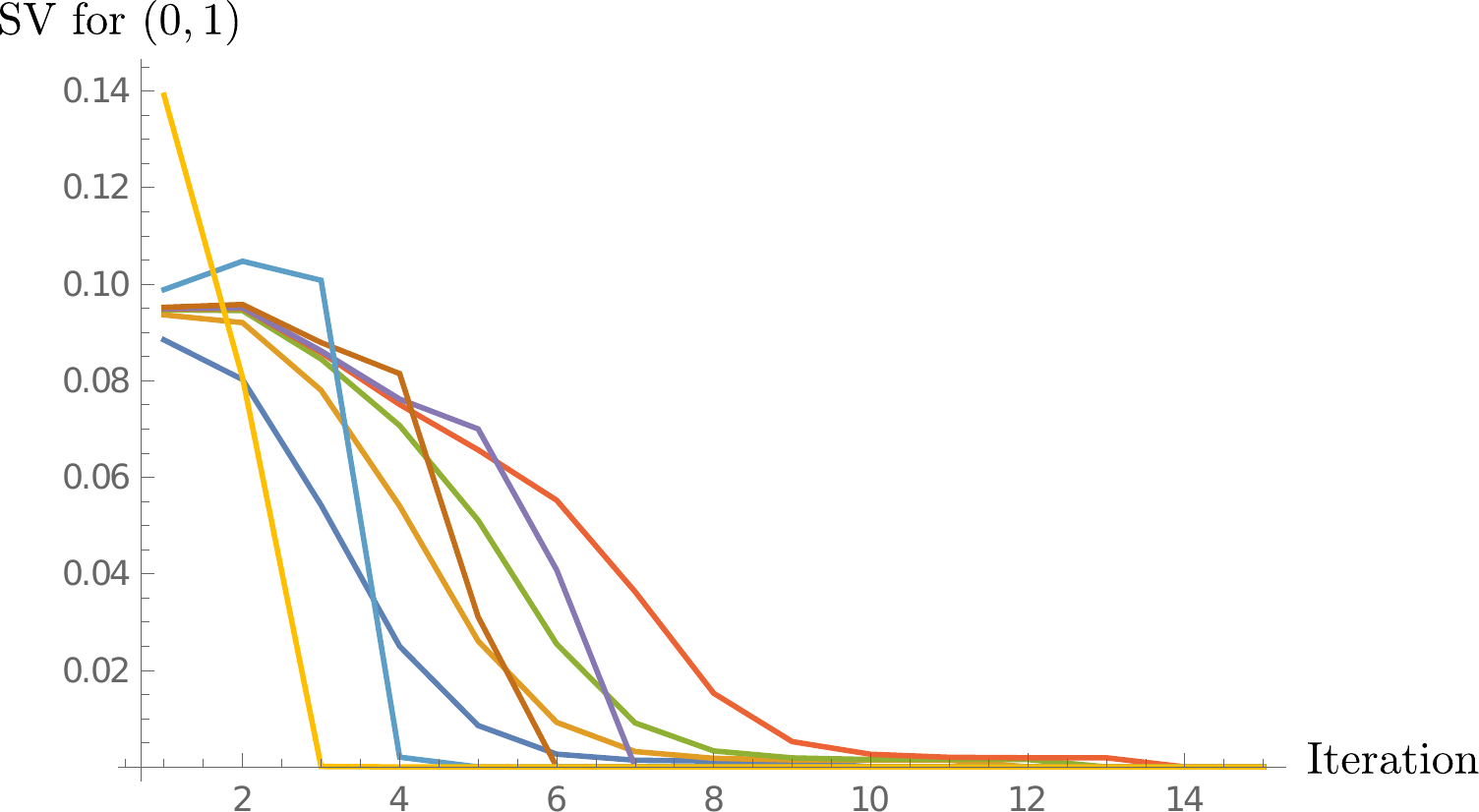}
  \caption{Singular values associated with effective punctures in the
  ${\rm k}=3$ integer-only model with the Torsion algorithm. Matching colours describe the
  same initial coupling constant $g$ in both plots.
  {\it Left:} Singular value for effective puncture $(1,1)$ over several
  iterations. From top to bottom, we have the following initial coupling constants $g$:
  $1.5$, $1.42$, $1.413$, $1.4125$, $1.4124$, $1.412$, $1.41$, $1.4$. {\it Right:} Plot of the
  singular value for effective puncture $(0,1)$ for several iterations.}
  \label{fig:int_k=3_torsion}
\end{figure*}

Next, we consider the integer-only models, where we allow only labels with integer representations $j$. For ${\rm k}=3$, this constrains $j=0$ and $j=1$, but includes a non-trivial coupling $1\otimes 1=0\oplus 1$. (This fusion category is also known as ``golden chain'' model.)

This model can be simulated in the Torsion and the NTP versions of the algorithm. As we will comment on in more detail in the next section, there is almost no difference between the flow in these two different versions.

Considering the singular values for the $\sigma=(1,1)$ excitation, we again see a flow to the weak coupling limit value for $g<g_c$ and a flow to the strong coupling limit for $g>g_c$. But, in contrast to the ${\rm k}=3$ model with half-integer labels, when tuning towards the critical coupling we do not see a plateau behaviour for the singular values. This suggests a first order phase transition. Correspondingly, the singular values for the torsion excitations decay relatively fast even near the critical coupling.

\begin{figure}
  \includegraphics[width=0.475\textwidth]{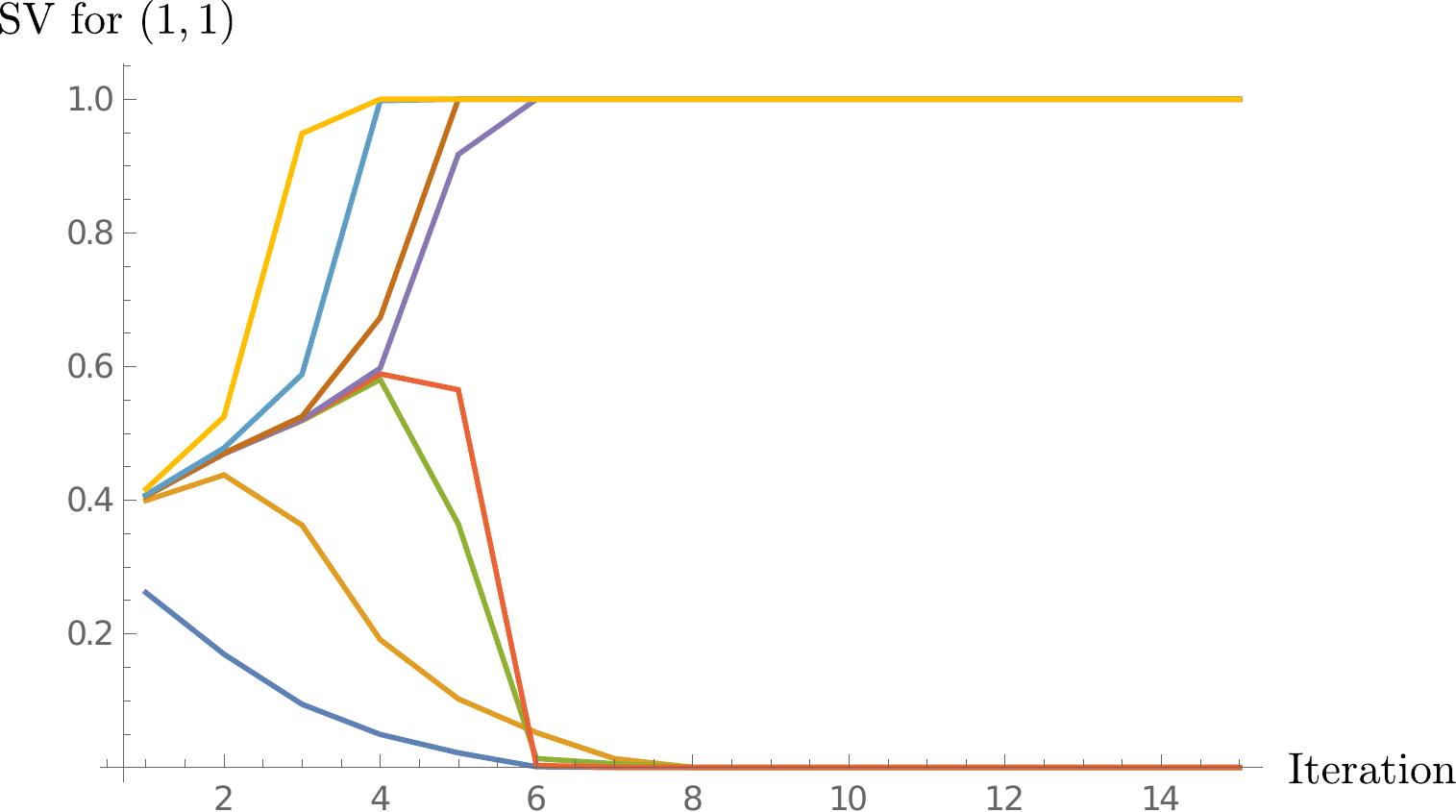}
  \caption{Singular values associated with the effective puncture $(1,1)$ in the
  ${\rm k}=4$ integer model with the NTP algorithm.
  From top to bottom, we have the following initial coupling constants $g$:
  $1.3$, $1.295$, $1.2941$, $1.29402$, $1.29401$, $1.294$, $1.29$, $1.2$.}
  \label{fig:int_k=4_no_torsion}
\end{figure}

The $k=4$ integer-only model behaves similarly (Fig.~\ref{fig:int_k=4_no_torsion}). Also, we do not observe a plateau behaviour around the critical coupling.

\subsubsection{On truncating torsion} \label{trunctorsion}

For the ${\rm k}=3$ integer-only model, we compare the results of the Torsion and NTP versions of the algorithm. The critical couplings show only a tiny difference in the fifth digit. The plots~\ref{fig:int_k=3_no_torsion} and~\ref{fig:int_k=3_torsion} show the flow of the singular values within the two different truncation schemes, and they do not show a visible difference. This does even include the singular values for the torsion excitations.

This is quite surprising, since such a close agreement is not suggested
by the size of the singular values: if we compare the singular values for the punctures
$(1,1)$ and $(0,1)$ in the first iteration, the latter's size is a quarter of the
former. Hence, we clearly cannot argue that these degrees of freedom are irrelevant
for the dynamics, yet it does not significantly change the singular values for pure curvature excitations
$(j,j)$ nor for torsion excitations. It seems that torsion
is mostly generated as curvature-induced torsion, that is, mostly from recoupling
between the pure curvature excitations. But, once these torsion excitations are ``transported'' (via fusion tree transformations) to the effective punctures, they do not significantly influence the behaviour in the next iteration.

\begin{figure}[h]
	\includegraphics[width=0.475\textwidth]{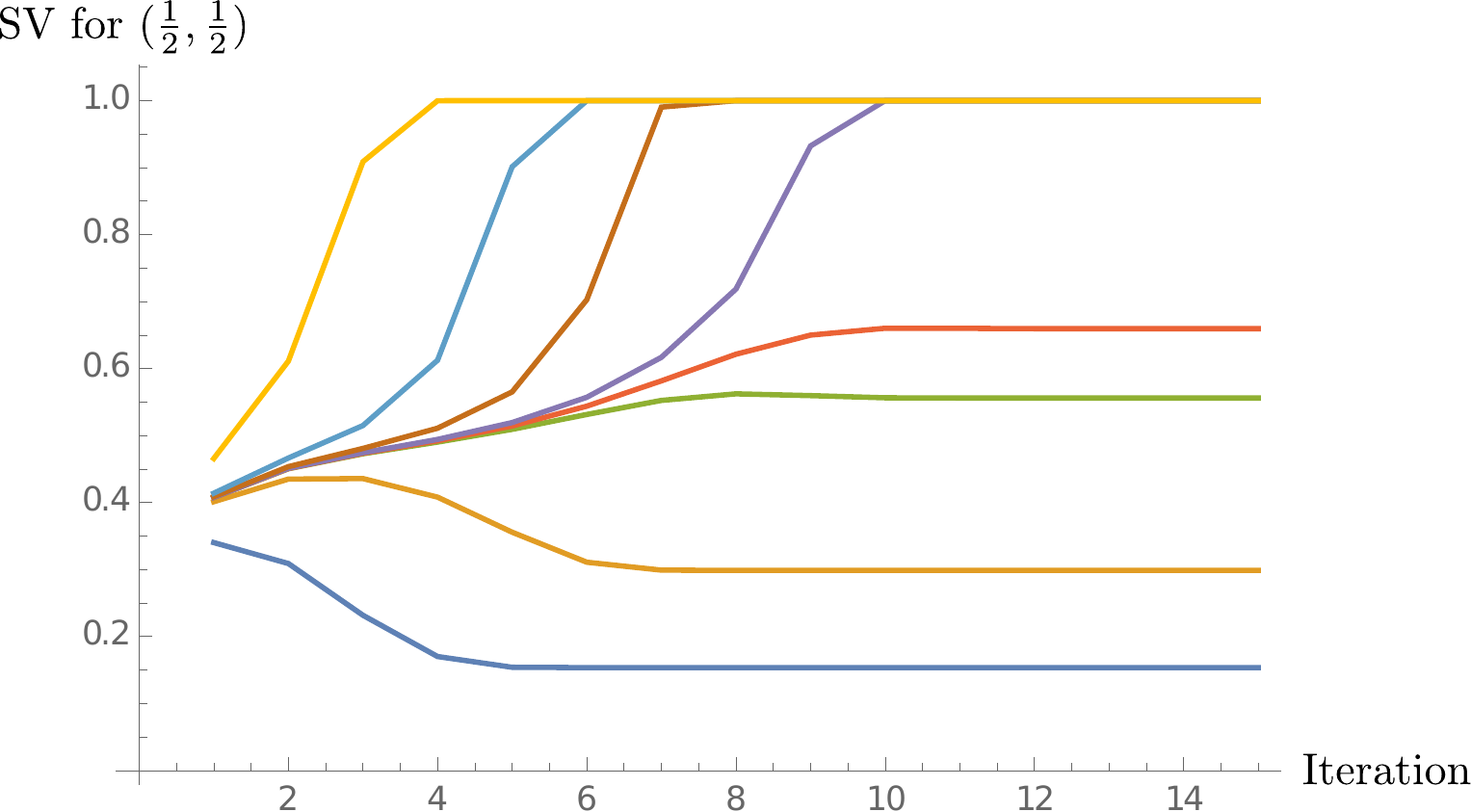}
	\caption{
		Singular value for $\sigma_l = (\frac{1}{2},\frac{1}{2})$ in the ${\rm k} = 3$ model using the FTT algorithm for several iterations. While one can still recognize a phase transition, the qualitative behaviour is different compared to the NTP algorithm (Fig.~\ref{fig:k=3_no_torsion}). For $g > g_c$ the system flows to the strong coupling fixed point; however, for $g < g_c$ it does not flow to the weak coupling fixed point. Instead, it reaches a fixed point which depends on the initial coupling constant $g$. From top to bottom, we plot $g=1.1$, $g=1.06$, $g=1.056$, $g=1.0552$, $g=1.0551$, $g=1.055$, $g=1.05$ and $g=1.0$.
		\label{fig:plot_ftt}
	}
\end{figure}

	However, truncating torsion entirely as in the FTT version of the algorithm shows quantitative and qualitative differences. While the FTT version allows us to identify a phase transition with a critical coupling $g_c$ still close to those found in the torsion and NTP algorithms (Table~\ref{tab:critical_g}), the flow of singular values, and thus the flow of amplitudes, is significantly different: compare Fig.~\ref{fig:k=3_no_torsion} showing the singular values for the NTP algorithm and Fig.~\ref{fig:plot_ftt} showing the singular values for the FTT one.

	With the FTT algorithm we have that for $g < g_c$ the system does not flow to the weak coupling fixed point; instead, it flows to a continuum line of fixed points. The singular values for these fixed points  depend on the initial coupling $g$.  Moreover, with the FTT algorithm one can no longer recognize a plateau in the singular values close to the phase transition as with the NTP algorithm, see Fig.~\ref{fig:plot_ftt} and Fig.~\ref{fig:k=3_no_torsion} respectively.

	Similar features occur when tensor network coarse-graining methods, like the tensor network renormalization group~\cite{GuWen}, are applied to models like the 2D Ising model. These algorithms possess fixed points of a particular form, dubbed ``corner double line tensors.'' These appear because these algorithms do not fully resolve entanglement among short scale degrees of freedom, and they promote short scale entanglement to larger scales under consecutive iterations. This caveat was addressed by entanglement filtering algorithms~\cite{GuWen,MERAPartitionF}, which include measures (e.g., disentanglers in~\cite{MERAPartitionF}) that are designed to remove short scale entanglement.

The main difference between the FTT version, where the continuum line of fixed points appear, and the NTP algorithm, where this continuum line does not appear, is that the FTT version also implements a truncation for the fusion basis transformations: in the NTP version torsion excitations are only removed at the effective punctures after the SVD procedure. In particular, the branch labels are of the more general form $\sigma_b=(i_b,\overline{i_b})$, allowing for a larger Hilbert space on which the fusion basis transformations can act.

	This  suggests to us that the fusion tree transformations do indeed act as disentangler maps, as we discussed in Sec.~\ref{sec:algorithm}. For the Torsion and NTP versions, the fusion tree transformation remove short scale entanglement  sufficiently well for the truncation employed, i.e., we keep only the largest singular value $\lambda_1(\sigma_{\tilde l})$ for each channel $\sigma_{\tilde l}$.

\subsubsection{Abelian models}

Among the $q$-deformed lattice gauge theories discussed above, there are two Abelian ones: the half-integer $\rm k = 1$ model and the integer $\rm k = 2$ model. Both models permit only two representation labels, $j=0$ and $j=j_{\text{max}} = \frac{\rm k}{2}$.  Therefore, the face weights (in the initial amplitude) can only take two values. Hence, we can identify our model with $\mathbb{Z}_2$ gauge theory by choosing parameters $g$ and $\beta$ such that the face weigths in both theories match. The plaquette weights in $\mathbb{Z}_2$ gauge theory are given by $\exp(- \beta \, k )$ with $k\in \{0,1\}$ (up to a constant).

Translating the critical couplings $g_c$ in Table~\ref{tab:critical_g} to critical couplings $\beta_c$ for the $\mathbb{Z}_2$ gauge theory, we find that both critical coupling constants translate to the same $\beta_c \sim 0.804$\footnote{There are slight deviations in the fourth digit of $\beta_c$ in both models due to the limited precision of $g_c$. We also ran simulations for $\mathbb{Z}_2$ gauge theory weights and found the same critical $\beta_c$ for both models at this value.}. Indeed the initial plaquette weights differ only by signs resulting from quantum dimension factors\footnote{Each puncture $l$ carries a dimension factor $v_{j_l}$, which is negative for half-integer representations, see~\eqref{HKAqd}.}, and these signs cancel out under gluing.

The value of $\beta_c\sim 0.804$ deviates by only a few percent from the value found in Monte Carlo simulations, $\beta_c \approx 0.761$.

\subsubsection{An example of anyon condensation}

We have encountered two different strong coupling fixed points: one in the models where half-integers allowed and another one where only integers allowed. In the fusion basis, the integer-only fixed point describes a state where all the magnetic charges with integer representation charges are excited, whereas the fixed point of the full model describes a state where all magnetic charges are excited.

The transition from the integer-only fixed point to the half-integer one can be interpreted as an anyon condensation~\cite{Burnell}, where the charges here are identified with anyons. The fixed points are interpreted as vacuum states. Thus, including half-integer charges means that additional types of anyons condense to form a new vacuum state.

Note that these strong coupling fixed points have a very simple amplitude in the spin network basis. There, the amplitude for the half-integer strong coupling fixed point allows only the trivial representation label $j=0$ to appear with non-vanishing amplitude. The integer-only strong coupling fixed point allows two representation labels $j=0$ and $j=j_{\rm max}$. Both representations have quantum dimension equal to one, and the allowed couplings are reproducing the ones for ${\mathbb Z}_2$ gauge theory (in the strong coupling representation): $0\otimes 0=0$ and $0\otimes j_{\rm max}=j_{\rm max}$ as well as $j_{\rm max} \otimes j_{\rm max}=0$.  Thus, we expect an Ising-type phase transition.

\begin{figure}[h]
  \includegraphics[width=0.475\textwidth]{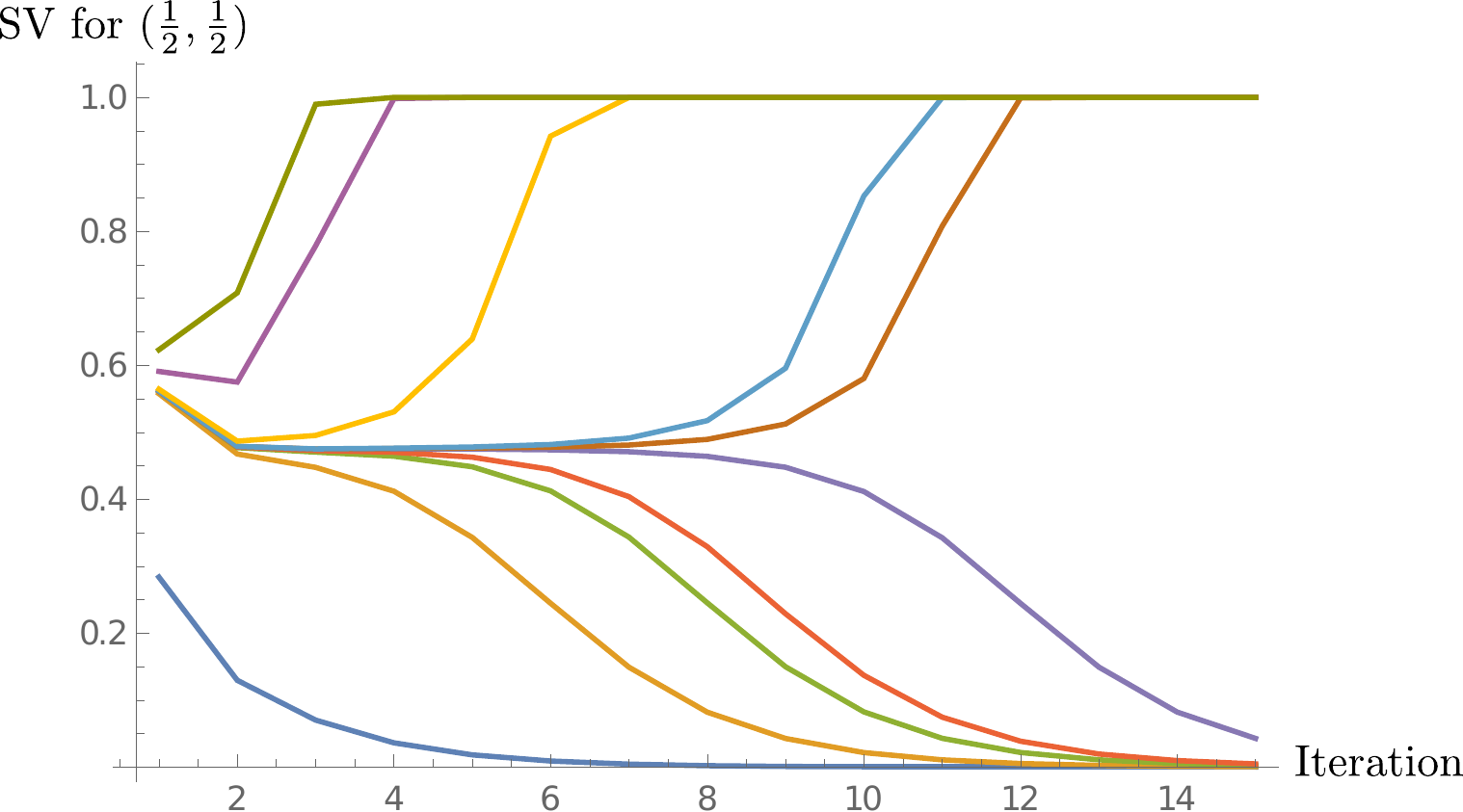}
  \caption{Singular value associated with the charge $(\frac{1}{2},\frac{1}{2})$ for
  several iterations and different values of $\alpha$. Singular value equal to $1$
  corresponds to the half-integer strong coupling limit, whereas the singular value vanishes in the integer-only strong coupling limit. From top to bottom, we have the following initial values for $\alpha$:
  $\alpha = 0.2$, $\alpha = 0.19$, $\alpha = 0.182$, $\alpha = 0.1812$, $\alpha = 0.18119$,
  $\alpha = 0.18118$, $\alpha = 0.1811$, $\alpha = 0.181$, $\alpha = 0.18$ and $\alpha = 0.1$.}
  \label{fig:half_int_flow}
\end{figure}

Here, we consider as initial amplitudes a simple superposition of the amplitudes for the integer-only and half-integer strong coupling fixed point
\begin{equation}
  \mathcal{A}(\alpha,\{I\}) = \alpha \, \mathcal{A}^{\text{SC}}(\{I\})
  + (1- \alpha) \, \mathcal{A}^{\text{SC-int}}(\{I\}) \;  \q .
\end{equation}
Varying the superposition parameter $\alpha$ from zero to one we expect to find a transition from the integer-only to the half-integer fixed point. We use the  ${\rm k}=2$ example  with the NTP version of the algorithm, which gives a phase transition at $\alpha_c \sim 0.18118$. Fig.~\ref{fig:half_int_flow} shows the singular values for the magnetic charge $(\tfrac{1}{2},\tfrac{1}{2})$, whose values distinguish clearly between the two fixed points. Tuning towards the critical parameter $\alpha_c$ we find a plateau-like behaviour indicative of a higher order phase transition.

\subsection{Observables}

The coarse-graining algorithm presented herein tracks a certain class of observables given by the ribbon operators. These ribbon operators are generalizations of Wilson loops that measure both the electric and magnetic charges in the encircled region.

The fusion basis diagonalizes a set of ribbon operators $\{R^{\tau}_\gamma\}$, where $\tau=(k,\overline{k})$ is also a charge label, and $\gamma$ denotes a loop.   This set always includes the ribbon operators around the single plaquettes.  If the charge labels of this plaquette are given by $\rho_l=(j,\overline{j})$, the eigenvalues of the ribbon operator are given by (see e.g.~\cite{ExTQFT} for details)
\ba
\lambda_{k\overline{k},j\overline{j}} =\frac{S_{kj}S_{\overline{k}\,\overline{j}}}{ S_{0j} S_{0\overline{j}}} \,=\, (-1)^{2(k+\overline{k})} \frac{\sin\left( \frac{\pi}{{\rm k}+2}  (2k+1)(2j+1)     \right) \sin\left( \frac{\pi}{{\rm k}+2}  (2\overline{k}+1)(2\overline{j}+1)     \right)  }{ \sin\left( \frac{\pi}{{\rm k}+2} (2j+1)     \right)  \sin\left( \frac{\pi}{{\rm k}+2}  (2\overline{j}+1)   \right)    }
\ea
where $S_{kj}$ is the S-matrix of the fusion category $\SU(2)_{\rm k}$. For $\tau=(k,0)$ and $j=\overline{j}$ we obtain
\ba
\lambda_{k0,jj} =(-1)^{2l} \frac{\sin\left( \frac{\pi}{{\rm k}+2}  (2k+1)(2j+1)     \right)  }{ \sin\left( \frac{\pi}{{\rm k}+2} (2j+1)     \right)    }  \q .
\ea
This can be compared to the eigenvalue $\sin((2k+1)\theta)/\sin(\theta)$ for the $k$--Wilson loop in the Lie group $\SU(2)$, where $\theta$ is the class angle of the group element $g$. Thus, for torsion-free plaquettes with label $(j,j)$, we conclude that the ribbon operators $R^{k0}$ are analogous to the Wilson loop operator in the $k$--representation, and they probe a (discretized) version of curvature angle (i.e., magnetic charge) $\theta=\pi(2j+1)/({\rm k}+2)$.

In our coarse-graining algorithm, after $N$ full iterations, an effective plaquette is  coarse-grained from $N \times N$ initial plaquettes. Therefore, ribbon operators around such effective plaquettes, arising from subsequent iterations, represent ribbons around a larger and larger number of initial lattice plaquettes.

We now consider the expectation value of such ribbon operators as computed from the partition function. The partition function ${\cal Z}$ is the one associated with the 3--torus, which can be obtained by identifying the three pairs of sides of a cubical building block. ${\cal Z}$ is then obtained by gluing the corresponding labels with each other  and summing the cube's amplitude over all remaining free labels.  Doing that for our amplitude ${\cal A}^N$ after $N$ iterations, our coarse-graining algorithm computes the (approximation to the) partition function for a 3--torus built from $3^N$ basic building blocks.

We next consider the expectation value of a  normalized ribbon operator $R^{k0}$ around one of the effective $N\times N$ plaquettes of the torus.  After normalization, this ribbon operator has eigenvalue one when applied to a puncture with label $\rho=(0,0)$.  The expectation value is obtained by inserting the operator into the partition function, and in the notation for the left cube in Sec.~\ref{Sec:gluing}, is given by
\ba
\langle R_N^{k0}\rangle \,=\,\frac{1}{\cal Z}  \sum_{\rho_1} \frac{\lambda_{k0,  o(\rho_1) u(\rho_1)}}{ v_k^2} \sum_{\rho_2,\rho_5,\sigma_1,\sigma_2,\sigma_3}   \frac{{\cal D}^6 \,\,  } {v_{o(\rho_1)} v_{u(\rho_1)}v_{o(\rho_2)} v_{u(\rho_2)}v_{o(\rho_5)} v_{u(\rho_5)} }
{\cal A}^N(\sigma_1,\sigma_2,\sigma_2; \rho_1,\rho_2,\rho_1,\rho_2,\rho_5,\rho_5)\q
\ea
where
\ba
{\cal Z}&=& \sum_{\rho_1,\rho_2,\rho_5,\sigma_1,\sigma_2,\sigma_3}   \frac{{\cal D}^6 \,\,  } {v_{o(\rho_1)} v_{u(\rho_1)}v_{o(\rho_2)} v_{u(\rho_2)}v_{o(\rho_5)} v_{u(\rho_5)} }
{\cal A}^N(\sigma_1,\sigma_2,\sigma_2; \rho_1,\rho_2,\rho_1,\rho_2,\rho_5,\rho_5) \q .
\ea

~\\

\subsubsection{Expectation values in the ${\rm k}=2$ model}

We now consider the expectation value of the ribbon operator $R_N^{\tiny{\frac{1}{2}} 0}$ in the ${\rm k}=2$ half-integer model using the NTP version of the algorithm. The ribbon operator $R^{\tiny{\frac{1}{2}}0}$ is analogous to a Wilson loop operator in the fundamental ($j=1/2$) representation.
\begin{figure}
  \includegraphics[width=0.475\textwidth]{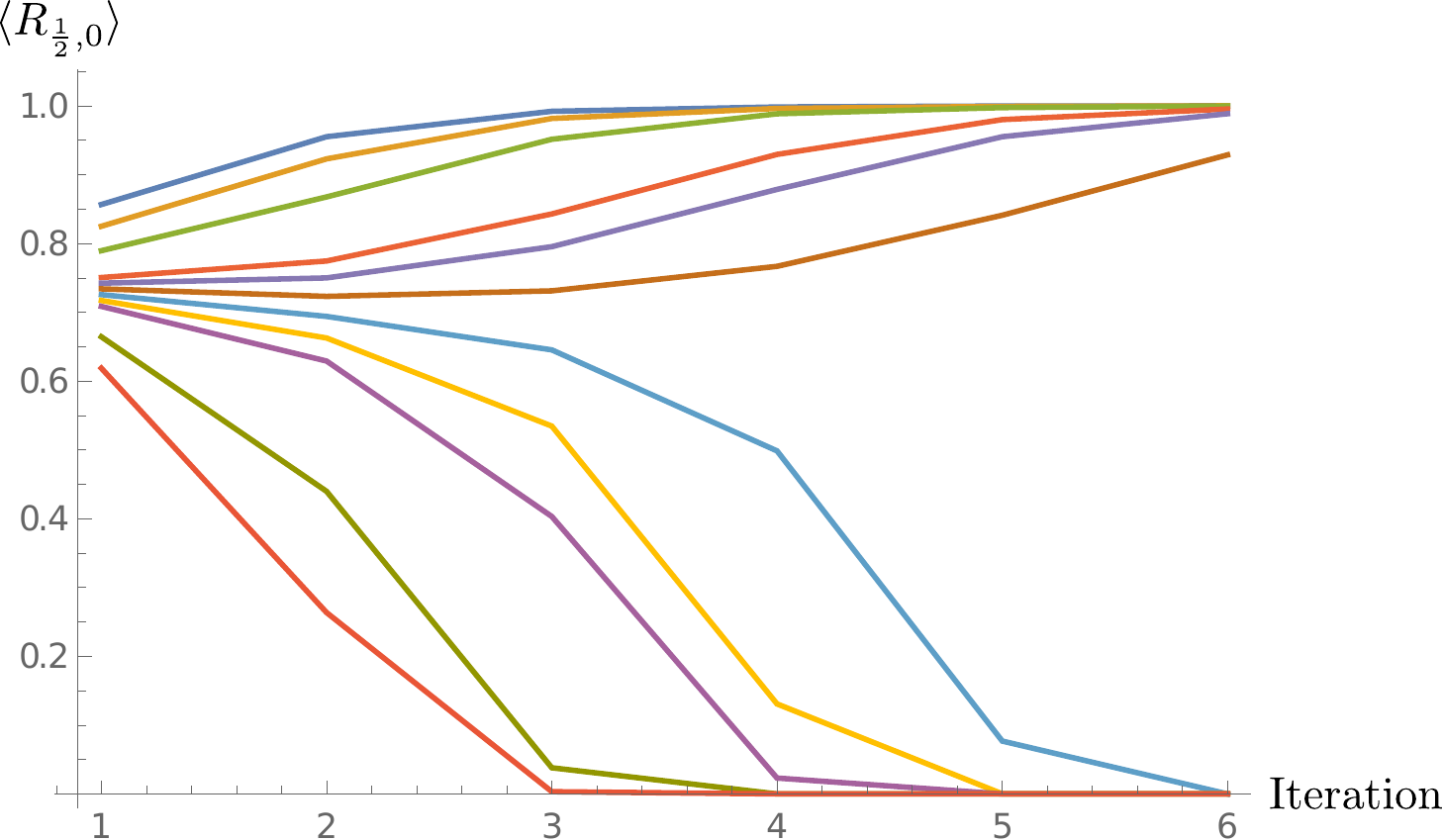}
  \caption{The expectation value $\langle R_N^{\tiny{\frac{1}{2}} 0} \rangle$
  over several iterations for different initial coupling constants $g$ in the ${\rm k}=2$ half-integer model. From top to
  bottom: $g=1.0$, $g=1.05$, $g=1.1$, $g=1.15$, $g=1.16$, $g=1.17$, $g=1.18$,
  $g=1.19$, $g=1.2$, $g=1.25$ and $g=1.3$. The phase transition occurs
  around $g_c \sim 1.1734$.
  \label{fig:Wilson_iter}
  }
\end{figure}

In Fig.~\ref{fig:Wilson_iter}, we show the expectation value $\langle R_N^{\tiny{\frac{1}{2}} 0} \rangle$
over several iterations $N$ and for different initial couplings $g$. We see that the expectation value is a good
order parameter, since it differentiates well the two fixed points and, therefore, phases.
In the weak coupling limit, the expectation value is non-vanishing, whereas, due to the averaging of the eigenvalues $\lambda_{k0,jj}$ over $j$, it vanishes in
the strong coupling limit. This is the same behaviour as we observed for Wilson loop operators in lattice gauge theory.
Closer to the phase transition, we require more iterations before the expectation
values converge to the fixed point values in agreement with the results for the
coarse-graining flow.
\begin{figure}
  \includegraphics[width=0.475\textwidth]{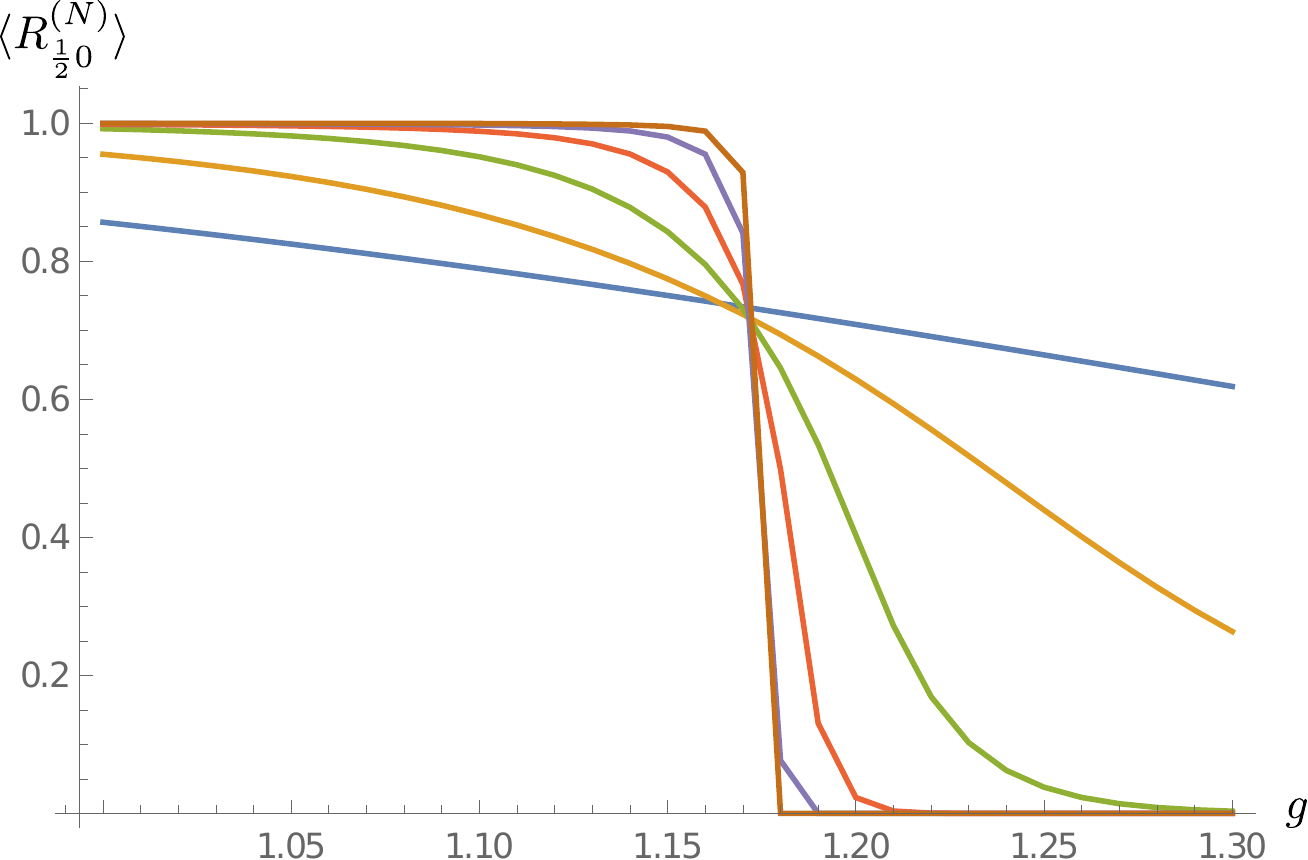}
  \caption{ We show $\langle R_N^{\tiny{\frac{1}{2}} 0} \rangle$ for several iterations $N$
  as we vary the initial coupling constant $g$, and we observe convergence to the expectation
  values for the respective fixed points. Hence, steeper curves imply more iterations were
  performed. All curves almost meet in a single point which marks the phase transition.
  \label{fig:phases}
  }
\end{figure}

The phase transition is further illustrated in Fig.~\ref{fig:phases}, where we plot
$\langle R_N^{\tiny{\frac{1}{2}} 0} \rangle$ as a function of the initial coupling constant
$g$ for different $N$.  Around the critical value $g_c$, the graphs quickly become steeper as we increase the number of iterations $N$. This reflects that for growing $N$ the system tends nearer and nearer to one of the two final fixed points. Moreover, the graphs intersect (almost) at the
same point, where the expectation values do not change with consecutive iterations,
i.e. length scale. This is an
indication for a higher order phase transition.
\begin{figure}
  \includegraphics[width=0.475\textwidth]{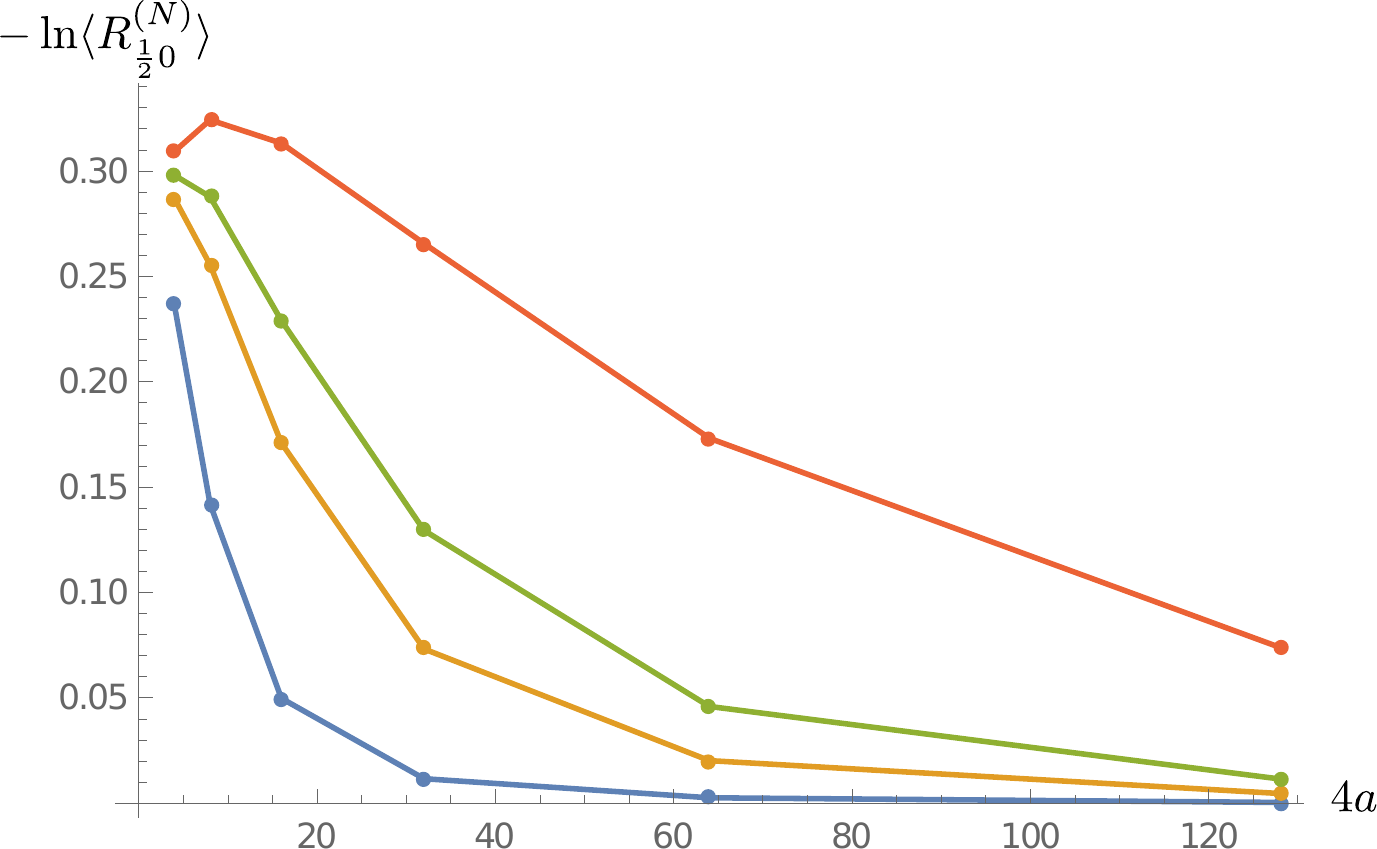}
  \caption{ We show $ -\ln \langle R_N^{\tiny{\frac{1}{2}} 0} \rangle$ for different initial
  couplings $g$ in the weak coupling phase as a function of $4N=4a$. From top to bottom, we have $g=1.17$, $g=1.16$,
  $g=1.15$ and $g=1.1$.
  \label{fig:weak-coupling-circum}
  }
\end{figure}

The observables also reveal properties of the system in the respective phases. In Fig.~\ref{fig:weak-coupling-circum}
we plot the logarithm of the expectation value, that is, $-\ln \langle R^{\frac{1}{2} 0}_N \rangle$, as a function of the number $N$ of iterations for different values of the initial couplings $g$ in the weak coupling phase $\left[0,g_c\right)$. Note that $N$ determines the  perimeter  $4N$  of the effective plaquette around which we take the ribbon operator.

In general, we observe that for larger perimeters the $-\ln \langle R^{\frac{1}{2} 0}_N\rangle$ values
decrease,  consistent with the fact that the expectation values for the Wilson loops reach the value one in the weak coupling fixed point.

As expected, the decrease is much more pronounced further away from the phase transition, e.g., for the blue curve ($g=1.1$) in Fig.~\ref{fig:weak-coupling-circum} the logarithm of the expectation value appears to drop off
$\sim \frac{1}{N}$, signalling a (hyper-) deconfining behaviour.
\begin{figure}
  \includegraphics[width=0.475\textwidth]{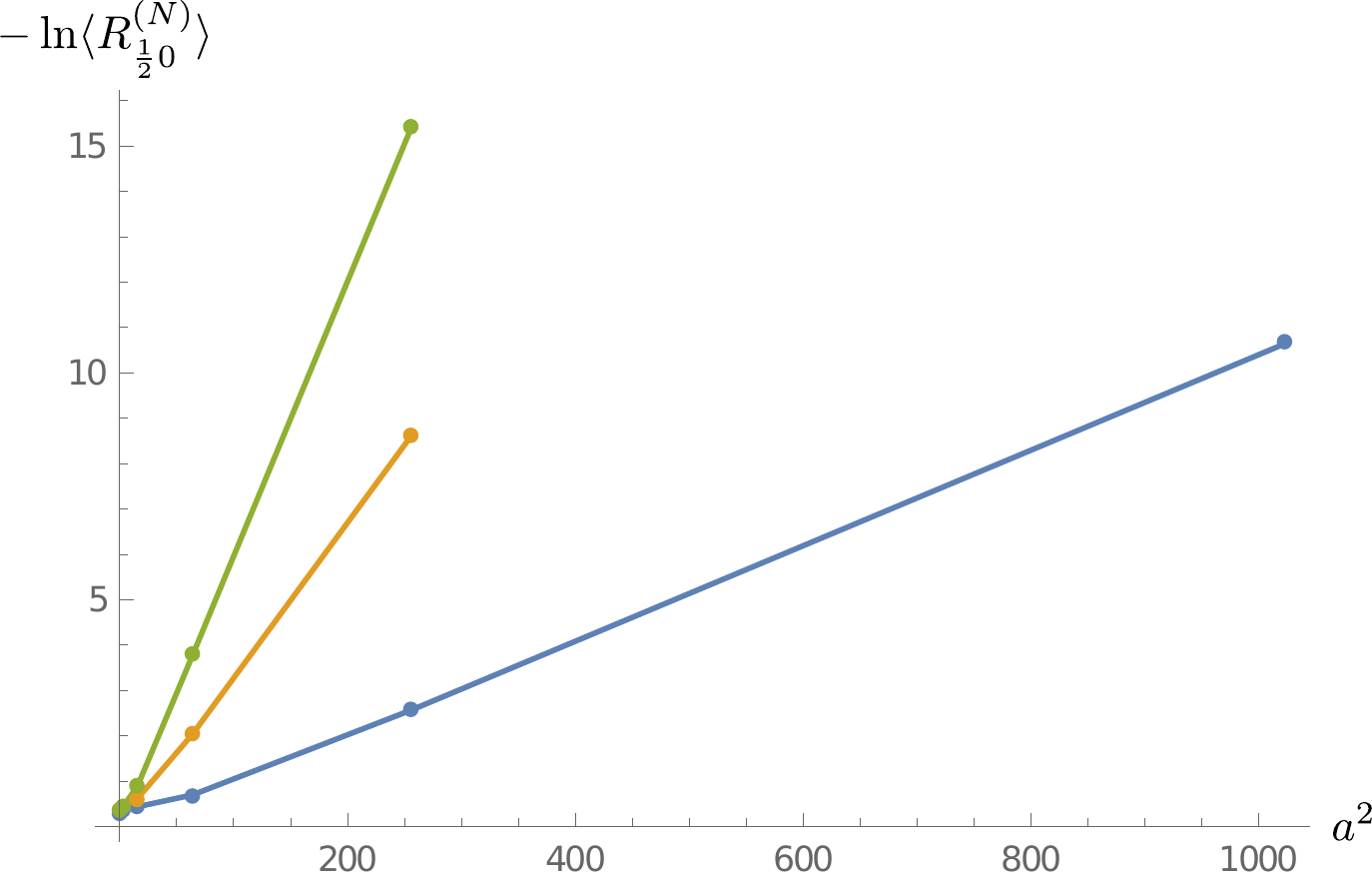}
  \caption{ We show $- \ln \langle R_N^{\tiny{\frac{1}{2}} 0} \rangle$ for different initial
  $g$ in the strong coupling phase over the area $N^2=a^2$ of the plaquette
  enclosed by the ribbon operator. From top to bottom, we have $g=1.2$, $g=1.19$
   and $g=1.18$.
  \label{fig:strong-coupling-area}
  }
\end{figure}

Similarly, we study the same expression in the strong coupling phase (Fig.~\ref{fig:strong-coupling-area}).
As expected, we observe a different behaviour: $- \ln \langle R^{\frac{1}{2} 0}_N \rangle$
increases linearly with the area $N^2$ enclosed by the ribbon operator. This illustrates the area law $\langle R^{\frac{1}{2} 0}_N \rangle \rangle \sim \exp\{-N^2\}$,  and thus the confining behaviour for the strong coupling regime.

\section{Discussion}\label{discussion}

In this article we presented a new tensor network renormalization algorithm which can be applied to $(2+1)$-dimensional lattice systems with (non-Abelian) gauge symmetry, and also used to study anyon condensation. This algorithm builds on the decorated tensor network algorithm~\cite{DecTNW}. The novel input here is that the tensor networks  encode the amplitudes of the system in the fusion basis. This produces an algorithm where observables which measure the magnetic and electric charges of a region are central, and they can be monitored throughout the coarse-graining process.  This is quite different from standard tensor network algorithms, in which the interpretation of the coarse-grained tensor network amplitudes requires additional work~\cite{GuWen}.

The fusion basis diagonalizes a hierarchically ordered set of these observables, and in this way it defines a coarse-graining scheme for the plaquettes of the basic and glued building blocks. Fusion basis transformation change this ordering, and thus reorganize the blocking of the degrees of freedom. The fusion basis transformations, therefore, play an essential part in the algorithm and function as pre-defined disentanglers.

A crucial advantage of employing the fusion basis, rather than the spin network basis~\cite{ DecTNW,DelcampSF,Milsted}, is that the fusion basis explicitly controls the electric charge excitations, which on the lattice level appear as violations of gauge invariance. Thus, we have constructed different versions of the coarse-graining algorithm. A first version allows electric charge excitations (and violations of gauge invariance) to appear on the coarse-grained lattice level. A second version truncates these electric charges. Yet, this only applies  to the charges of the effective plaquettes, and not to the electric charges which might appear on pairs of effective plaquettes, for instance.  This version is equivalent to a truncation to the gauge invariant Hilbert space after each coarse-graining step. A third version of the algorithm does not allow for any electric charges. This truncation reduces even the gauge invariant Hilbert space.

We have seen that the first and second version lead to almost identical results for the class of lattice gauge theories considered herein. The third version led to the accumulation of short range entanglement under coarse-graining. Because this version of the algorithm restricts the fusion basis transformations, we can take this as a hint that these transformation do indeed act as pre-defined disentanglers. Furthermore, one should rather keep the degrees of freedom encoded in the electric charges associated with coarser regions. That is, for Yang-Mills lattice gauge theory, it is safe to truncate back to the gauge invariant Hilbert space  after each coarse-graining step, but one should not further restrict this gauge invariant Hilbert space.

We have applied this coarse-graining algorithm to lattice gauge theory with the quantum-deformed structure group $\SU(2)_{\rm k}$. The level ${\rm k}$ in some sense determines the size of the structure group, and one regains $\SU(2)$ for ${\rm k}\rightarrow \infty$. Using this quantum deformation of $\SU(2)$, we can  work with a finite-dimensional system -- which is imperative for tensor network algorithms -- and, nevertheless, keep an exact gauge symmetry.  The coarse-graining algorithm allowed us to extract the critical couplings for a certain range of ${\rm k}$. Despite being able to access only a relatively small range of ${\rm k}$, we saw that the critical couplings do seem to approach zero surprisingly fast. Of course, it would be good to confirm this behaviour for larger  ${\rm k}$.  We have also found certain differences in the coarse-graining behaviour between the two different models, the half-integer and integer-only models, which we considered.  The fusion basis algorithm allowed us to extract in a straightforward way the scaling behaviour of the Wilson loops, and to monitor the appearance and relevance of the electrical charges.

In future work, we plan to apply this algorithm to study anyon condensation~\cite{Burnell} as well as transitions between (weak coupling limit) vacua for different ${\rm k}$~\cite{DittrichLambda}. The latter scenario can be interpreted as a phase transition between different values of the discretized cosmological constant.  Anyon condensation is much less understood than, e.g., lattice gauge theory, and systems with low ${\rm k}$ already exhibit a rich structure. The algorithm is ideally suited to track whether some anyons have condensed or not. These anyons are described by the magnetic and electric charges $\rho=(j,\overline{j})$ we employed. Condensation of an anyon type $\rho$ would be indicated by the singular values for this channel $\rho$ approaching a non-vanishing fixed point value. It would be also interesting to see whether there is a relation between the statistical properties of the $(2+1)$--dimensional  systems and their $(1+1)$--dimensional  descriptions~\cite{TopIntertwiner,QGIntertwiners}.

Tensor network algorithms rely on an explicit summing procedure, so they require considerable computing resources, and much more so for higher (than two-) dimensional systems.  Although the required memory and computational time scale polynomially with the system size (in our example with the level ${\rm k}$), the power with which it scales is quite large; see the discussion in Sec.~\ref{numcost}.  Here, we have only performed first tests of the algorithm. There are certainly a number of optimizations that can be applied, such as vectorization, multi-core parallelization on the CPU, and parallelization on multi-GPU systems.

Another possible direction to reach higher levels ${\rm k}$ is to simplify the current algorithm. One option is to use smaller building blocks, as illustrated in~\cite{DecTNW}. Another is offered by the central role observables take in our algorithm: we can aim to extract coarse-graining rules in terms of these observables. These would lead to much simpler coarse-graining descriptions, which could be compared with the results of the full algorithm, at least for small level ${\rm k}$.

Such simpler algorithms can then be suitable to treat four-dimensional systems, e.g., four-dimensional quantum gravity models, which are not amenable to Monte Carlo simulations. There are a number of long-standing open questions, e.g., the understanding of the large scale limit of spin foam models, the nature of the quantum gravity vacuum~\cite{Dittrich12,TimeEvol,DittrichReview14}, a possible restoration of diffeomorphism symmetry~\cite{Diff08,Improved,DittrichReview12, BahrSteinhausPRD}, and the fixing of free parameters in the models~\cite{DittrichMeasure, BahrSteinhausPRL}, which we hope can be answered by better coarse-graining algorithms.

\begin{acknowledgements}
The authors thank William Donnelly, Stefan K\"uhn, Karl Jansen, Aldo Riello and Andreas Wipf for discussions.
Research at Perimeter Institute is supported in part by the Government of Canada through the Department of Innovation, Science and Economic Development Canada and by the Province of Ontario through the Ministry of Economic Development, Job Creation and Trade. SSt is funded by the Deutsche Forschungsgemeinschaft (DFG, German Research Foundation) - Projektnummer / project number 422809950. Computations were made on the supercomputer Mammouth-MP2B from Universit\'e de Sherbrooke, managed by Calcul Qu\'ebec and Compute Canada. The operation of this supercomputer is funded by the Canada Foundation for Innovation (CFI), the Minist\`ere de l'\'Economie, de la Science et de l'Innovation du Qu\'ebec (MESI) and the Fonds de recherche du Qu\'ebec - Nature et technologies (FRQ-NT).
\end{acknowledgements}

\appendix

\section{$\SU(2)_{\rm k}$ basics}\label{app-basics}

We provide here some basic facts about the fusion category ${\rm SU}(2)_{\rm k}$. An extensive introduction can be found in~\cite{Yellowbook,Biedenharn}.
 ${\rm SU}(2)_{\rm k}$ can also be understood to arise from a deformation of the group $\SU(2)$ and its category of representations. The deformation parameter is given by a root of unity $q= \exp(2\pi \i /({\rm k}+2))$ where ${\rm k}$ is a positive integer.
With this deformation parameter, we define the quantum numbers
\be\label{qNumber}
[n]\,:=\,\frac{q^{n/2}-q^{-n/2}}{q^{1/2}-q^{-1/2}}=\frac{\sin\left(\tfrac{\pi}{{\rm k}+2}\,n\right)}{\sin\left(\tfrac{\pi}{{\rm k}+2}\right)},
\q
\forall\,n\in\mathbb{N} \q .
\ee

The irreducible objects  of the fusion category ${\rm SU}(2)_{\rm k}$ are given by the admissible irreducible and unitary representations (irreps) of the corresponding quantum-deformed group. These admissible irreps are labeled by ``spins'' $j\in\{0,1/2,1,\dots,{\rm k}/2\}$.

The quantum dimensions are given by $d_j=[2j+1]$. Admissible representations, that is, irreps from the set $\{0,1/2,1,\dots,{\rm k}/2\}$ have positive, non-vanishing quantum dimension.  We will often use the signed quantum dimensions
\ba
v_j^2\,:=\,(-1)^{2j} d_j
\ea
and their square roots $v_j$ (fixing once and for all one root).  The total quantum dimension is given by
\ba\label{qdimension}
{\cal D}&\,:=\,&\sqrt{\sum_jv_j^4}\,\,=\,\,\sqrt{\frac{{\rm k}+2}{2}}  \frac{1}{\sin\left(\frac{\pi}{{\rm k}+2}\right)}   \q .
\ea

 As with the group case, we can form the tensor product between $\SU(2)_{\rm k}$ representations. In the terms of the fusion category, this defines a ``fusion'' product.  Admissible triples are triples $(i,j,l)$ of irreps that include the trivial representation in their tensor product. Such triples $(i,j,l)$ are defined by the usual $\SU(2)$ coupling conditions
\be\label{Admissibility2}
i\leq j+l,
\q
j\leq i+l,
\q
l\leq i+j,
\q
i+j+l\in\mathbb{N},
\ee
as well as the $\SU(2)_{\rm k}$--specific condition
\be
i+j+l\leq{\rm k} \q .
\ee
The fusion symbol  $N^l_{ij}=\delta_{ijl}$ is equal to one if $(i,j,l)$ is an admissible triple and vanishes otherwise.

The data of a fusion category can be used to define a Hilbert space of graph-based states~\cite{ExTQFT}.
States in this Hilbert space are given by superpositions of labelled three-valent graphs embedded in a surface with fixed topology, which here will be a punctured sphere. The strands of the graphs are labelled by the (irreducible) objects of the category, that is, by the spins $j$. The labels of three strands meeting at a vertex have to satisfy the coupling conditions~(\ref{Admissibility2}).

One imposes a number of equivalences between the graph states. One of these equivalence relations is given by the F-move
\ba\label{FmoveApp1}
  \includegraphics[scale=0.5,valign=c]{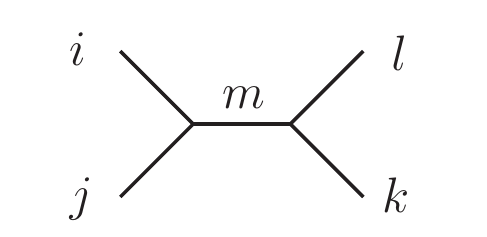}
\,&=&\,\,\sum_n F^{ijm}_{kln}     \includegraphics[scale=0.5,valign=c]{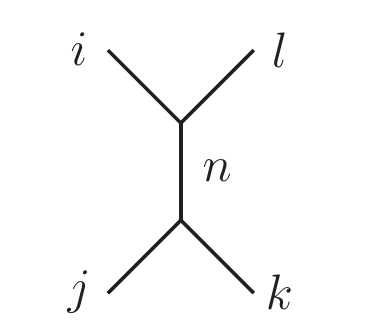} \q ,\\
\ea
where we specify the $F$-symbol in Sec.~\ref{FRS}.
Additionally, we have the equivalence
\ba\label{eq31}
 \includegraphics[scale=0.5,valign=c]{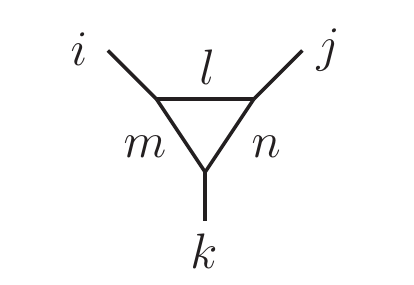}
&=&\frac{v_nv_l}{v_j}F^{ikj}_{nlm} \includegraphics[scale=0.5,valign=c]{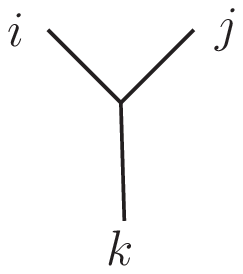} \q .
\ea
Note that this is only an equivalence if the triangular loop of the graph on the left side of~(\ref{eq31}) does not include a puncture. We indicate such punctures with fat black dots. The set of equivalences also allow us to deform strands of the graph state, as long as one does not move these strands over a puncture. But one can move a given strand over or under another strand. Furthermore, one can omit strands which carry a $j=0$ label, but also add such strands freely to a graph state.

The strands are also allowed to cross each other, but one does need to differentiate between under-crossings and over-crossings. Such crossings can be resolved using a so-called braiding, which is defined by
\be\label{braiding}
  \includegraphics[scale=0.5,valign=c]{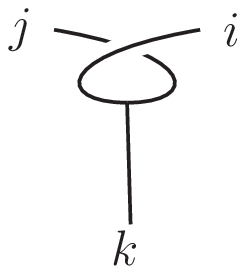}
=\,\,R^{ij}_k \includegraphics[scale=0.5,valign=c]{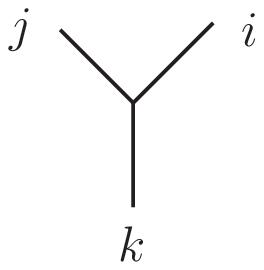},
\q\q\q\q
 \includegraphics[scale=0.5,valign=c]{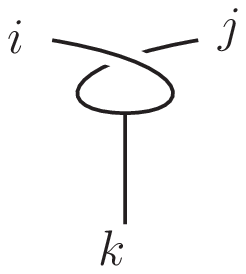}=\,\,\big(R^{ij}_k\big)^*\includegraphics[scale=0.5,valign=c]{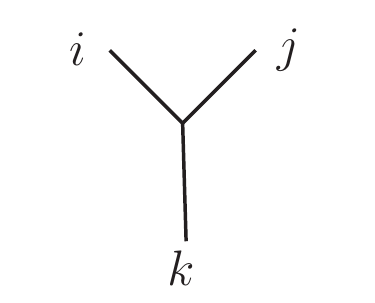}.
\ee
where we specify the $R$-matrix below.  From these definitions, one can derive the following equivalences
\ba\label{resolution of crossing}
  \includegraphics[scale=0.5,valign=c]{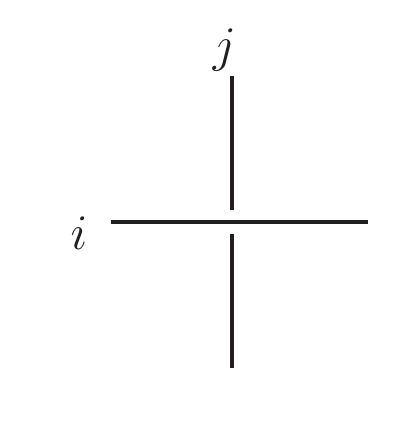}\,=\,\,\,\sum_k\frac{v_k}{v_iv_j}R^{ij}_k  \includegraphics[scale=0.5,valign=c]{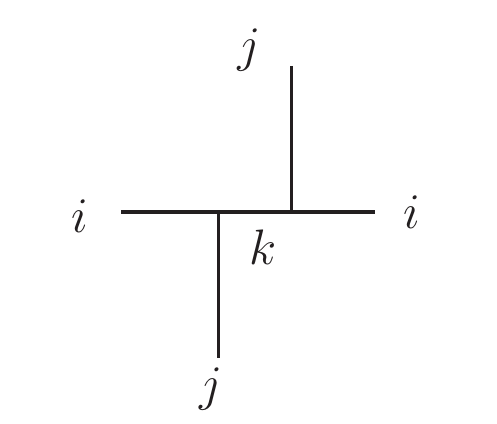},
\q\q\q
  \includegraphics[scale=0.5,valign=c]{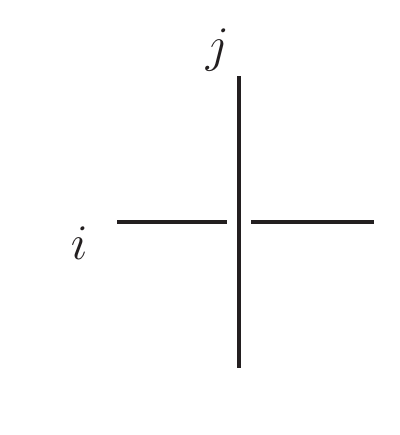}\,=\,\,\,\sum_k\frac{v_k}{v_iv_j}\big(R^{ij}_k\big)^*  \includegraphics[scale=0.5,valign=c]{Drawings/Uncrossing}.\q\q
\ea

Allowing over- and under-crossings, we can have also a strand encircling another strand.  Resolving the crossings according to~(\ref{resolution of crossing}), one finds the equivalence
\ba\label{smatrix1}
 \includegraphics[scale=0.5,valign=c]{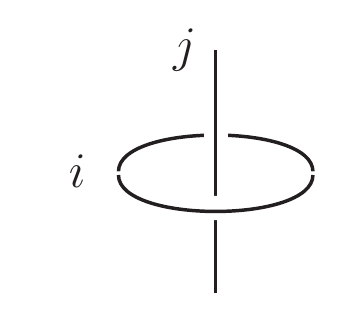}=\,\, \frac{S_{ij}}{S_{0j}} \includegraphics[scale=0.5,valign=c]{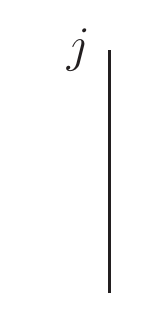}\q .
\ea
where the $S$-matrix is given by
\ba\label{Smatrix1}
S_{ij}\,=\,\frac{1}{\cal D}\sum_l v_l^2R^{ij}_lR^{ji}_l \q .
\ea

Another graphical notation which we need is the vacuum line defined as
\be\label{vacuums}
 \includegraphics[scale=0.5,valign=c]{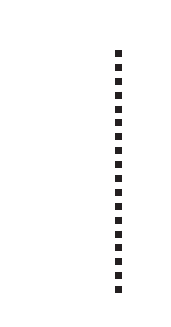}\,:=\, \,\frac{1}{\cal D} \sum_j v_j^2 \includegraphics[scale=0.5,valign=c]{Drawings/VLine-j}.
\ee

A vacuum loop surrounding a puncture allows us to pull a strand across this puncture:
\ba\label{sliding}
\includegraphics[scale=0.5,valign=c]{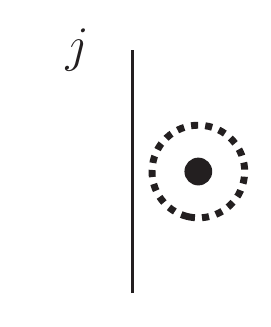}\,\,\,=\,\includegraphics[scale=0.5,valign=c]{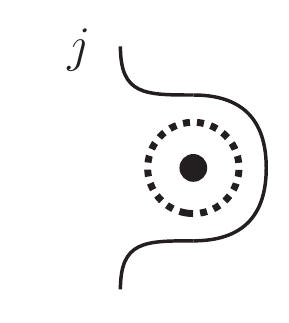}.
\ea
This identity can be proven using the definition of the vacuum line as well as the F-move equivalence.

\subsection{Explicit definition of the $F$-symbol, $R$-matrix and $S$-matrix}\label{FRS}

 To define the $F$--symbols we first introduce for any admissible triple $(i,j,k)$ the quantity
\be
\Delta(i,j,k)\,:=\,\delta_{ijk}\sqrt{\frac{[i+j-k]![i-j+k]![-i+j+k]!}{[i+j+k+1]!}},
\ee
where $[n]!:=[n][n-1]\dots[2][1]$.

 The (Racah--Wigner) quantum $\{6j\}$-symbol is then given by the formula
\ba
\left\{
\begin{array}{ccc}
i&j&m\\
k&l&n
\end{array}\right\}
&:=&\Delta(i,j,m)\Delta(i,l,n)\Delta(k,j,n)\Delta(k,l,m)\sum_z(-1)^z[z+1]!\nonumber\\
&&\times\frac{\Big([i+j+k+l-z]![i+k+m+n-z]![j+l+m+n-z]!\Big)^{-1}}{[z-i-j-m]![z-i-l-n]![z-k-j-n]![z-k-l-m]!},\q
\ea
where the sum runs over
\be
\max(i\!+\!j\!+\!m,i\!+\!l\!+\!n,k\!+\!j\!+\!n,k\!+\!l\!+\!m)\leq z\leq\min(i\!+\!j\!+\!k\!+\!l,i\!+\!k\!+\!m\!+\!n,j\!+\!l\!+\!m\!+\!n).
\ee
The $F$--symbols can then be defined as
\be\label{FDefinition}
F^{ijm}_{kln}
\,:=\,
(-1)^{i+j+k+l} \,\sqrt{[2m+1][2n+1]}\,
\left\{
\begin{array}{ccc}
i&j&m\\
k&l&n
\end{array}\right\}.
\ee

The $F$-symbol satisfies a number of consistency conditions and properties, which include
\begin{subequations}
\ba
\text{Physicality:}\q&F^{ijm}_{kln}\,=\,F^{ijm}_{kln}\delta_{ijm}\delta_{iln}\delta_{kjn}\delta_{klm},\\
\text{Tetrahedral symmetry:}\q&F^{ijm}_{kln}\,=\,F^{jim}_{lkn}\,=\,F^{lkm}_{jin}\,=\,F^{imj}_{knl}\frac{v_mv_n}{v_jv_l},\\
\text{Orthogonality:}\q&\sum_nF^{ijm}_{kln}F^{ijp}_{kln}\,=\,\delta_{mp}\delta_{ijm}\delta_{klm},\label{F-orthogonality}\\
\text{Reality:}\q&\big(F^{ijm}_{kln}\big)^*\,=\,F^{ijm}_{kln},\\
\ea
\end{subequations}

Furthermore, we need the $R$--matrix, which for  ${\rm SU}(2)_{\rm k}$  is given by
\ba
R^{ij}_k&=&(-1)^{k-i-j}\,\left(q^{k(k+1)-i(i+1)-j(j+1)}\right)^{1/2}  \q .
\ea

This gives for the $S$-matrix
\ba\label{s matrix in terms of R}
S_{ij}\,=\,=\,\frac{1}{\cal D}\sum_l v_l^2R^{ij}_lR^{ji}_l\,=\,\frac{(-1)^{2(i+j)}}{\cal D}[(2i+1)(2j+1)].
\ea

The $S$-matrix for ${\rm SU}(2)_{\rm k}$ is invertible and unitary, making $\SU(2)_{\rm k}$ into a modular fusion category. Note that the $S$--matrix is also real and symmetric:
\be\label{s matrix identities}
S_{ij}=S_{ji},
\q\q
\sum_lS_{il}S_{lj}=\delta_{ij}\q .
\ee

\section{On the $\SU(2)_{\rm k}$ fusion basis} \label{app-FB}

The equivalences which are imposed on the graph states make the task of finding a basis of independent states more involved. One such basis is the fusion (or Ocneanu) basis. One motivation for this basis arises from considering two-punctured spheres, which are topologically equivalent to tubes. These tubes can be glued to each other, yielding another tube. The gluing can be extended to graph states defined on these tubes. Moreover, a tube can be glued to any puncture of an $N$-punctured surface, giving again an $N$-punctured surface. Now, if one chooses a fusion basis state for the tube, this gluing turns out to define a projection operator onto the subspace of states that around the given puncture do agree with the fusion basis state of the tube. This allows to interpret the fusion basis as describing excitations which are  localized at the punctures~\cite{Lan:2013wia}.

In the following, we give a translation of the labelled fusion tree into the graph states that define the fusion basis (for an $N$-punctured sphere). For more details and motivation, see~\cite{Koenig, ExTQFT}.

The leaves of the fusion tree translate as follows into parts of graph states:
\ba
\includegraphics[scale=0.75,valign=c]{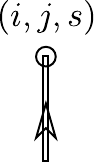} \; = \; \includegraphics[scale=0.58,valign=c]{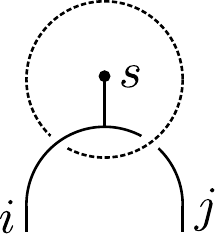} \q , \q\q\q\q\q
\includegraphics[scale=0.75,valign=c]{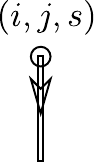} \; = \; \includegraphics[scale=0.58,valign=c]{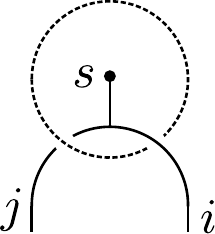} \q .
\ea
Here, (to simplify notion, in a slight difference to the main text) we use leaf labels $\rho=(i,j,s)$ and branches labels $\sigma=(a,b)$.

The three-valent vertices of the fusion tree denote the following parts of a fusion basis state
\ba
\includegraphics[scale=0.75,valign=c]{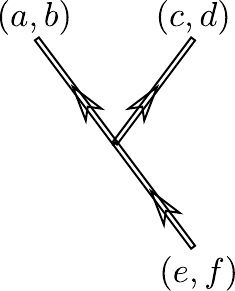} \; = \; \includegraphics[scale=0.59,valign=c]{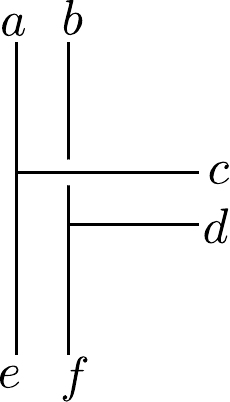} \q , \q\q\q\q\q
\includegraphics[scale=0.75,valign=c]{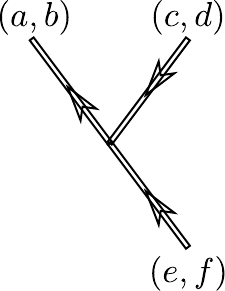} \; = \; \includegraphics[scale=0.59,valign=c]{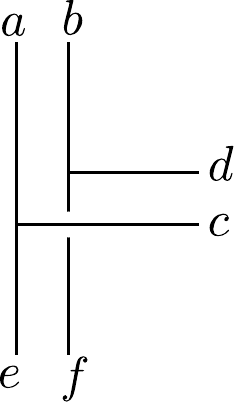} \q .
\ea

\subsection{Fusion tree transformations}

Here we provide the derivation of the fusion basis transformations~(\ref{Fmove}) and~(\ref{eq:pulling_trafo}). We start with the F-move
\begin{equation}
  \includegraphics[scale=0.5,valign=c]{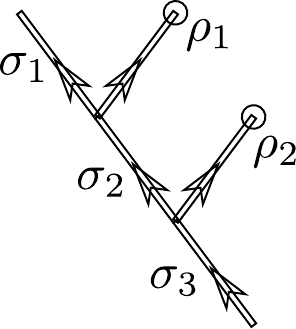} =\sum_{\sigma_4}
 \mathbb{F}^{\rho_1 \, \sigma_1 \, \sigma_2}_{\sigma_3 \, \rho_2 \, \sigma_4}
  \; \includegraphics[scale=0.5,valign=c]{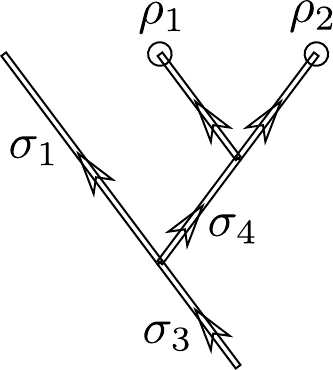} \; .
\end{equation}
where $\sigma_1=(a_1,b_1), \sigma_2=(a_2,b_2),\sigma_3=(a_3,b_3)$ and $\rho_1=( i_1,j_1,s_1),\rho_2=(i_2,j_2,s_2)$ as well as $\sigma_4=(k_1,l_1)$. With these definitions we have
\ba
\mathbb{F}^{\rho_1 \, \sigma_1 \, \sigma_2}_{\sigma_3 \, \rho_2 \, \sigma_4} \,=\, F^{i_1 \, a_1 \, a_2}_{a_3 \, i_2 \, k_1} \, F^{j_1 \, b_1 \, b_2}_{b_3 \, j_2 \, l_1} \q .
\ea

This transformation follows from the following sequence of equivalences between graph states, which involve deformations of strands, including moving strands across other strands, as well as F-moves (where some strands carry a $j=0$ label and are therefore not drawn):
\begin{flalign*}
   & \includegraphics[scale=0.6,valign=c]{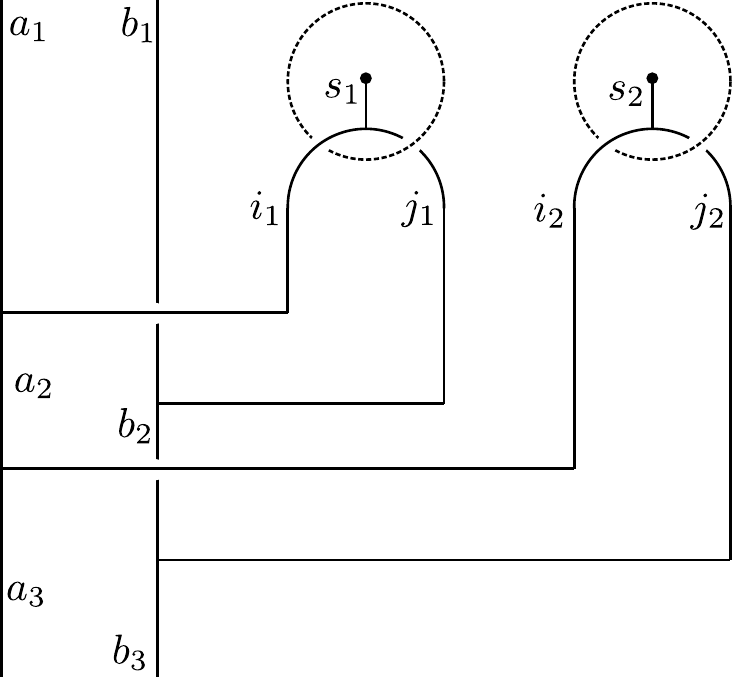} \; = \;
  \includegraphics[scale=0.6,valign=c]{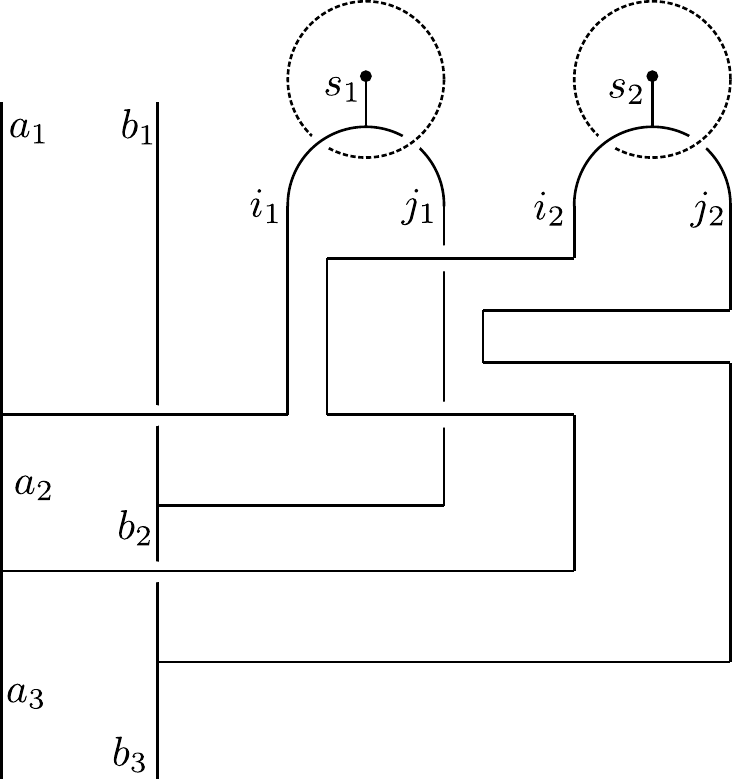} \; = \;
  \sum_{k_1,l_1} \frac{v_{k_1} v_{l_1}}{v_{i_1} v_{i_2} v_{j_1} v_{j_2}} \;
   \includegraphics[scale=0.6,valign=c]{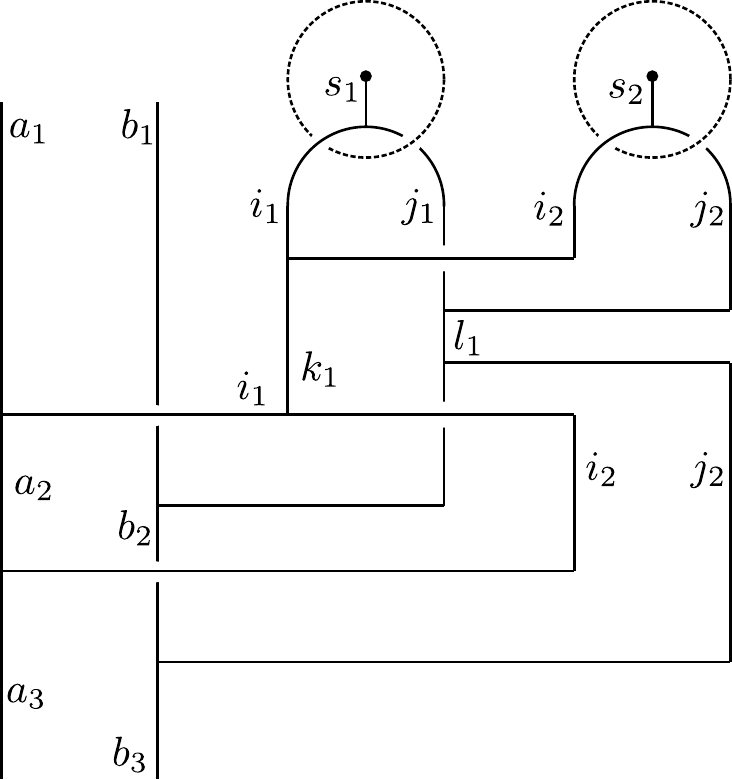} &&
\end{flalign*}
\begin{flalign*}
   & = \; \sum_{k_1,l_1} \frac{v_{k_1} v_{l_1}}{v_{i_1} v_{i_2} v_{j_1} v_{j_2}} \;
   \includegraphics[scale=0.6,valign=c]{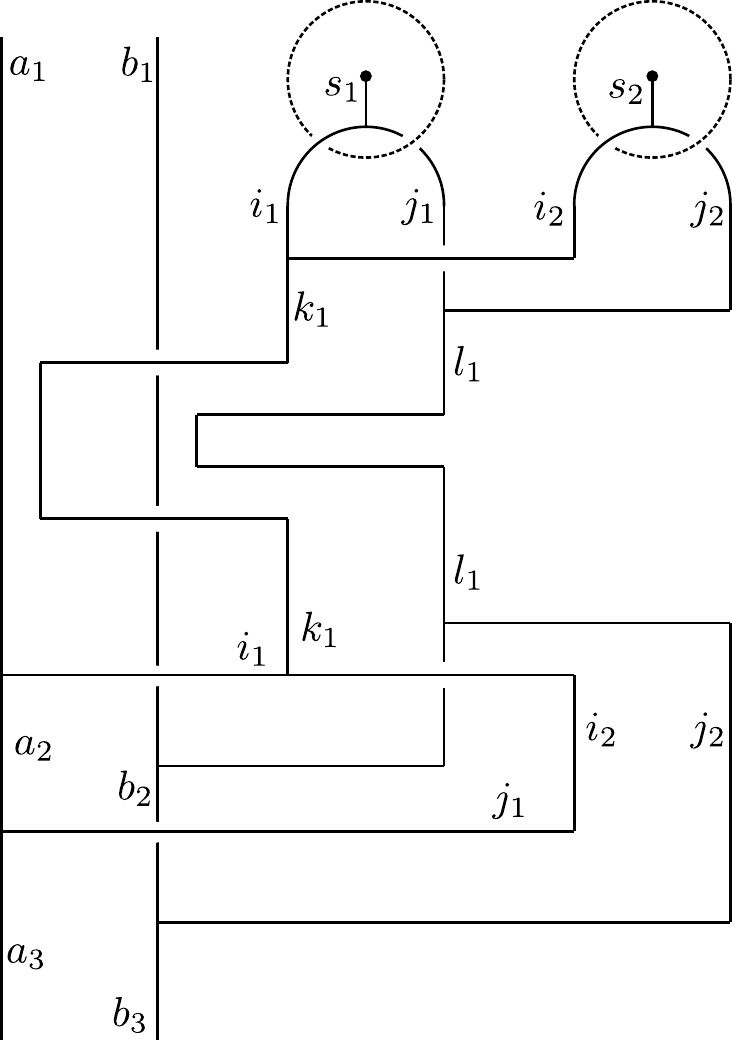} \; = \;
    \sum_{k_1,l_1,k_2,l_2} \frac{v_{k_2} v_{l_2}}{v_{i_1} v_{i_2} v_{a_1} v_{j_1} v_{j_2} v_{b_1}} \;
    \includegraphics[scale=0.6,valign=c]{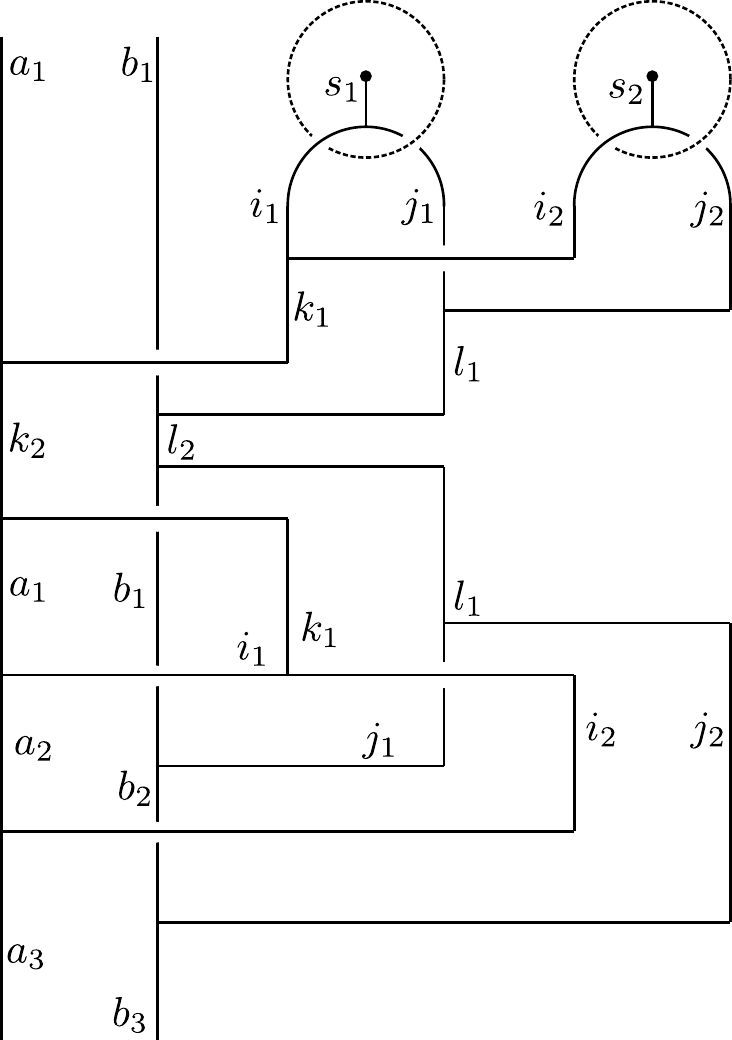} \, . &&
\end{flalign*}
We proceed by simplifying the lower part of the graph state (using also 3-1 moves~(\ref{eq31})), and thus will only depict this part:
\begin{flalign*}
  & \includegraphics[scale=0.6,valign=c]{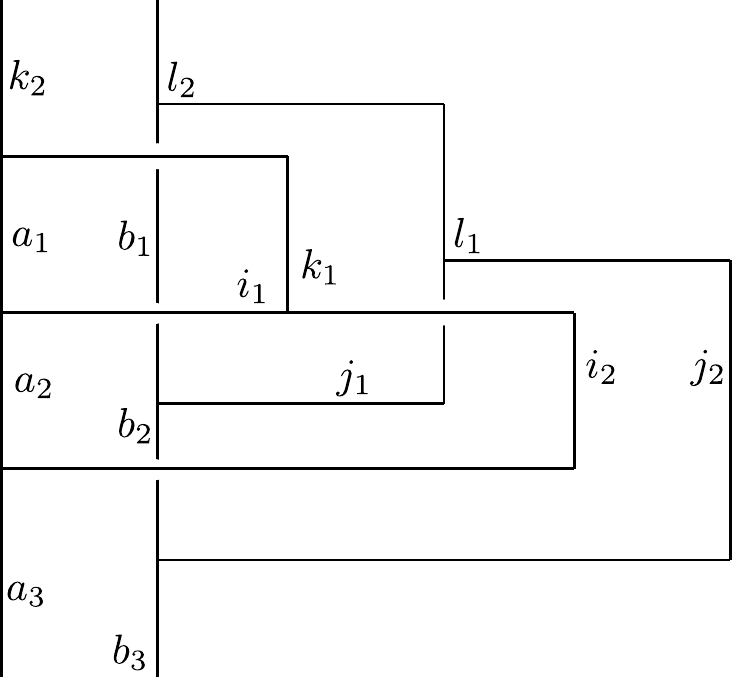} \; = \;
  \includegraphics[scale=0.6,valign=c]{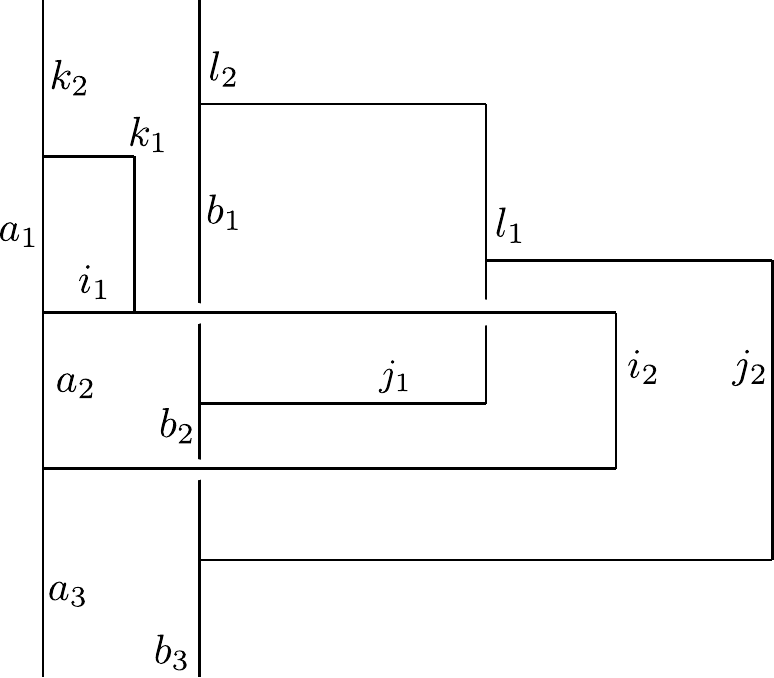} \; = \;
  \includegraphics[scale=0.6,valign=c]{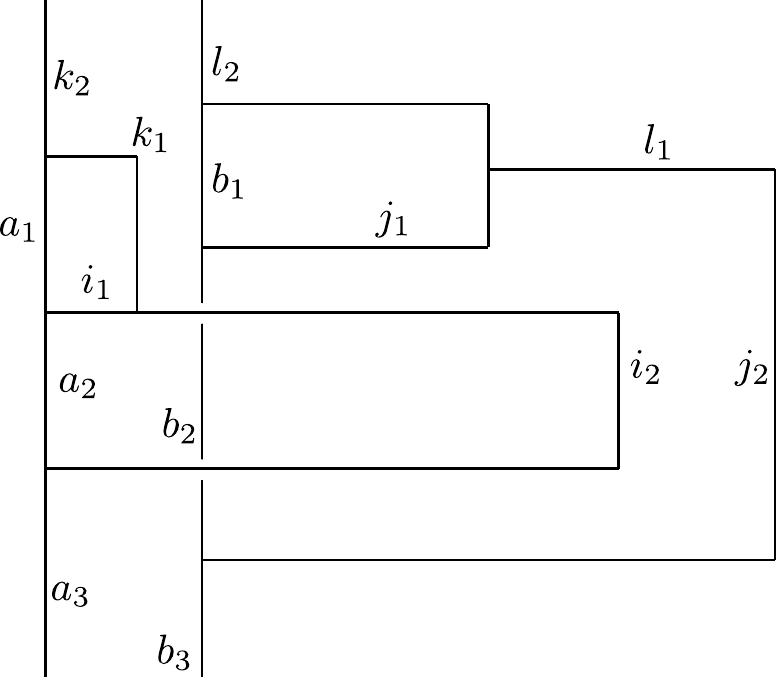} &&
 \end{flalign*}%\nonumber \\
 \begin{flalign*}
  & = \; \includegraphics[scale=0.6,valign=c]{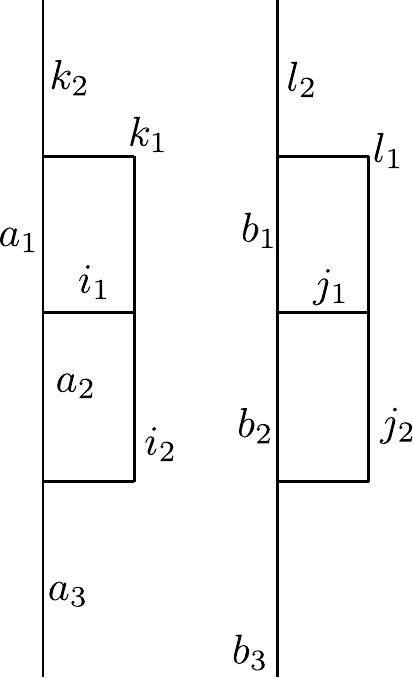} \; = \;
  \sum_{k_3,l_3} F^{a_1 \, a_2 \, i_1}_{i_2 \, k_1 \, k_3} F^{b_1 \, b_2 \, j_1}_{j_2 \, l_1 \, l_3}
   \; \includegraphics[scale=0.6,valign=c]{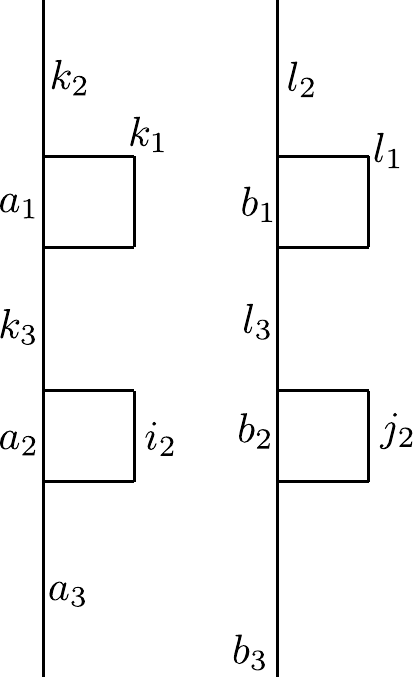} &&
\end{flalign*}
\begin{flalign*}
  & = \;  \sum_{k_3,l_3} \frac{v_{a_1} v_{k_1} v_{a_2} v_{i_2}}{v_{k_2} v_{k_3}}
   \frac{v_{b_1} v_{l_1} v_{b_2} v_{j_2}}{v_{l_2} v_{l_3}} \delta_{k_2,k_3} \delta_{k_2,a_3}
   \delta_{l_2,l_3} \delta_{l_2,b_3} \delta_{k_2 a_1 k_1} \delta_{k_3 a_2 i_2}
   \delta_{l_2 b_1 l_1} \delta_{l_3 b_2 j_2}
   F^{a_1 \, a_2 \, i_1}_{i_2 \, k_1 \, k_3} F^{b_1 \, b_2 \, j_1}_{j_2 \, l_1 \, l_3} \;
   \includegraphics[scale=0.6,valign=c]{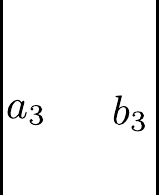} &&
   \end{flalign*} %\nonumber \\
   \begin{flalign*}
   & = \;  \frac{v_{a_1} v_{i_1} v_{i_2}}{v_{a_3}}
    \frac{v_{b_1} v_{j_1} v_{j_2}}{v_{b_3}} \delta_{a_3 a_1 k_1} \delta_{a_3 a_2 i_2}
    \delta_{b_3 b_1 l_1} \delta_{b_3 b_2 j_2}
    F^{i_1 \, a_1 \, a_2}_{a_3 \, i_2 \, k_1}
    F^{j_1 \, b_1 \, b_2}_{b_3 \, j_2 \, l_1} \;
    \includegraphics[scale=0.6,valign=c]{Drawings/fusion_basis_changes11} \; && \end{flalign*}%\nonumber \\
    \begin{flalign}
     = \;
    \frac{v_{a_1} v_{i_1} v_{i_2}}{v_{a_3}}
     \frac{v_{b_1} v_{j_1} v_{j_2}}{v_{b_3}} \delta_{k_2,a_3} \delta_{l_2,b_3}
     F^{i_1 \, a_1 \, a_2}_{a_3 \, i_2 \, k_1}
     F^{j_1 \, b_1 \, b_2}_{b_3 \, j_2 \, l_1} \;
     \includegraphics[scale=0.6,valign=c]{Drawings/fusion_basis_changes11} \, . &&
\end{flalign}
In summary, we obtain
\begin{flalign*}
  & \includegraphics[scale=0.6,valign=c]{Drawings/fusion_basis_changes1} &&
  \end{flalign*}% \nonumber \\
  \begin{flalign*}
 & \; =   \sum_{k_1,l_1,k_2,l_2} \frac{v_{k_2} v_{l_2}}{v_{i_1} v_{i_2} v_{a_1} v_{j_1} v_{j_2} v_{b_1}} \;
    \frac{v_{a_1} v_{i_1} v_{i_2}}{v_{a_3}}
     \frac{v_{b_1} v_{j_1} v_{j_2}}{v_{b_3}} \delta_{k_2,a_3} \delta_{l_2,b_3}
     F^{i_1 \, a_1 \, a_2}_{a_3 \, i_2 \, k_1}
     F^{j_1 \, b_1 \, b_2}_{b_3 \, j_2 \, l_1} \;
    \includegraphics[scale=0.6,valign=c]{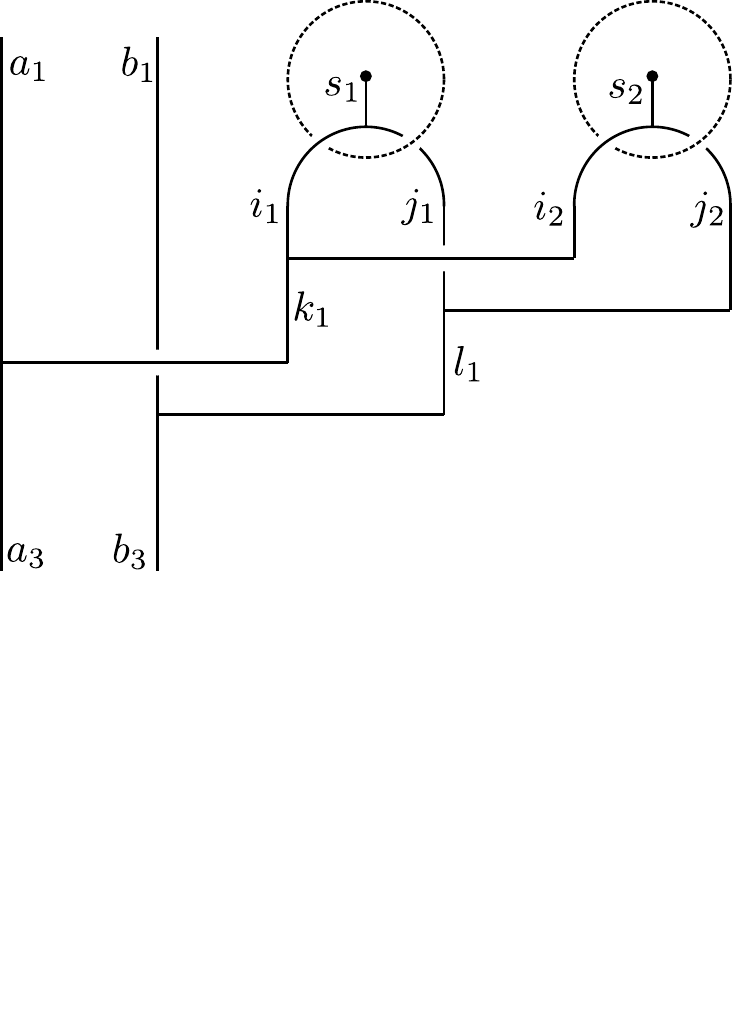} &&
    \end{flalign*} %\nonumber \\
    \begin{flalign}
    & \; =  \sum_{k_1,l_1}
        F^{i_1 \, a_1 \, a_2}_{a_3 \, i_2 \, k_1} \;
        F^{j_1 \, b_1 \, b_2}_{b_3 \, j_2 \, l_1} \;
       \includegraphics[scale=0.6,valign=c]{Drawings/fusion_basis_changes12} \; . &&
\end{flalign}

Additionally, we need to derive the R-move~\eqref{eq:pulling_trafo}:
\begin{equation}
  \includegraphics[scale=0.5,valign=c]{Drawings/pulling_trafo} =
   \mathbb{R}^{\rho_1 \sigma_1}_{\sigma_2}
  \includegraphics[scale=0.5,valign=c]{Drawings/pulling_trafo_2} \; .
\end{equation}
where with $\sigma_1=(a_1,b_1),\sigma_2=(a_2,b_2)$ and $\rho_1=(i_1,j_1,s_1)$, we have $ \mathbb{R}^{\rho_1 \sigma_1}_{\sigma_2}=  \left(R^{i_1 a_1}_{a_2}\right)^* R^{j_1 b_1}_{b_2}$.

We start by using the deformation of strands as well as~(\ref{sliding}). This equivalence allows us to pull strands over punctures, which are surrounded by a vacuum line. Next, we use~(\ref{resolution of crossing}) to resolve crossings, as well as the 3-1 move (where some labels are $j=0$):
\begin{flalign*}
  & \includegraphics[scale=0.6,valign=c]{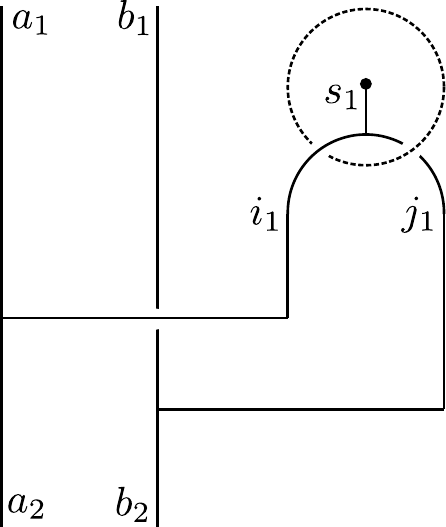} \; = \;
  \includegraphics[scale=0.6,valign=c]{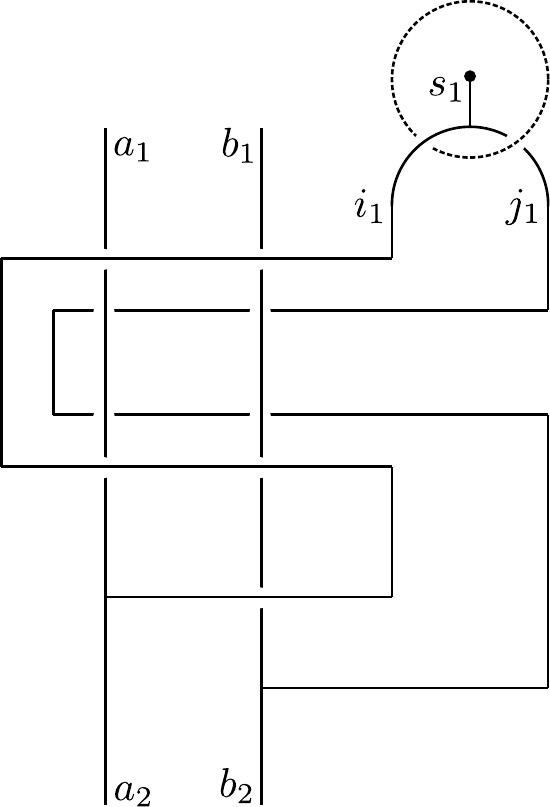} \; = \;
  \includegraphics[scale=0.6,valign=c]{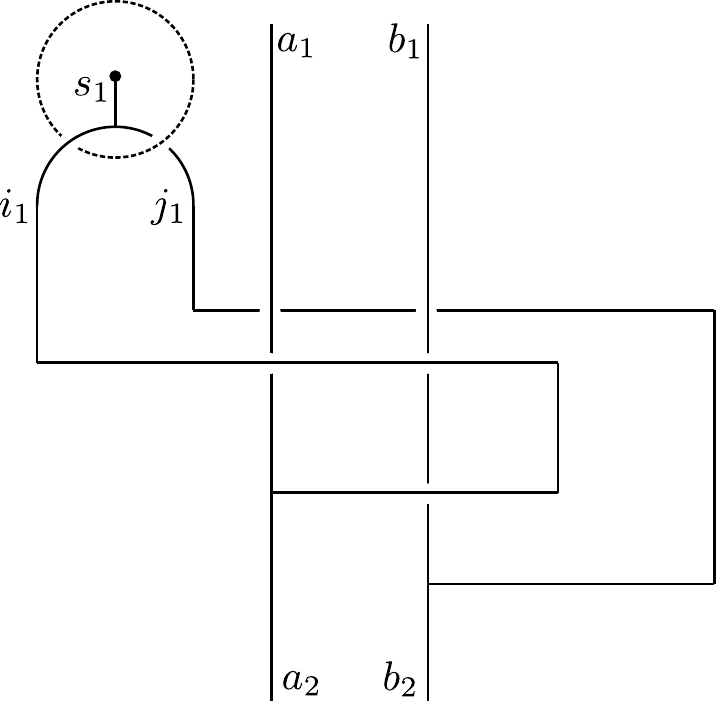} &&
  \end{flalign*}%\nonumber \\
  \begin{flalign*}
  & = \;  \sum_{k_1,l_1} \frac{v_{k_1} v_{l_1}}{v_{a_1} v_{i_1} v_{b_1} v_{j_1}}
  \includegraphics[scale=0.6,valign=c]{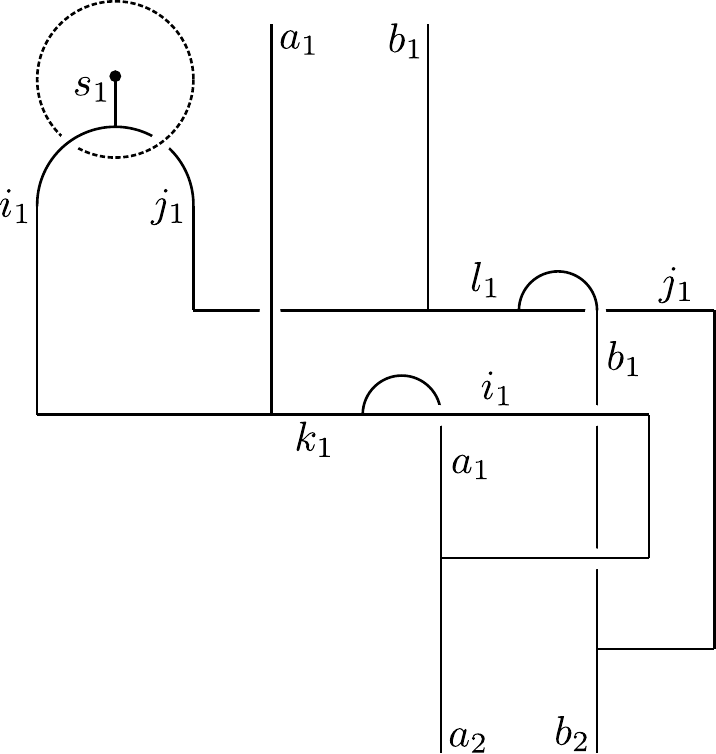} \; = \;
  \sum_{k_1,l_1} \frac{v_{k_1} v_{l_1}}{v_{a_1} v_{i_1} v_{b_1} v_{j_1}}
  ({R}^{i_1 a_1}_{k_1})^*{R}^{j_1 b_1}_{l_1}
  \includegraphics[scale=0.6,valign=c]{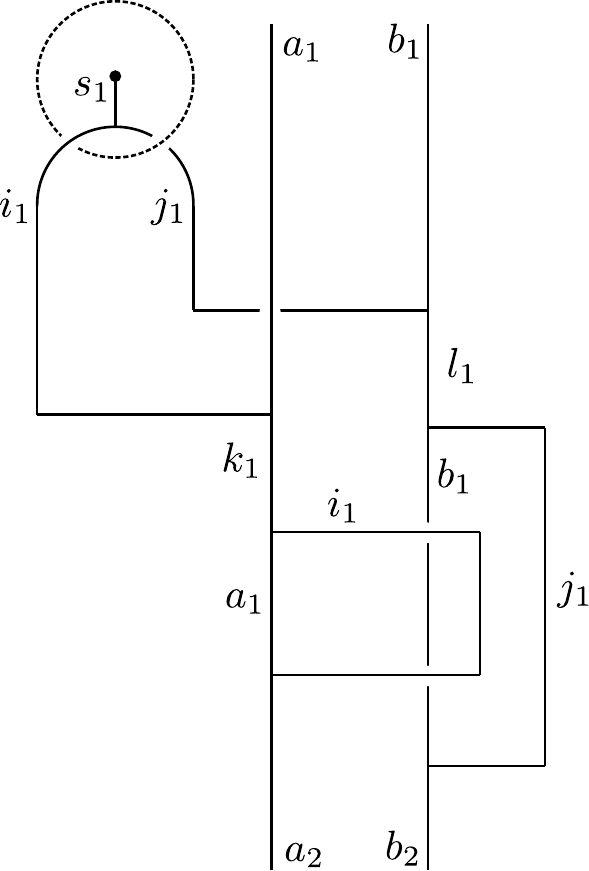} &&
\end{flalign*}
 % \nonumber \\
 \begin{flalign*}
  & = \;  \sum_{k_1,l_1} \frac{v_{k_1} v_{l_1}}{v_{a_1} v_{i_1} v_{b_1} v_{j_1}}
  ({R}^{i_1 a_1}_{k_1})^* {R}^{j_1 b_1}_{l_1}
  \includegraphics[scale=0.6,valign=c]{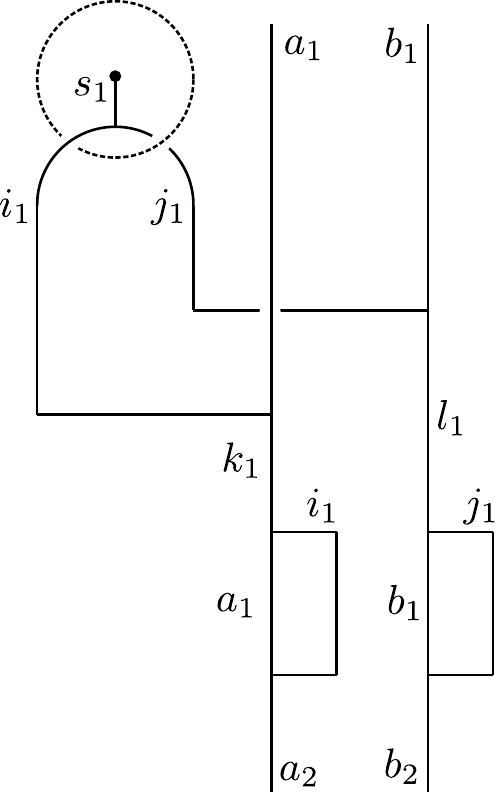} &&
  \end{flalign*}% \nonumber \\
  \begin{flalign*}
  & = \;  \sum_{k_1,l_1} \frac{v_{k_1} v_{l_1}}{v_{a_1} v_{i_1} v_{b_1} v_{j_1}}
  \frac{v_{a_1} v_{i_1}}{v_{a_2}} \frac{v_{b_1} v_{j_1}}{v_{b_2}} \delta_{k_1,a_2}
  \delta_{l_1,b_2} \delta_{a_1 i_1 a_2} \delta_{b_1 j_1 b_2}
  ({R}^{i_1 a_1}_{k_1})^*{R}^{j_1 b_1}_{l_1} && \q .
  \end{flalign*}
   Thus we have
  \begin{equation}
 % \implies \;
  \includegraphics[scale=0.6,valign=c]{Drawings/fusion_basis_pulling1}
  = \;  ({R}^{i_1 a_1}_{k_1})^* {R}^{j_1 b_1}_{l_1}
  \includegraphics[scale=0.6,valign=c]{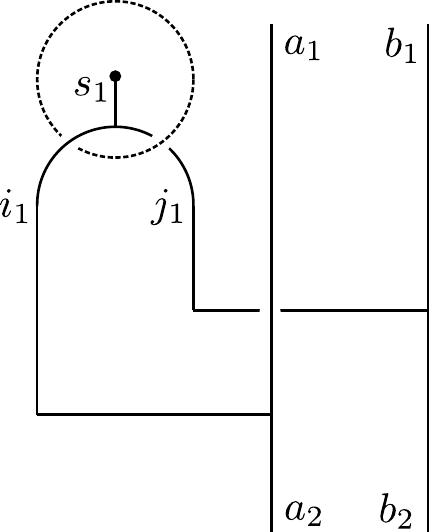} \q .
\end{equation}

\bibliography{biblio}

\end{document}